\documentclass[aps,prb,twocolumn,superscriptaddress,amsmath,amssymb,showkeys,longbibliography,english]{revtex4-2}

\usepackage{graphicx}
\usepackage{epstopdf}
\usepackage{dcolumn}
\usepackage{bm}
\usepackage{color} 
\usepackage{ulem}
\usepackage{soul}
\usepackage{xspace}
\usepackage{multirow}
\usepackage[hypertexnames=false]{hyperref}
\usepackage{physics}
\usepackage{here}
\usepackage{titlesec}
\usepackage{bibunits}
\usepackage[caption=false]{subfig}

\newcommand{\AppliedPhys}{Department of Applied Physics, University of Tokyo, Tokyo 113-8656, Japan}
\newcommand{\RCAST}{Research Center for Advanced Science and Technology, University of Tokyo, Tokyo 153-8904, Japan}
\newcommand{\RIKEN}{RIKEN Center for Emergent Matter Science, 2-1 Hirosawa, Wako, 351-0198, Japan}

\titleformat*{\section}{\normalsize\bfseries\sffamily\large\raggedright}
\titleformat*{\subsection}{\normalsize\bfseries\sffamily\raggedright}
\renewcommand\thesection{\arabic{section}}
\renewcommand\thesubsection{\arabic{subsection}}

\makeatletter
    \def\@seccntformat#1{\@ifundefined{#1@cntformat}%
       {\csname the#1\endcsname\space}
       {\csname #1@cntformat\endcsname}}
    \def\section@cntformat{\thesection.\space}
    \def\subsection@cntformat{\thesection.\thesubsection.\space} 
\makeatother

\renewcommand{\figurename}{\textsf{\textbf{Figure }}}
\renewcommand{\thefigure}{\textsf{\textbf{{\arabic{figure}}}}}\renewcommand{\theHfigure}{\arabic{figure}}

\def\partitle#1{\medskip \noindent {\textsf{\textbf{#1}}} }

\begin{document}
\begin{bibunit}

\title{\textsf{Interstitial-Electron-Induced Topological Molecular Crystals}}

\author{Tonghua Yu}             \affiliation{\AppliedPhys} 
\author{Ryotaro Arita}          \email{arita@riken.jp}   \affiliation{\RCAST} \affiliation{\RIKEN} 
\author{Motoaki Hirayama}       \email{hirayama@ap.t.u-tokyo.ac.jp}   \affiliation{\AppliedPhys} \affiliation{\RIKEN}

\begin{abstract}
\textsf{\textbf{Topological phases usually are unreachable in molecular solids, which are characteristic of weakly dispersed energy bands with a large gap, in contrast to topological materials. In this work, however, we propose that nontrivial electronic topology may ubiquitously emerge in a class of molecular crystals that contain interstitial electronic states, the bands of which are prone to be inverted with those of molecular orbitals. We provide guidelines hunt for such interstitial-electron-induced topological molecular crystals, especially in the topological insulating state. They exhibit a variety of exceptional qualities, as brought about by the intrinsic interplay of molecular crystals, interstitial electrons, and topological nature: (1) They may host cleavable surfaces along multiple orientations, with pronounced topological boundary states free from dangling bonds. (2) Strong response to moderate mechanical perturbations, whereby topological phase transition would occur under relatively low pressure. (3) Inherent high-efficiency thermoelectricity as jointly contributed by the non-parabolic band structure (therewith high thermopower), highly mobile interstitial electrons (high electrical conductivity), and soft phonons (small lattice thermal conductivity). (4) Ultralow work function owing to the active interstitial electrons. We utilize first-principles calculations to demonstrate these properties with the representative candidate K$_4$Ba$_2$[SnBi$_4$]. Our work suggests a pathway of realizing topological phases in bulk molecular systems, which may advance the interdisciplinary research between topological and molecular materials.}}
\end{abstract}


\maketitle

\section{Introduction}
\label{sec-intro}
The peculiarity of molecular crystals lies in the building blocks, which are primarily discrete zero-dimensional (0D) molecules instead of isolated atoms, bound together by the comparatively weak noncovalent interactions (e.g., van der Waals bonding for neutral molecules, ionic bonding for charged molecules) \cite{kitaigorodsky2012molecular, kronik2016excited}. Quintessential examples include fullerides \cite{claridge2009cluster}, 0D halide perovskites \cite{yin2017molecular, ju2018zero}, and versatile organic crystals \cite{smits2008bottom}. In comparison with atoms whose internal characteristics cannot be altered, molecules that are in essence tightly bonded atomic clusters, exhibit more diverse chemistry and higher flexibility through engineering the clusters at the bottom level, offering the promise of bottom-up materials design with customized functionalities \cite{smits2008bottom, claridge2009cluster, jena2018super, han2019two}. The molecular crystals also motivate us to define the structural dimensionality of a solid if its primary building blocks are multi-atom substructures. Namely, molecular solids are considered 0D, and systems comprised of infinite chain-like motifs show one-dimensional (1D) character, and so forth.

\begin{figure*}[pt]
	\centering
	\includegraphics[width=0.7\linewidth]{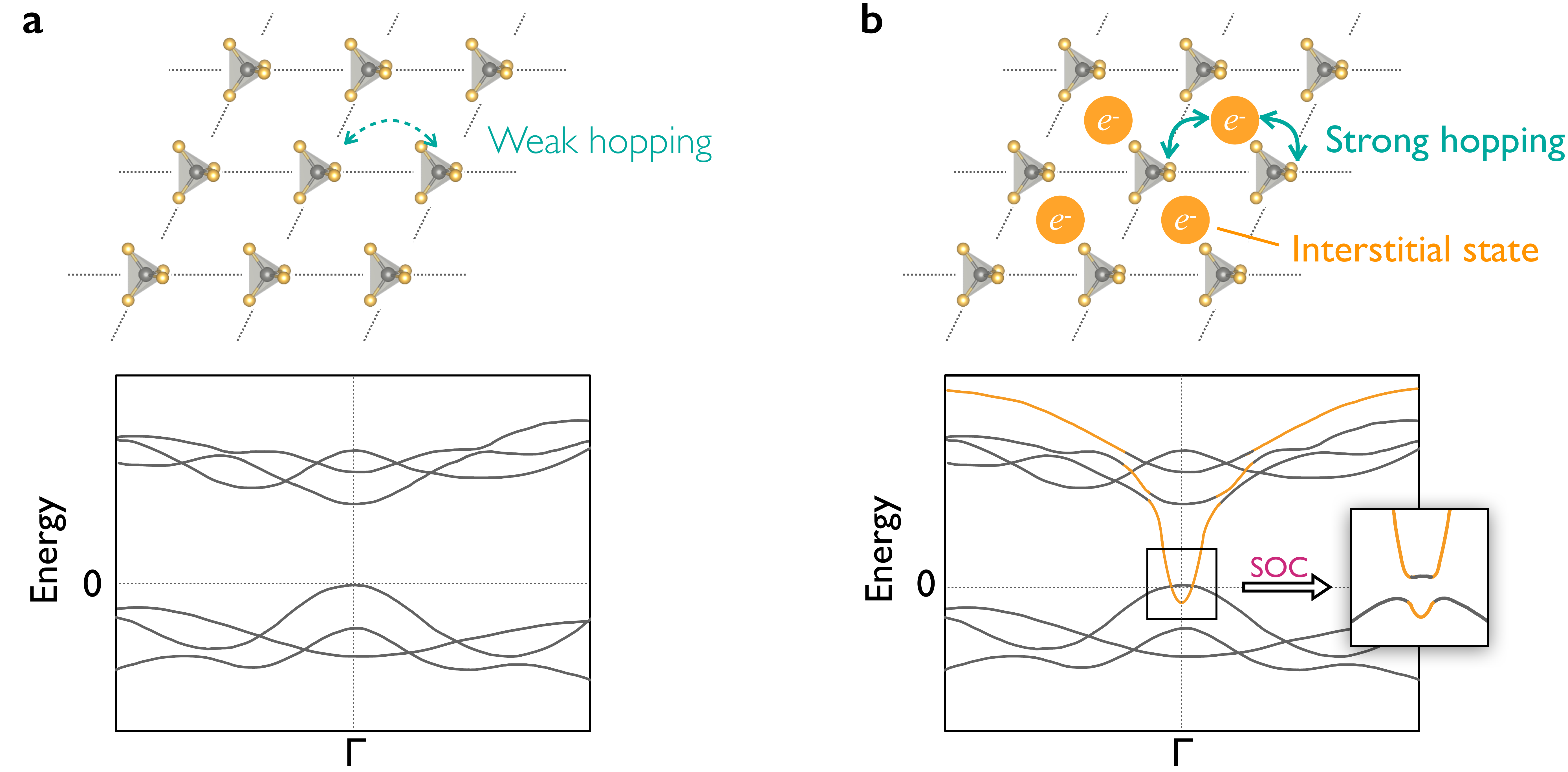}
	\caption{\sf Illustrative diagram of the ITMC. a) A schematic molecular crystal and the corresponding electronic band structure. The dashed green arrow signifies weak hopping of electrons between molecules. The resulting bands are relatively dispersionless. b) A molecular crystal with excess electrons at the interstitial sites. The intersite hopping is greatly enhanced, as shown by solid green arrows. The band stemming from interstitial states is therewith dispersive, as colored in orange. The inset depicts band inversion after introducing SOC.}
	\label{schematic_ITMC}
\end{figure*}

The molecules typically have a closed electronic shell to reach the stability, which entails a wide gap between the highest occupied molecular orbital (HOMO) and the lowest unoccupied molecular orbital (LUMO) \cite{jena2018super}. Resulting from the small overlap of intersite wavefunctions as the molecules form a crystal, the band structure exhibits weak dispersions and therewith retains the large HOMO-LUMO gap (Figure \ref{schematic_ITMC}a). Reflected in the real space, the electrons are fairly localized at each molecular site, as if constrained by the structural dimensionality. Such electronic structure is in stark contrast to that in topological materials, which have aroused paradigm renovations in condensed matter physics \cite{RevModPhys.82.3045, ando2013topological, RevModPhys.90.015001, hirayama2018topological} and many other fields \cite{RevModPhys.80.1083, RevModPhys.91.015006, RevModPhys.93.015005} over the last decade. A prerequisite of nontrivial electronic topology is gap closing or further reopening \cite{RevModPhys.82.3045, ando2013topological}. In this regard, molecular crystals tend to stand on the opposite side against topological materials \footnote{In this article, we concentrate on topological materials within the scope of topological semimetals \cite{RevModPhys.90.015001, hirayama2018topological} or first-order topological insulators \cite{RevModPhys.82.3045, ando2013topological}, with lately proposed phases such as higher-order topological insulating states characterized by protected gapless hinge or corner modes \cite{schindler2018higher} out of the discussions and awaiting for future exploration.}, which has restricted the realm of molecular materials from gaining consecutive profits through the recent bloom of topological physics. Indeed, the molecular crystals carrying nontrivial topology are very limited so far, and for some members the topological phases do not show up unless driven by external pressure in order to amplify the intersite interactions (e.g., $\alpha$-(BEDT-TTF)$_2$I$_3$ at 1.5 GPa \cite{kajita2014molecular}, [Pd(dddt)$_2$] at 12.6 GPa \cite{kato2017emergence}). Besides, most of these molecular materials host graphene-type Dirac points \cite{kajita2014molecular, kato2017emergence} or nodal lines \cite{zhou2019single}, with the novel relativistic phenomena (e.g., spin-polarized boundary states, Rashba effect) absent.

In this article, however, we show that topological phases could universally arise in certain molecular crystals, given some electron orbitals located in interstitial regions. As discussed above, weak interactions between molecules are the main obstacle to topological nontriviality. To address this issue, herein the interstitial electrons, which are relatively unbound from nuclei, could work as a bridge, extensively overlapping with orbitals of surrounding molecules, and hence greatly enhancing the intersite hopping. The resulting dispersive band would narrow down or even close the gap (Figure \ref{schematic_ITMC}b), through which the system may reach topological point- or line-nodal semimetal states. We take one step further to focus on systems containing heavy elements, so that the strong spin-orbit coupling (SOC) could mediate topological insulator (TI) phase. We also put forward rules of thumb to seek for such interstitial-electron-induced topological molecular crystals (ITMCs). 

Intertwined with interstitial electrons and topological phases, molecular crystals may exhibit extraordinary properties. First, the ITMCs hold multiple dangling-bond-free surfaces, each of which allows the access to the topological surface states (TSS), because of the 0D character. For the same reason, they tend to show evident response to mechanical perturbations. An important consequence is, for example, that topological phase transition may be achieved under comparatively low pressure. Moreover, we see that interstitial orbitals enable the breakthrough of electronic dimensionality against the constraint imposed by the lattice. In other words, while the periodically aligned molecules remain localized and loosely coupled, the interstitial electrons could be extended throughout the crystal. As a result, ITMCs fulfill the ``phonon-glass electron-crystal'' criteria \cite{slack1995handbook}, intrinsically exhibiting high thermoelectric (TE) performance. Also due to the active interstitial electrons, the work function of ITMCs is generally lower than that of conventional TIs, enabling the doping of TSS to large-gap insulators or molecules and other promising applications.

In the following, we start with the introduction of interstitial orbitals to delineate the idea of ITMCs, and then describe how we can find candidate systems in materials databases. For clarity and concreteness, we use first-principles modeling to identify the existing material K$_4$Ba$_2$[SnBi$_4$] as an ITMC candidate, and take it as the primary example to demonstrate the striking signatures of ITMCs. The same idea is applied to other candidates additionally.

\section{Results}
\subsection{Interstitial-electron-induced topological molecular crystals}
\label{subsec-ITMCs}
Interstitial electrons are the hallmark of an unusual class of solids named electrides, where some excess electrons are confined within the interstitial space and stabilized by the surrounding cations, not attached to any atom, bond, or molecule \cite{dye2003electrons, hosono2021advances, PhysRevMaterials.5.044203}. Since the interstitial orbital loosely interacts with nuclei, the corresponding band usually has a high energy and is thus distributed around the Fermi level ($E_\mathrm{F}$). Furthermore, by virtue of the large spatial spread ($s$-orbital-like) \cite{matsuishi2003high}, the interstitial band is rendered a large width, capable of admixing the highest occupied and the lowest unoccupied states \cite{PhysRevX.8.031067}. Now provided the presence of interstitial orbitals in a molecular crystal, the gap is possible to be closed by the dispersive interstitial band (Figure \ref{schematic_ITMC}b). The stability of interstitial electrons is accompanied by the surrounding cationic units, which could be positively charged molecules or intercalated ions. Finally, strong SOC may be obtained if heavy atoms sit beside the interstitial sites, since we concentrate on the relativistic topological states.

\begin{figure*}[pt]
	\centering
	\includegraphics[width=\linewidth]{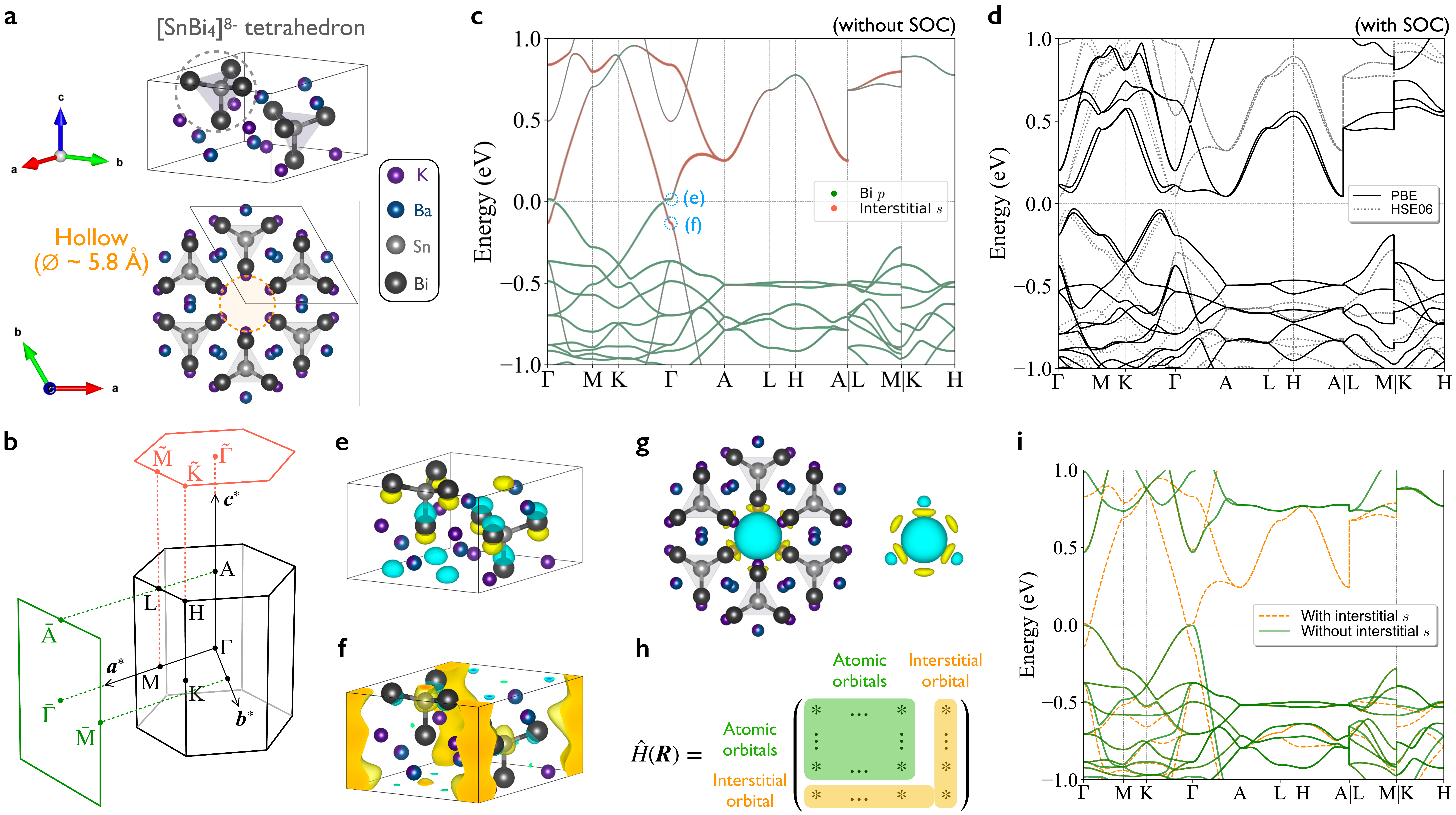}
	\caption{\sf Crystal and electronic structures of K$_4$Ba$_2$[SnBi$_4$]. a) Front and top view of the unit cell of K$_4$Ba$_2$[SnBi$_4$]. The Wyckoff position $6c$ with the coordinate (0.517, 0.483, 0.226) is statistically filled by K$^+$ ($1/3$) and Ba$^{2+}$ ($2/3$), as indicated in the corresponding color. An extended hollow with a diameter of 5.8 {\AA} is marked by a dashed orange circle. b) Brillouin zone of K$_4$Ba$_2$[SnBi$_4$], with the high-symmetry $\boldsymbol{k}$ points explicitly labeled. Two projected Brillouin zones along the $\boldsymbol{a}^*$ and $\boldsymbol{c}^*$ directions are indicated too. c) Orbital-resolved electronic band structure of K$_4$Ba$_2$[SnBi$_4$] in the absence of SOC, as predicted by the PBE pure functional within the DFT framework. The zero of energy is shifted to $E_\mathrm{F}$. High-symmetry $\boldsymbol{k}$ points are given in (b). d) Electronic band structures of K$_4$Ba$_2$[SnBi$_4$] in the presence of SOC, as predicted by the PBE pure functional (solid black line) and by the HSE06 hybrid functional (dashed grey line), with the band gap 75.4 meV and 85.7 meV, respectively. e,f) Wave functions of the lowest unoccupied state (e) and the highest occupied state (f) at the $\Gamma$ point with no SOC. They are dominated by the [SnBi$_4$]$^{8-}$ orbitals and interstitial states, respectively. Their positions in the spectrum are labeled in (c). Yellow and light blue bubbles respectively stand for the positive and negative parts of the wave function. g) Wannier orbital of the interstitial state centered at the hollow in the absence of SOC, as plotted with (left panel) and without (right panel) the atoms. Yellow and light blue respectively denote the positive and negative parts of the orbital. h) Diagrammatic representation of the tight-binding Hamiltonian built from Wannier orbitals without SOC. The matrix components related to the atomic and interstitial orbitals are colored in green and orange, respectively. i) The band structures produced by the tight-binding Hamiltonian with (dashed orange line) and without (solid green line) the inclusion of interstitial orbitals. The zero of energy and high-symmetry $\boldsymbol{k}$ points are the same with those in (c).}
	\label{bulk}
\end{figure*}

We draw two remarks before proceeding. First, one should differentiate the interstitial electrons from the ordinary overlap of adjacent atomic electrons, with the latter commonly seen in solids. In comparison, interstitial electrons are more independent and off the parent atoms, although it is difficult to draw an absolute border line between these two concepts merely based on the density distribution. Exotic phenomena of electrides (e.g., high mobility, large nonlinear optical response) \cite{hosono2021advances} are a manifestation of the uniquity of interstitial electrons. Ideally, contribution from interstitial orbitals to the electronic structure cannot be replaced by a superposition of atomic orbitals \cite{PhysRevB.103.205133}. See Section \ref{sec-electrides} in the Supporting Information for more details. Second, although we borrow the idea from electrides, ITMCs need not to be typically electrides, but are more a generalized form (see Section \ref{sec-electrides}, Supporting Information). In addition, electrides are rarely found among bulk molecular solids. While some electrides have been proposed as a platform for topological materials recently \cite{PhysRevX.8.031067}, none of them consist of 0D molecular units to exhibit the unique qualities of ITMCs.

Based on the preceding analyses, we provide the following guidelines to seek for the ITMCs. (1) The target materials are mainly composed of discrete molecular units, and permissively additional intercalated ions; the latter could be indispensable to maintain the structural stability. (2) The lattice has vacancies environed by cations with small electronegativity, which donate and stabilize interstitial electrons. (3) Heavy atoms are present and near the vacancies. Following the rules above, we concentrate on a few Zintl phase compounds as the ITMC candidates in this article. Zintl phases are a family of semiconductors where electropositive cations (generally alkali or alkaline earth cations) donate electrons to support the filled valence (e.g., following the octet rule, namely, eight electrons in the valence shell) of more electronegative elements \cite{kauzlarich2007zintl, kauzlarich2017zintl}. When the donated electrons are insufficient, the electronegative atoms would form polyanionic structures via covalent bonds in order to close the shell. These polyanions behave as molecular units if they are spatially localized. Here, we choose the polyanions constituted by Sn and Bi to be the molecular units. The heavy nucleus of Bi could offer strong relativistic effects; additionally, the large radius of its valence electron orbitals ($6p$) is conducive to intersite electron hopping compared to that of light elements. Next, we analyze the example material K$_4$Ba$_2$[SnBi$_4$].

\subsection{Crystal and electronic structures of K$_4$Ba$_2$[SnBi$_4$]}
\label{subsec-bulk}

The experimental synthesis of K$_4$Ba$_2$[SnBi$_4$] was reported in 2000 \cite{eisenmann2000pniktogenidostannate}. It crystallizes in a hexagonal lattice ($a=$ 11.395(2) {\AA} and $c=$ 8.320(2) {\AA} at ambient pressure), with the noncentrosymmetric space group $P6_3mc$ (No. 186), as shown in Figure \ref{bulk}a. The primary structural motifs are discrete [SnBi$_4$]$^{8-}$ tetrahedra, located at the Wyckoff site $2b$ and oriented along the [0001] direction, separated by K$^+$ and Ba$^{2+}$ cations at Wyckoff sites $6c$. The crystal thus shows the 0D character (see Section \ref{sec-chembonds} in the Supporting Information for comprehensive analyses) that we will echo and make use of frequently in the subsequent discussions. Importantly, the K$^+$ cations form a hollow at the edge of unit cell, extending along [0001], with a diameter about 5.8 {\AA} (the lower panel of Figure \ref{bulk}a), which provides sufficient space to accommodate possible interstitial electronic states.

We determine the electronic structure of K$_4$Ba$_2$[SnBi$_4$] by means of the density functional theory (DFT) with the Perdew-Burke-Ernzerhof (PBE) type functional \cite{PhysRevLett.77.3865}; see Section \ref{sec-methods} Methods for details. Figure \ref{bulk}c depicts the nonrelativistic band structure, where the high-symmetry $\boldsymbol{k}$ points are given in Figure \ref{bulk}b. Although dispersions of most bands are comparatively weak, in line with the 0D feature, we immediately notice a highly dispersive band around $E_\mathrm{F}$, emanating from the interstitial orbital (see the orbital-resolved bands in Figure \ref{bulk}c) and mostly unoccupied. Only a tiny portion at the bottom of the interstitial band is below $E_\mathrm{F}$ through inversion with the Bi $6p$ band (Figures \ref{bulk}e,f); the tiny occupation at the interstitial site indicates that the electron transfer from cations to polyanions is not complete, and therewith offers an opportunity of nontrivial topology in the electronic states. It is also noteworthy that the band structure near $E_\mathrm{F}$ is dictated by the orbitals of [SnBi$_4$]$^{8-}$ and interstitial states (Figure \ref{bulk}c), while the inserted ions have negligible impact and merely play the role of preserving the electroneutrality and lattice stability, as molecular crystals ought to be.

To further demystify the signature of interstitial states, we construct Wannier functions \cite{PhysRevB.56.12847, PhysRevB.65.035109} of the atomic and interstitial orbitals. The interstitial orbital is shown to be centered at the hollow with a large spread (Figure \ref{bulk}g), as it should be. A tight-binding Hamiltonian built from these localized orbitals (Figure \ref{bulk}h) can immediately replicate the DFT band structure (Figure \ref{bulk}i). However, supposing that the interstitial states were absent from this crystal, as imitated by excluding the components related to the interstitial orbitals from the Hamiltonian (Figure \ref{bulk}h), the band edge would shrink substantially and leave a comparatively large gap (around 0.5 eV), as unveiled by the solid green lines in Figure \ref{bulk}i. This simulation evidences the decisive role of interstitial states for the strong band dispersion and inversion in K$_4$Ba$_2$[SnBi$_4$].

\begin{figure}[t]
	\centering
	\includegraphics[width=0.98\linewidth]{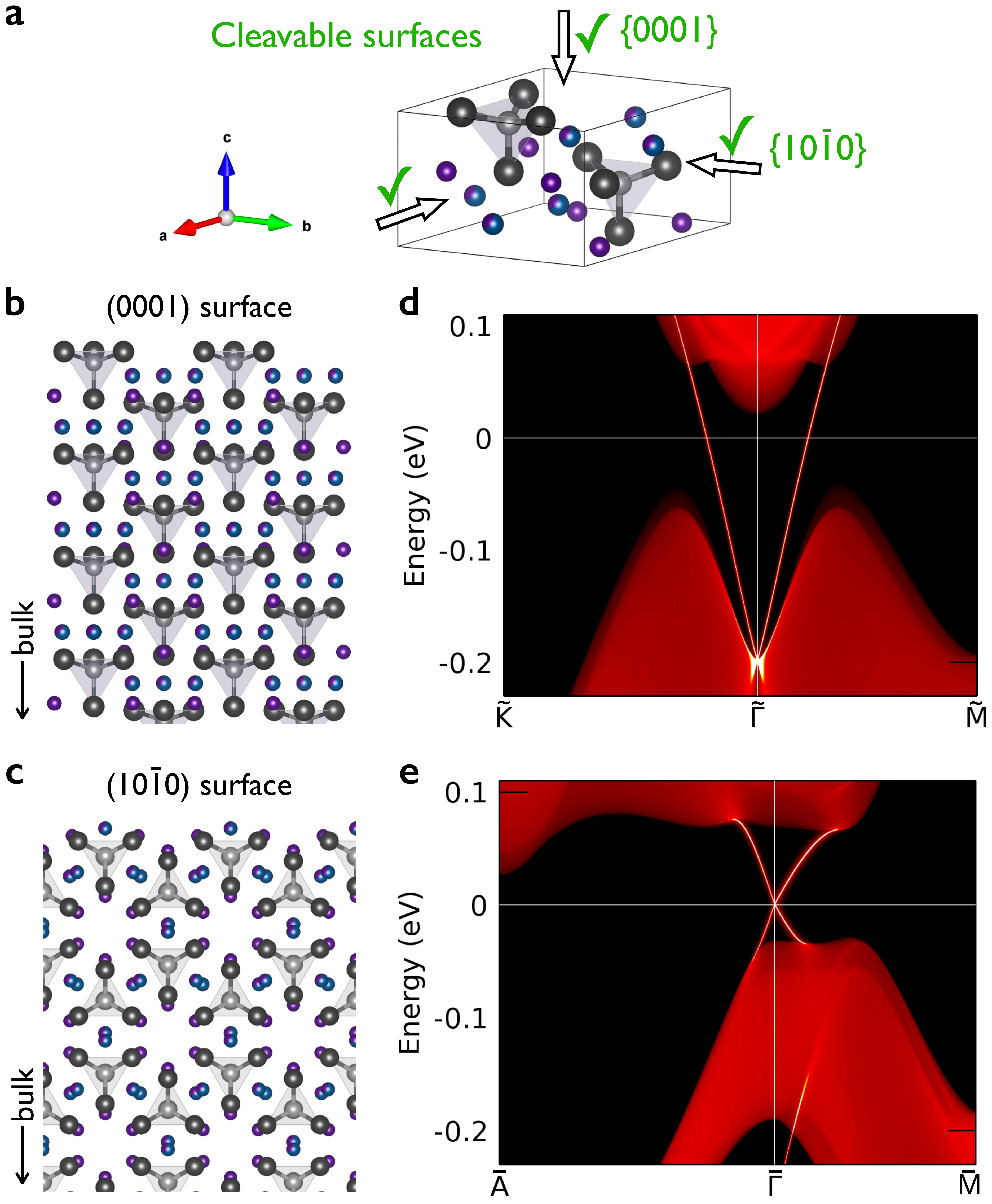}
	\caption{\sf Cleavage and surface states of K$_4$Ba$_2$[SnBi$_4$]. a) Dangling-bond-free surfaces of K$_4$Ba$_2$[SnBi$_4$], as indicated by black arrows and green ticks. b,c) Slabs of K$_4$Ba$_2$[SnBi$_4$] terminated with the (0001) surface (b) and $(10\bar{1}0)$ surface (c). d,e) Momentum-resolved density of states of a semi-infinite (0001) slab (d) and $(10\bar{1}0)$ slab (e) predicted based on the Wannier functions with SOC. High-symmetry $\boldsymbol{k}$ points in the projected Brillouin zone are given in Figure \ref{bulk}b. Yellow and red correspond to high and low density, respectively.}
	\label{cleavage}
\end{figure}

\begin{figure*}[pt]
	\centering
	\includegraphics[width=0.95\linewidth]{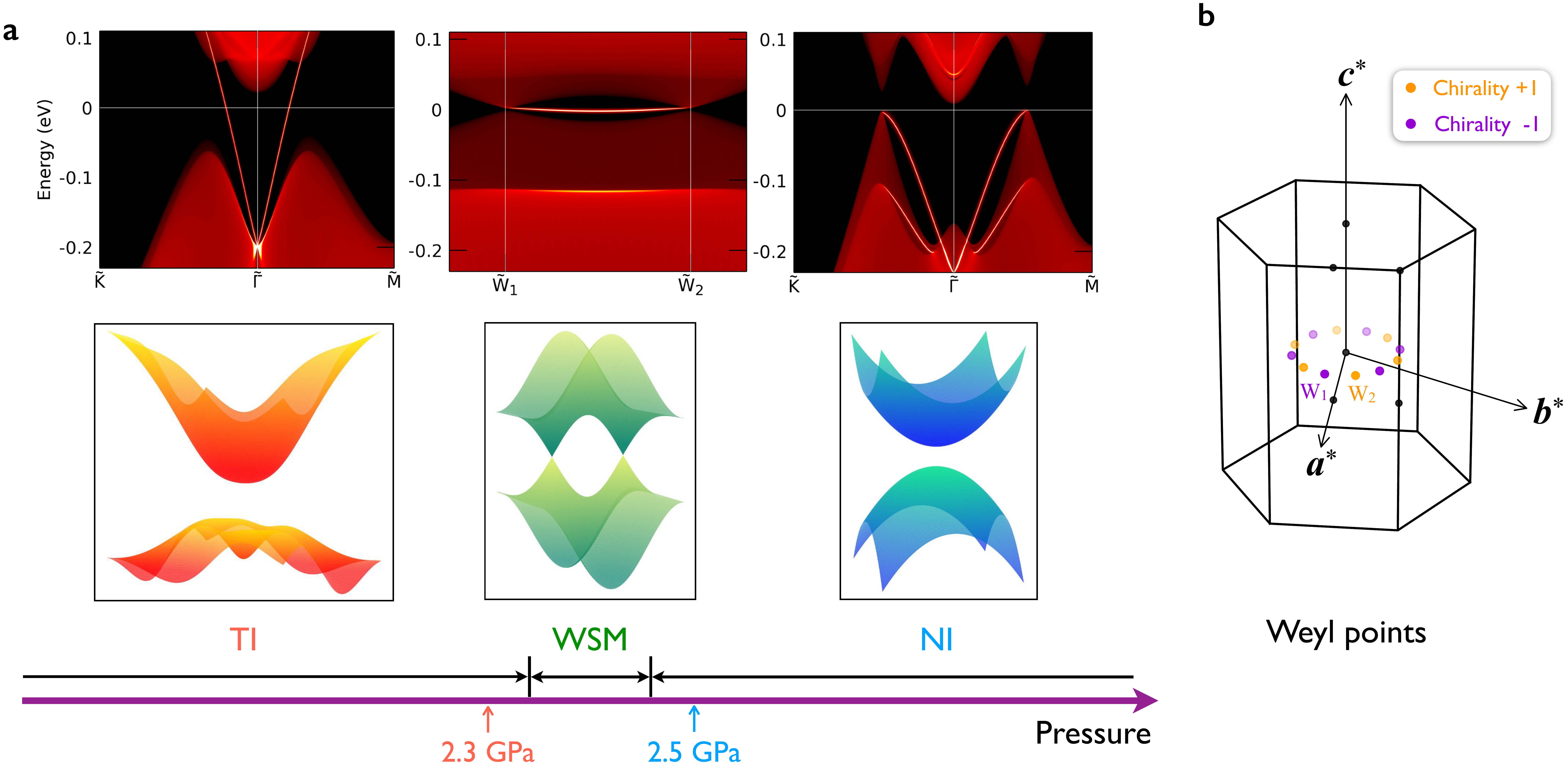}
	\caption{\sf Different topological phases in K$_4$Ba$_2$[SnBi$_4$]. a) TI-WSM-NI phase transition in K$_4$Ba$_2$[SnBi$_4$] with the increasing external pressure. Momentum-resolved density of states of a semi-infinite (0001) slab (upper panel) with a schematic bulk band structure (lower panel) is depicted for each phase. The WSM phase arises within the range of 2.3 - 2.5 GPa. b) Distribution of the Weyl points for the WSM phase. The chirality of each Weyl point is indicated.}
	\label{pressure}
\end{figure*}

At last, by including SOC in the calculations, we predict that K$_4$Ba$_2$[SnBi$_4$] is a strong TI with an intermediate gap (75.4 meV) around $\Gamma$ (Figure \ref{bulk}d). To avoid the false-positive prediction of TIs caused by the well-known gap underestimation in DFT \cite{PhysRevB.84.041109}, we provide the band structure at the level of the nonlocal Heyd-Scuseria-Ernzerhof (HSE06) hybrid functional \cite{heyd2003hybrid} as reference (Figure \ref{bulk}d), that shows no critical disparity and preserves the odd $\mathbb{Z}_2$ and a slightly larger gap (85.7 meV). Interestingly, the band gap of K$_4$Ba$_2$[SnBi$_4$], to our knowledge, tops all the noncentrosymmetric TIs hitherto known at ambient pressure within the level of DFT (see Section \ref{sec-TI_gaps}, Supporting Information), which may let the surface transport of K$_4$Ba$_2$[SnBi$_4$] fairly free from the bulk carrier disturbance and trigger a variety of applications.

\subsection{Multiple cleavable surfaces} 
\label{subsec-cleavage}

It is known that the gapless Dirac cone could be manifested on any surface of a strong TI in principle \cite{RevModPhys.82.3045, ando2013topological}. In most realistic materials, nevertheless, the examination of such TSS is hindered by dangling bonds for some terminations (e.g., lateral cleavage of the layered TI Bi$_2$Te$_3$ \cite{chen2009experimental, heremans2017tetradymites}). On the contrary, by virtue of the noncovalent interactions between constituent molecules and ions, the ITMC K$_4$Ba$_2$[SnBi$_4$] can be cut along multiple surfaces, namely, $\lbrace0001\rbrace$ and $\lbrace10\bar{1}0\rbrace$, free from breaking the covalent bonds, as indicated in Figure \ref{cleavage}a. Figures \ref{cleavage}d,e display the clean pronounced TSS on the $(0001)$ and $(10\bar{1}0)$ surfaces, respectively, as predicted by the Wannier function model. This is compatible with the DFT prediction (see Section \ref{sec-DFTslabband}, Supporting Information), which involves a more reasonable treatment of charge redistribution at the surface. We remark that the availability of cleavage along multiple directions is of even greater importance if the ITMC is a weak TI, in which case the TSS appear only on a certain surface \cite{RevModPhys.82.3045, ando2013topological, PhysRevX.11.031042}.
\vfill

\subsection{Pressure-induced topological phase transition}
\label{subsec-pressure}

Molecular solids generally exhibit large compressibility owing to the absence of a global covalently bonded network. As a direct result, a different topological phase might appear under intermediate mechanical perturbations \cite{PhysRevB.90.155316, PhysRevB.93.195149, hirayama2018topological}. We fully relax the crystal structure of K$_4$Ba$_2$[SnBi$_4$] at a series of increasing pressures within DFT, and monitor the $\mathbb{Z}_2$ index if gapped. The gap opens at 2.3 GPa and 2.5 GPa, but carrying an odd and an even $\mathbb{Z}_2$, respectively (Figure \ref{pressure}a), indicative of topological phase transition midway. In common with the universal phase diagram for noncentrosymmetric TIs \cite{PhysRevB.90.155316, hirayama2018topological}, a Weyl semimetal (WSM) phase is shown to lie in between the lower-pressure TI and the higher-pressure normal insulator (NI) states (Figure \ref{pressure}a). Figure \ref{pressure}b presents 6 pairs of Weyl points distributed on the $k_z$= 0 plane due to the time reversal and six-fold screw symmetries. Notice that the range of pressure for the WSM is within the interval of 2.3 - 2.5 GPa, which is a moderate condition and fairly accessible in the laboratory. We make a detour to mention that the isostructural compound K$_4$Ba$_2$[SnSb$_4$] (where Sb replaces Bi) is found to be an NI. Therefore, the phase transition can be realized also by doping Sb to K$_4$Ba$_2$[SnBi$_4$], just like the alloy Bi$_{1-x}$Sb$_x$ \cite{hsieh2008topological}. In short, this system can be modulated into a WSM either under comparatively low pressure, or through chemical doping of Sb, suggesting the role of a single ITMC as a testbed of multiple topological phases.
\vfill

\begin{figure*}[pt]
	\centering
	\includegraphics[width=\linewidth]{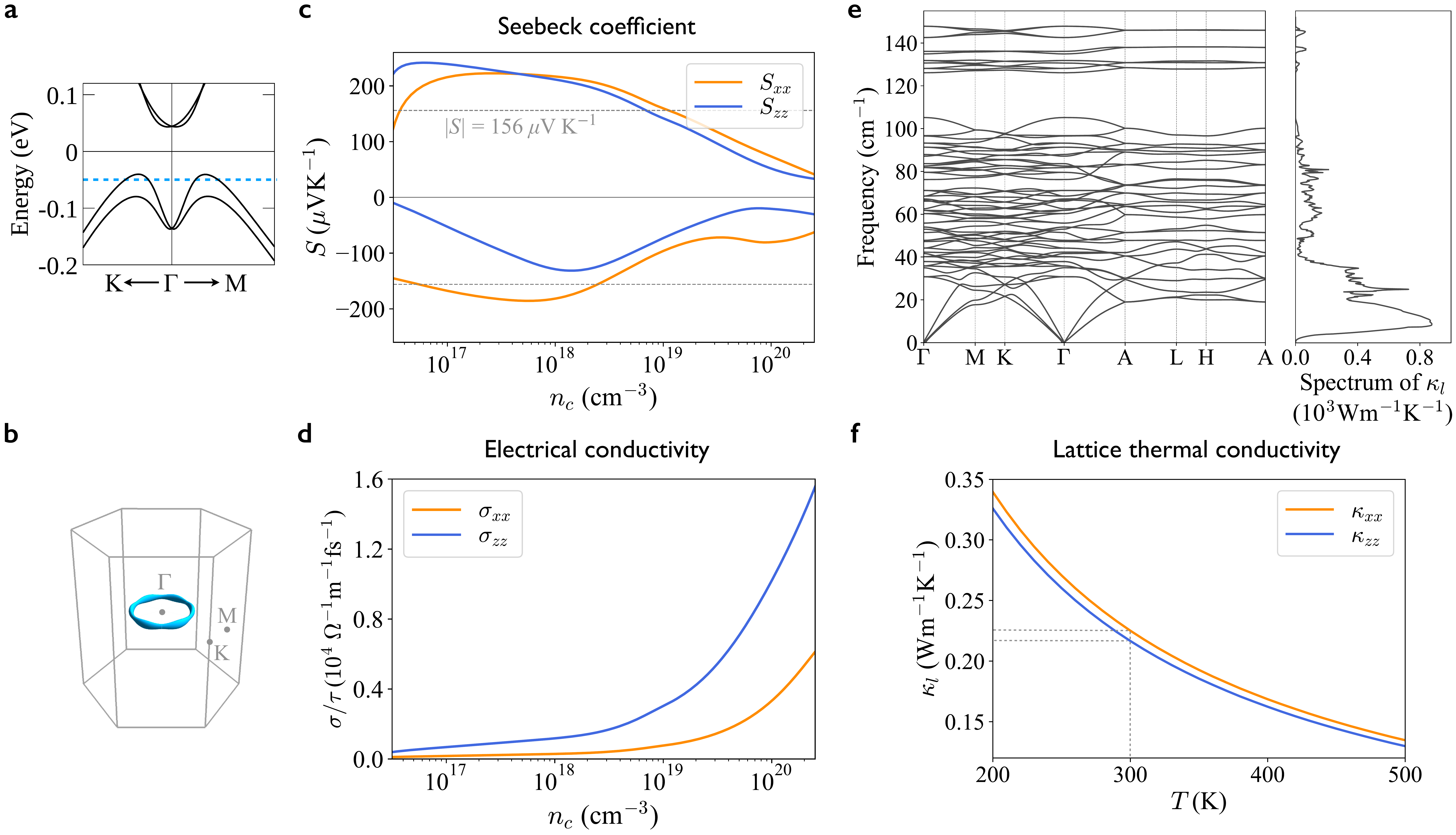}
	\caption{\sf Thermoelectricity of K$_4$Ba$_2$[SnBi$_4$]. a) Band structure of K$_4$Ba$_2$[SnBi$_4$] with SOC in the vicinity of the $\Gamma$ point, which is calculated with the HSE06 functional. This figure is a magnified view of Figure \ref{bulk}d. b) Isoenergy surface centered at the $\Gamma$ point, with the corresponding energy marked in (a). The size is exaggerated for easy visualization. c) Seebeck coefficient $S$ of K$_4$Ba$_2$[SnBi$_4$] versus the carrier concentration $n_c$ at 300 K. Both $p$-type and $n$-type doping are plotted. The dashed grey lines indicate the threshold of $|S|=156\ \mu$V K$^{-1}$. d) Electrical conductivity $\sigma$ against $n_c$ at 300 K. $\sigma$ is evaluated in the unit of the scattering time $\tau$ under the constant scattering time approximation. Only $p$-type doping is shown. e) Phononic band structure (left panel) and frequency-resolved lattice thermal conductivity (right panel). High-symmetry $\boldsymbol{k}$ points are given in Figure \ref{bulk}b. For the spectrum of $\kappa_l$ (right panel), the $z$ component is shown alone. Note that the frequencies of the low-lying optical modes, which produce scattering via anharmonic coupling with acoustic modes, are merely around 30 cm$^{-1}$. f) Lattice thermal conductivity $\kappa_l$ versus the temperature, based on the calculation of anharmonic phonons.}
	\label{thermoelectricity}
\end{figure*}

\subsection{Thermoelectricity}
\label{subsec-thermoelectricity}

TE materials can realize energy conversion between heat and electricity \cite{goldsmid2010thermoelectricity}, the capability of which is characterized by the dimensionless figure of merit, $zT = \sigma S^2T/(\kappa_e + \kappa_l)$, where $\sigma$, $S$, $T$, $\kappa_e$ and $\kappa_l$ represent the electrical conductivity, Seebeck coefficient, absolute temperature, electronic and lattice component of the thermal conductivity, respectively. A large $zT$ entails high thermopower (i.e., the absolute value of $S$) and electrical conductivity, as well as low thermal conductivity at the given temperature. Just like many TIs (e.g., Bi$_2$Te$_3$, one of the best-performance room-temperature TE materials) \cite{goldsmid2010thermoelectricity, heremans2017tetradymites, xu2017topological, PhysRevApplied.3.014004}, K$_4$Ba$_2$[SnBi$_4$] exhibits the potential of extraordinary TE efficiency, considering the fact that the non-parabolic dispersion at the valence band top (Figure \ref{thermoelectricity}a), or the accompanying ring-like Fermi surface at low-level $p$-type doping (Figure \ref{thermoelectricity}b), can greatly enhance the density-of-states effective mass and therewith the thermopower, similar to the scenario in other TIs \cite{PhysRevApplied.3.014004, heremans2017tetradymites, goldsmid2010thermoelectricity}. The enhancement is supported by our computations based on Boltzmann transport theory (see Section \ref{sec-methods} Methods). As reflected in Figure \ref{thermoelectricity}c, $|S|$ at $p$-type doping is greater than at $n$-type at $300$ K, close to room temperature. On top of that, the components of $p$-doped $S$ are above 156 $\mu$V K$^{-1}$ (a baseline to potentially have $zT > 1$ at room temperature, as derived from the Wiedemann-Franz law \cite{goldsmid2010thermoelectricity}) over a broad range of $n_c$, suggestive of a salient $zT$. At the optimal doping, $S_{zz}$ (the $z$ direction is parallel to [0001]) even reaches 240 $\mu$V K$^{-1}$, already comparable to Bi$_2$Te$_3$, with the maximal $S$ about 250 $\mu$V K$^{-1}$ \cite{goldsmid2010thermoelectricity}.

\begin{figure*}[pt]
	\centering
	\includegraphics[width=\linewidth]{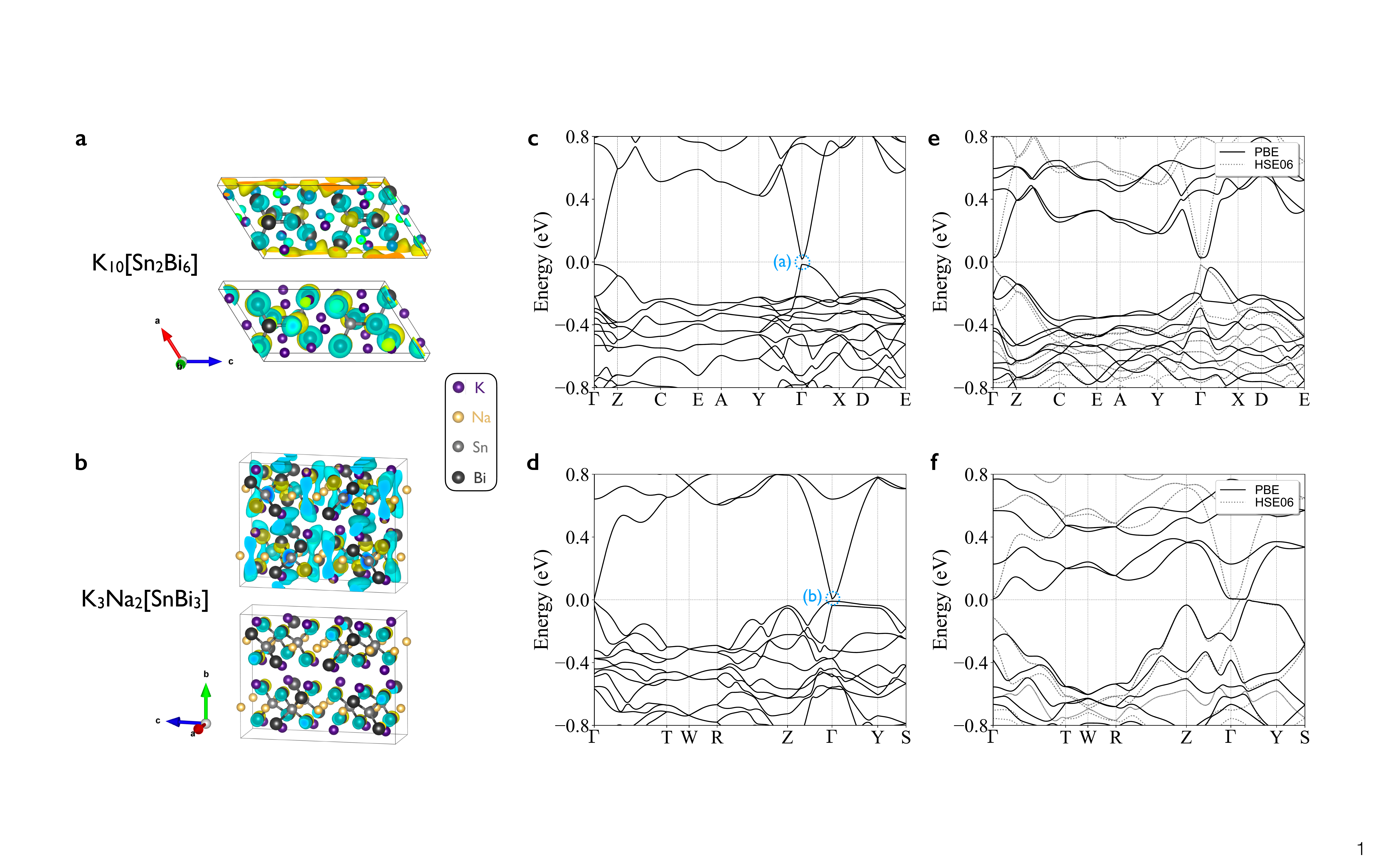}
	\caption{\sf Other ITMC systems. a,b) Crystal structure of K$_{10}$[Sn$_2$Bi$_6$] (a) and K$_3$Na$_2$[SnBi$_3$] (b) together with their wave functions of the lowest unoccupied state (upper panel) and highest occupied state (lower panel) at the $\Gamma$ point with no SOC. The corresponding position in the spectrum is also marked in (c) or (d). Yellow and light blue bubbles represent the positive and negative components of the wave function, respectively. c,d) Electronic band structure of K$_{10}$[Sn$_2$Bi$_6$] (c) and K$_3$Na$_2$[SnBi$_3$] (d) in the absence of SOC. The energy is measured from the Fermi level. e,f) Band structure of K$_{10}$[Sn$_2$Bi$_6$] (e) and K$_3$Na$_2$[SnBi$_3$] (f) with SOC. The HSE06 result is offered in addition to the PBE. Note that the band structure in (d) or (f) is based on the primitive cell (not shown) of K$_3$Na$_2$[SnBi$_3$], while (b) presents the conventional cell.}
	\label{otherITMCs}
\end{figure*}

While the intimate TI-TE relation is embodied in ITMCs too, it is much beyond the extent in conventional TIs. One distinction consists in the interstitial electrons that are weakly constrained by the crystal framework and therewith prone to high mobility. Subsequently, a considerable $\sigma$, in particular along the [0001] direction (the extended vacant channel), can be expected for K$_4$Ba$_2$[SnBi$_4$]. Figure \ref{thermoelectricity}d shows that at 300 K, $\sigma_{zz}$ prominently exceeds $\sigma_{xx}$ overall. Therefore, the cooccurrence of high $S$ and high $\sigma$, which have conflicting requirements to band dispersions \cite{goldsmid2010thermoelectricity,kuroki2007pudding}, can be successfully achieved in ITMCs. In addition, being a molecular crystal, K$_4$Ba$_2$[SnBi$_4$] bears a structural complexity that can trap heat, akin to the other Zintl compounds for thermoelectrics \cite{PhysRevB.81.075117, kauzlarich2007zintl, kauzlarich2017zintl}. The loosely coupled heavy [SnBi$_4$]$^{8-}$ clusters and intercalated ions result in soft phonons, namely, low-velocity acoustic modes and low-lying optical modes that govern the phonon-phonon scattering (Figure \ref{thermoelectricity}e), whereby $\kappa_l$ can be drastically suppressed \cite{PhysRevB.81.075117}. Indeed, by calculating anharmonic phonons (see Section \ref{sec-methods} Methods), we evaluate $\kappa_l$ of K$_4$Ba$_2$[SnBi$_4$] to be 0.22 - 0.23 W m$^{-1}$ K$^{-1}$ at 300 K (Figure \ref{thermoelectricity}f), even lower than that of Bi$_2$Te$_3$ (with in-plane and out-of-plane components 1.2 W m$^{-1}$ K$^{-1}$ and 0.4 W m$^{-1}$ K$^{-1}$, respectively \cite{PhysRevB.90.134309}); also notice the high isotropy of the low $\kappa_l$ in K$_4$Ba$_2$[SnBi$_4$] relative to that in the layered Bi$_2$Te$_3$, which again justifies the 0D nature of K$_4$Ba$_2$[SnBi$_4$] from the viewpoint of phonons. Hence, we conclude that while the phonons behave as being trapped, the electrons remain mobile across the crystal by means of the comparatively independent interstitial orbitals, giving birth to the distinguishing ``phonon-glass electron-crystal'' behavior \cite{slack1995handbook} for ITMCs. Moreover, the linearized TSS that are robust against backscattering can further enhance $zT$ if thin films are exploited \cite{xu2017topological}. In short, the triple identity (i.e., TIs, generalized form of electrides, and molecular crystals) could endow ITMCs with an inherent state-of-the-art TE performance.

\subsection{Work function}
\label{subsec-workfunc}
The knowledge of work function is crucial to heterostructures, with the mismatch between the contact systems dominating the interface properties through the electron transfer, band bending, etc. In the ITMCs, the work function is routinely small due to the active interstitial electrons. As an example, the work function of K$_4$Ba$_2$[SnBi$_4$] for the $(10\bar10)$ surface is calculated to be 2.2 eV in the presence of SOC (see Section \ref{sec-workfunc}, Supporting Information), akin to alkali metals, which is unprecedented in TIs so far (e.g., the work function of Bi$_2$Te$_3$ is around 5.3 eV \cite{takane2016work} on the contrary). Consequently, the helical surface Dirac carries could easily transfer to the contact system by constructing heterostructures (see Section \ref{sec-interfaces}, Supporting Information). We propose a promising application in catalysis. TSS have been shown to have the potential of boosting certain chemical reactions by serving as an electron bath \cite{PhysRevLett.107.056804}. Here the ITMCs, further equipped with a low work function, are expected to activate very inert molecules that has a shallow LUMO, thus being prospective high-efficiency catalysts \cite{hosono2021advances}.

\subsection{Other candidate systems}
\label{subsec-otherITMCs}

We briefly present another [Sn, Bi]-polyanion-based ITMC, K$_{10}$[Sn$_2$Bi$_6$] \cite{asbrand1993crystal}. Formed in a monoclinic lattice (space group $P2_1/c$, No. 14), the unit cell of K$_{10}$[Sn$_2$Bi$_6$] contains two discrete [Sn$_2$Bi$_6$]$^{10-}$ molecular units in addition to the inserted K$^+$ ions (Figure \ref{otherITMCs}a), being 0D. The interlayer region offers space for interstitial states, as indicated by the yellow bubbles at the cell edge (the upper panel of Figure \ref{otherITMCs}a). The electronic bands are highly dispersive around the $\Gamma$ point (Figure \ref{otherITMCs}c) owing to the interstitial states. Further driven by SOC, the valence band maximum and conduction band minimum are inverted as unveiled by the PBE functional (Figure \ref{otherITMCs}e), and the system turns out to be a strong TI, with a gap around 60.7 meV. Although the additional HSE06 calculation predicts a trivial non-inverted band structure (Figure \ref{otherITMCs}e), the gap is merely about 33.2 meV. Therefore, we can still expect a topological phase only through exposing a fairly low pressure, by making use of the weak intermolecular coupling in ITMCs. The work function is around 2.3 eV for the (100) surface, similar to that of K$_4$Ba$_2$[SnBi$_4$].

We also show the application of the ITMC idea to 1D crystals, which is quite straightforward. K$_3$Na$_2$[SnBi$_3$] (space group $Ibca$, No. 73) is a 1D Zintl material \cite{asbrand1998dimere}, with the [Sn$_2$Bi$_6$]$^{10-}$ chain extending along the [100] direction (Figure \ref{otherITMCs}b). The intercalation of K$^+$ and Na$^+$ ions provides excess electrons at interstitial spaces (light blue bubbles in the upper panel of Figure \ref{otherITMCs}b). Consequently, band inversion arises at $\Gamma$, between a molecular orbital and the interstitial state (Figures \ref{otherITMCs}d,f), resulting in a strong TI phase, as supported by either a pure or a hybrid functional calculation, notwithstanding a small indirect gap ($<10$ meV). Cleavage or strain on the surfaces (010) and (001) is available, though the operation on the (100) surface is obstructed by the [Sn$_2$Bi$_6$]$^{10-}$ chain. We find that the work function along the (010) surface is as low as 2.1 eV.

\section{Conclusions and Discussion}
We have elucidated the concept of ITMCs based on first-principles calculations, and demonstrated their remarkable properties as a result of the characteristic unification of molecular crystals, interstitial states, and topological phases, namely, multiple dangling-bond-free surfaces, topological phase transition under moderate mechanical perturbations, high-performance thermoelectricity, and ultralow work function. The discussed phenomena and applications are within current experimental reach.

We address more comments on molecular crystals to highlight their superiority in terms of materials design. As initially noted in Section \ref{sec-intro}, molecular crystals stand out owing to molecules as building blocks. Indeed, the individual molecules, if stably existing when extracted from the bulk, can exhibit physical and chemical properties akin to or even beyond elements in the periodic table \cite{claridge2009cluster}, like the superhalogenic nature with an extremely shallow LUMO (e.g., NO$_3$ \cite{behera2013nitrate}). In this context, such atomic clusters are particularly dubbed superatoms, to emphasize the crucial fact that they may mimic atoms \cite{claridge2009cluster, jena2018super}. Bulk solids built from superatoms show more versatile features as inherited from these clusters, than those with purely atomic components. Furthermore, unlike atoms with fixed attributes, the characters of a superatom may be precisely tuned by engineering the cluster (e.g., adding, removing, or replacing a single atom). Hence, materials hierarchically assembled from superatoms allow tailored properties and functionalities \cite{jena2018super}, like lattice constants and band gap \cite{claridge2009cluster}.

Interstitial states are key to the topological phases of ITMCs, the mechanism of which can be heuristically understood from the gap closing caused by the dispersive interstitial band, as elaborated in Sections \ref{sec-intro} and \ref{subsec-ITMCs}. Indeed, we can gain deeper insight by focusing on real space. The ordinary molecular crystals, with each molecular orbital exponentially localized at the site, can be adiabatically connected to the atomic limit (more technically, the obstructed atomic limit \cite{bradlyn2017topological}), where the the molecular units are at infinite separation. Consequently, molecular crystals are topologically trivial in general. With electron orbitals located at vacancies, however, the system is naturally driven away from the atomic limit \cite{PhysRevX.8.031067, PhysRevB.103.205133}, leading to a high propensity for nontrivial topology.

Now it is natural to emphasize that the character of being a molecular crystal stands on the foremost position for the novelty of ITMCs, which is indispensable for their diverse prominent features, and for the promise of tailoring the functions of the materials as assembled from molecules. Then followed by the participation of interstitial electrons that are tightly connected with topological phases, the systems become distinguishing from normal molecular solids. This is the vital logic lying in the name ITMC. Therefore, ITMCs are immediately differentiated from other families whose main components are not discrete molecular units, such as ionic solids comprised of isolated atoms alone, metals with globally delocalized bondings, and crystals with extended networks of covalent bonds. As an example, some 1D solids were reported to hold topological band structure, such as $A$(Mo$X$)$_3$ ($A=$ Na, K, Rb, In, Tl; $X=$ S, Se, Te) \cite{PhysRevX.7.021019}, Bi$_4$I$_4$ \cite{noguchi2019weak, PhysRevX.11.031042}, and (TaSe$_4$)$_2$I \cite{shi2021charge}. But they lose some degrees of freedom and fexibility as hindered by infinite chains compared to molecular solids that show the 0D nature. On top of that, the energy bands originating from the extended covalent bonds can be easily dispersed, and importantly contribute to the topological band structure of these 1D systems \cite{PhysRevX.7.021019, shi2021charge}. In particular, networks of hypervalent bonding, a type of electron-rich multicenter bonding beyond the octet rule \cite{papoian2000hypervalent}, are characteristic of folded half-filled bands, and therewith show high tendency towards band crossing or inversion \cite{khoury2021chemical}. In contrast, for molecular solids, due to the lack of any global covalent network, alternative ways to nontrivial band topology are desired -- this is the very moment at which interstitial states come in handy.

Although we have chosen to demonstrate the idea and superiority of ITMCs as realized in the selected Zintl compounds in this work, the signatures of ITMCs are distinctive from those that are of substantial study for the general Zintl family. For instance, in spite of topological phases predicted in a couple of Zintl compounds with polyanions, the polyanionic frameworks are two-dimensional (2D) sheets (e.g., $A$In$_2$As$_2$ with $A=$ Ca, Sr, Eu \cite{tan201714mgbi11}) or 1D chains (e.g. Eu$_5$In$_2$Bi$_6$ \cite{PhysRevB.105.235128}), with isolated anionic clusters not reported yet. They are distinct from ITMCs just like the 1D topological materials as discussed above. It is widely known that Zintl materials are also considered suitable for thermoelectrics, since they tend to exhibit the ``phonon-glass electron-crystal'' behavior \cite{kauzlarich2007zintl, kauzlarich2017zintl}. Nonetheless, the ``electron-crystal'' electronic structure commonly relies on the global structural motifs of covalent bonds (i.e., 1D or 2D polyanions) \cite{kauzlarich2007zintl}, despite the fact that the weak coupling of polyanions produces the ``phonon-glass'' character in the same way with ITMCs. The electronic and phononic transports are thus more interdependent. In ITMCs, on the contrary, interstitial electrons and molecular units, that are loosely bound to each other, may work as relatively independent components, govern the electrical mobility and lattice thermal conductivity more separately, further enabling a large $\sigma$ while preserving a small $\kappa_l$. Besides, the interstitial electrons provide one additional tunable parameter for the molecular materials design, which may be controlled by engineering the confinement geometry via strain or substitution of atoms.

Following the searching guidelines as summarized in Section \ref{subsec-ITMCs}, more ITMC candidates can be picked out from materials databases (possibly in conjunction with high-throughput techniques \cite{setyawan2010high-throughput}), or even artificially designed, which remains to be an open question. For example, a variety of Bi clusters \cite{ruck2015between} can take the role of polyanions in order for strong SOC; the option of anions is even broader if the relativistic topological states are not of particular interest. Other than isolated atoms, alkali superatoms (e.g., K$_3$O \cite{reber2007superatom}), the properties of which can be flexibly controlled, also have the potential to provide and stabilize the interstitial electrons on account of their ultra small electronegativity. Compared to atomic cations, they may enlarge the spacings and reduce interactions between molecules, so that the solid maintains more of the molecular character. Our results suggest that some long-known molecular solids would recapture attention once recognized as ITMCs, meanwhile the well-established theories and technologies in the community of molecular crystals (e.g., flexible bottom-up materials design\cite{smits2008bottom, claridge2009cluster, han2019two}) might enable functionality tailoring for topological materials.

\section{Methods}
\label{sec-methods}
\partitle{DFT calculations.} The DFT calculations of the ITMC candidate materials are performed based on the projector augmented wave scheme \cite{PhysRevB.59.1758} as implemented in the Vienna \textit{ab initio} simulation package (\texttt{VASP}) \cite{PhysRevB.54.11169}. For static calculations (i.e., self-consistent electronic relaxation, band structure, etc.), we employ the exchange-correlation functional given by the generalized gradient approximation (GGA) with the PBE parametrization \cite{PhysRevLett.77.3865}. The nonlocal HSE06 hybrid functional \cite{heyd2003hybrid} is supplemented for the check of band topology at ambient pressure. The energy cutoff of the plane wave is chosen as $30\%$ above the maximum value recommended in the pseudopotential library. The Brillouin zone is sampled by a $\Gamma$-centered grid with the resolution of $2\pi \times 0.024 \text{ \AA}^{-1}$ such that not only the total energy but also the band gap are converged (see Section \ref{sec-kgrid}, Supporting Information). The PBEsol \cite{PhysRevB.79.155107} instead of PBE parametrization is applied for the lattice relaxation, since the former can reproduce the lattice constants closer to the experimental values at ambient pressure (the lattice constants predicted by PEBsol are $a=$ 11.379 {\AA}, $c=$ 8.282 {\AA}, and by PBE are $a=$ 11.555 {\AA}, $c=$ 8.519 {\AA}, while recalling that the experimental values are $a=$ 11.395(2) {\AA}, $c=$ 8.320(2) {\AA}). The internal coordinates of all atoms are fully relaxed with a force criterion of 0.01 eV/{\AA}. Since the molecules of candidate ITMCs are not neutral but charged (e.g., [SnBi$_4$]$^{8-}$), ionic bonding governs the intermolecular interactions. As a result, van der Waals forces, that are comparatively much weaker, are not included in the calculations. Indeed, the empirical dispersion correction to DFT (i.e., DFT-D) \cite{grimme2010consistent, grimme2011effect} is tested for K$_4$Ba$_2$[SnBi$_4$], which shows no further improvements, and produces identical band structure and gap to those without DFT-D. We use pre-/post-processing tools and utilities \cite{wang2021vaspkit, setyawan2010high-throughput, curtarolo2012aflow, herath2020pyprocar, togo2018texttt, momma2008vesta, kawamura2019fermisurfer} to aid the calculations.

Virtual crystal approximation (VCA) \cite{nordheim1931elektronentheorie, PhysRevB.61.7877} is exploited for the fractionally occupied Wyckoff position $6c$ (0.517, 0.483, 0.226) in K$_4$Ba$_2$[SnBi$_4$], with the ratios of K and Ba being 1/3 and 2/3 respectively in the pseudopotential, since they possess similar ionic radii (K$^+$: 1.33 {\AA}, Ba$^{2+}$: 1.35 {\AA}) and electronic shells. The VCA is further validated by combinations of other pairs of alkali and alkaline earth elements. Namely, we also calculate the band structure using the mixing Rb-Sr, Rb-Ba, Cs-Sr, and Cs-Ba to approximate the mixed atom, in addition to K-Ba. In all cases, the bands near $E_{\mathrm{F}}$ remained almost unaffected either with or without SOC (see Section \ref{sec-vca}, Supporting Information). This is consistent with the fact that the intercalated ions show negligible influence on the electronic structure around $E_{\mathrm{F}}$ (as pointed out in Section \ref{subsec-bulk}), indicative of the expedient exemption from more sophisticated treatments (e.g., special quasi-random structures \cite{PhysRevLett.65.353} or coherent potential approximation \cite{PhysRev.156.809, wimmer2021mn}) for the fractional occupation.

To assess the weight of interstitial orbitals in Bloch states in K$_4$Ba$_2$[SnBi$_4$], we set empty spheres with the radius 1.45 {\AA} at the Wyckoff site $2a$ $(0, 0, 0)$. Projection of Bloch states onto the $s$-orbital within the empty spheres indicates the contribution from interstitial states. 

\partitle{Wannier functions and band topology.} Wannier functions are built via the code \texttt{Wannier90} \cite{PhysRevB.56.12847, PhysRevB.65.035109, pizzi2020wannier90}. The maximal localization procedure is not applied, and a post-processing symmetrization \cite{PhysRevMaterials.2.103805} is conducted, so that the Wannier functions can respect the symmetries of the crystal and atomic orbitals, which is crucial to high-quality band interpolation and surface state prediction. The trial projectors for K$_4$Ba$_2$[SnBi$_4$] are chosen as interstitial $s$, Sn $sp^3$, Bi $p$, and mixed-atom $d_{z^2}$ orbitals; the $d_{z^2}$ orbital is selected because the bottom of the corresponding band is entangled with the top of Bi $p$ band (Figure \ref{bulk}c), albeit with a small overlap. The $\mathbb{Z}_2$ invariants are evaluated through Wannier charge centers over the occupied bands \cite{PhysRevB.83.235401}. The chirality of Weyl points is computed by integrating the Berry flux over a surface enclosing the Weyl point \cite{hirayama2018topological, RevModPhys.90.015001}. Both calculations ($\mathbb{Z}_2$ and chirality) are performed based on the Wannier tight-binding Hamiltonian as implemented in the \texttt{WannierTools} package \cite{wu2018wanniertools}.

\partitle{Surface states.} We utilize the iterative Green's function method \cite{sancho1985highly} to compute the $(10\bar10)$ and (0001) surface states of K$_4$Ba$_2$[SnBi$_4$], by constructing a semi-infinite slab on the basis of Wannier functions. The calculation is conducted with \texttt{WannierTools}. Note that we may need to shift the positions of all Wannier functions in order to keep the [SnBi$_4$]$^{8-}$ tetrahedron as an entirety inside a unit cell, otherwise the Sn-Bi covalent bond would be cut and thus produce trivial surface states.

To take the charge redistribution at the surface into account, we also perform a DFT calculation for an 8-layer $(10\bar10)$ slab (thickness $\approx$ 79 {\AA}) in the presence of SOC. 20 {\AA} of vacuum is adopted so that the vacuum level is tested to converge. The outermost K$^+$/Ba$^{2+}$ ions and [SnBi$_4$]$^{8-}$ tetrahedra are relaxed; no significant improvement in the band structure can be seen if relaxing more atoms in the deeper layers.

\partitle{Thermoelectricity.} The Seebeck coefficient $S$ and electrical conductivity $\sigma$ are assessed in the framework of Boltzmann transport theory under the rigid band picture, with the help of \texttt{\textsc{BoltzWann}} module \cite{pizzi2014boltzwann} in \texttt{Wannier90}. The constant scattering time approximation is adopted, so that the scattering time $\tau$ (that is unknown for K$_4$Ba$_2$[SnBi$_4$], not measured in experiment yet and highly intricate to computationally estimate) can be extracted from $\sigma$, and cancels in the expression of $S$ \cite{pizzi2014boltzwann}. $\sigma$ is estimated in the unit of $\tau$. The Wannier functions are obtained from the DFT results with SOC, under the HSE06 hybrid functional, since the hybrid functional is believed to describe the band edge more accurately in general, which is crucial to thermoelectrics \cite{goldsmid2010thermoelectricity}. A $50\times50\times50$ fine $\boldsymbol{k}$-grid is used.

To characterize the phononic transport properties of K$_4$Ba$_2$[SnBi$_4$], we switch from \texttt{VASP} to the \texttt{\textsc{Quantum ESPRESSO}} package \cite{giannozzi2009quantum, giannozzi2017advanced}, because computing interatomic force constants within the VCA is not supported in \texttt{VASP}. Since the lattice dynamics is relatively insensitive to the details of the electronic band edge, it suffices to use a pure functional (PBEsol) instead of the expensive hybrid functional (HSE06). The PBEsol pseudopotentials from \texttt{pslibrary} \cite{dal2014pseudopotentials} are applied. SOC has small impact on the lattice dynamics and is ignored to save the computational cost. The atomic coordinates are further relaxed with a force criterion of 0.1 meV/{\AA}. A $1\times1\times2$ supercell is built for the evaluations of phononic spectrum and lattice thermal conductivity $\kappa_l$, since the lattice constants $a$ and $b$ are already quite large and over 10 {\AA}. In particular, anharmonic phonons are computed to access $\kappa_l$, through the Boltzmann transport theory for phonons under the scattering time approximation, namely,
\begin{align}
    \kappa_l(T) = \frac{1}{VN_q} \sum_{\boldsymbol{q},j} c_{\boldsymbol{q},j}(T) \boldsymbol{v}_{\boldsymbol{q},j} \otimes \boldsymbol{v}_{\boldsymbol{q},j} \tau_{\boldsymbol{q},j}(T).
    \nonumber
\end{align}
Here $T$ denotes the temperature, $\boldsymbol{q}$ the wave vector, $j$ the polarization, $V$ the unit cell volume, $N_q$ the number of $\boldsymbol{q}$ points, $c_{\boldsymbol{q},j} = \hbar\omega_{\boldsymbol{q},j}\,\partial n_{\boldsymbol{q},j}/\partial T$, with $\omega_{\boldsymbol{q},j}$ the phonon frequency and $n_{\boldsymbol{q},j}$ the occupation number, $\boldsymbol{v}_{\boldsymbol{q},j}$ the phonon velocity, $\tau_{\boldsymbol{q},j}$ the phonon lifetime. The anharmonic interatomic force constants are computed up to the spacings around 10 {\AA}. The calculations of $\kappa_l$ are done through the code \texttt{ALAMODE} \cite{tadano2014anharmonic}, sampled on a $12\times12\times12$ fine $\boldsymbol{q}$-grid.

\partitle{Work function.} We construct a slab along a given direction and calculate the Fermi energy ($E_{\text{F,slab}}$) as well as the vacuum potential energy ($\phi_{\text{vac}}$) outside the material under the formalism of DFT. The work function ($W$) for the corresponding surface is thus expressed as $W = \phi_{\text{vac}} - E_{\text{F,slab}}$ \cite{cahen2003electron}. Since the bulk is insulating but the surface is metallic in TIs, we determine $E_{\text{F,slab}}$ directly through the slab calculation, instead of deriving it from the bulk. In fact, the latter value, which is obtained by shifting the bulk Fermi level ($E_{\text{F,bulk}}$) to align the averaged Hartree potential of the slab middle layers with that of the bulk \cite{fall1999deriving}, is tested to be almost equal to the former, within the range of the band gap (i.e., a few 10 meV).

The 8-layer $(10\bar10)$ slab of K$_4$Ba$_2$[SnBi$_4$] relaxed in the same way as in the surface state calculation is adopted. SOC is switched on. For K$_{10}$[Sn$_2$Bi$_6$] and K$_3$Na$_2$[SnBi$_3$], a 6-layer (100) slab and a 6-layer (010) slab are built, respectively. The thickness of the vacuum is 20 {\AA} in both cases. The outermost ions and molecular units are relaxed.

\begin{acknowledgments}
The authors thank S. Matsuishi, H. Hosono, C.-L. Zhang, and Y. Tokura for valuable discussions.
T. Y. acknowledges financial support from RIKEN Junior Research Associate Program and JSPS KAKENHI Grants (No. 22J23068).
R. A. was supported by a Grant-in-Aid for Scientific Research (No. 19H05825) from the Ministry of Education, Culture, Sports, Science and Technology.
M. H. acknowledges financial support from JSPS KAKENHI Grants (No. 20K14390) and PRESTO, JST (JPMJPR21Q6).
\end{acknowledgments}

\section*{References}

\vfill
\end{bibunit}


\begin{bibunit}

\onecolumngrid
\pagebreak
\phantom{\color{white}\pagebreak}
\twocolumngrid
\newpage
\onecolumngrid

\titleformat*{\section}{\normalsize\bfseries\sffamily\large\raggedright}
\renewcommand\thesection{S\arabic{section}}
\renewcommand{\figurename}{\textsf{\textbf{Figure }}}
\renewcommand{\thefigure}{\textsf{\textbf{{S\arabic{figure}}}}}\renewcommand{\theHfigure}{S\arabic{figure}}

\renewcommand{\tablename}{\textsf{\textbf{Table }}}
\renewcommand{\thetable}{\textsf{\textbf{{S\arabic{table}}}}}\renewcommand{\theHtable}{S\arabic{table}}

\setcounter{page}{1}
\setcounter{figure}{0}
\setcounter{section}{0}

\makeatletter
\def\l@subsection#1#2{}
\def\l@subsubsection#1#2{}
\renewcommand{\p@subsection}{}
\makeatother

\title{\textsf{Supporting Information for \texorpdfstring{\\ \vspace{0.2cm}}{} Interstitial-Electron-Induced Topological Molecular Crystals}}

\author{Tonghua Yu}             \affiliation{\AppliedPhys} 
\author{Ryotaro Arita}          \email{arita@riken.jp}   \affiliation{\RCAST} \affiliation{\RIKEN} 
\author{Motoaki Hirayama}       \email{hirayama@ap.t.u-tokyo.ac.jp}   \affiliation{\AppliedPhys} \affiliation{\RIKEN}

\setcounter{page}{1}
\maketitle
\vfill
\vfill
\twocolumngrid
\newpage
\twocolumngrid

\section{Electrides and interstitial orbitals}
\label{sec-electrides}
Electrides are a unique family of ionic compounds widely studied by chemists, less known to other communities. In this supplementary section, we elucidate electrides and the characteristic interstitial orbitals in a more specific way, and shed light on the difference between our candidate ITMCs and classic electrides.

We have presented a general notion of electrides at the beginning of Section \ref{subsec-ITMCs}, the pivotal trait of which is that electrons located at vacant space act as anions \cite{dye2003electrons}. Ideally, just like the scene that electrons are trapped at anionic vacancies in a crystal (known as F-centers), one may view electrides as stoichiometric F-centers in which anionic sites trap electrons \cite{dye2003electrons}. In practice, however, one will find that the interstitial electrons need not to be completely off the parent atoms and localized at the vacancy (for instance, by calculating the electron density). As a result, there seemingly exists ambiguity between the interstitial electrons of electrides and the ordinary atomic orbital hybridization often seen in solids (e.g., metallic or covalent bonding states).

Nevertheless, interstitial electrons stand out because of the novel phenomena that they lead to for electrides, and are not characteristic of normal materials with trivial electron density overlap. We support this point of view by showing the example of the pressurized Na. As is universally known, Na at ambient pressure is a silvery-white metal, of course with hybridization of neighboring $s$ electrons at non-atomic regions. But under high pressure (around 200 GPa), the system was experimentally found to be optically transparent and electrically insulating \cite{ma2009transparent}, which is counterintuitive since adding pressure normally gives rise to metallic phases. It turns out that the crystal transits from the original face-centered cubic structure to a double hexagonal closed packed structure with interstitial cavities; the cavities accommodate significant electron density that is responsible for the peculiar optical and electronic properties in the high-pressure phase \cite{ma2009transparent}. Namely, the pressurized Na is an electride that exhibits ionicity with the confined excess electrons being anions.

The nontriviality of interstitial electrons can be testified by the magnetic property too. Electrides can potentially show Stoner ferromagnetic or Mott insulating states, with the magnetic moments located at interstitial sites solely \cite{hosono2021advances, PhysRevB.69.195106, PhysRevLett.107.087201, PhysRevB.98.125128, sui2019prediction, PhysRevB.102.180407, PhysRevB.103.235126}. This motivates one to define interstitial orbitals in the tight-binding approximation, which are often chosen as $s$ orbitals considering the large spread of interstitial electrons, as we do for K$_4$Ba$_2$[SnBi$_4$] (see Section \ref{subsec-bulk}). Although the $s$-character of interstitial orbitals have origin from the $s$ electrons of surrounding parent atoms, the physical and chemical consequences are not simply explained by those atomic $s$ orbitals (recall the pressurized Na relative to the normal Na). The interstitial orbital is also evidenced by the band representation (BR) analysis in topological quantum chemistry \cite{bradlyn2017topological}. In conventional ionic crystals, the occupied bands can be expressed as a sum of atomic-orbital-induced BRs, whereas such expression is not applicable for electrides, unless BRs centered at vacancies are involved \cite{PhysRevB.103.205133}. In a nutshell, interstitial orbitals are built upon physical essence, more than a simplified picture.

In the meantime, one has to realize the implicit precondition for interstitial orbitals to take effect in electrides, that is, they are largely occupied in the ground state (see, for example, [Ca$_{24}$Al$_{28}$O$_{64}$]$^{4+}\cdot 4e^-$ \cite{matsuishi2003high} and [Ca$_2$N]$^+\cdot e^-$ \cite{lee2013dicalcium}). Otherwise, electrons are bound to atoms or molecules, with no excess ones serving as anions, and hence the electride characters are not manifested, at least not fully. In this respect, our ITMC candidates do not typically belong to the electride family. The interstitial band of ITMCs can be nearly unoccupied, as long as a minor portion is below $E_\mathrm{F}$ via band inversion -- in order for topological nontriviality (see Sections \ref{subsec-bulk} and \ref{subsec-otherITMCs}). Thus, these ITMC materials should be distinguished from the conventional electrides: First, as stated above, if the interstitial states are mostly above $E_\mathrm{F}$, the system does not show electride features completely (e.g., no exotic dielectric or optical properties \cite{hosono2021advances}), which could be mediated via external electron doping though. Second, interstitial orbitals are energetically less stable that atomic orbitals, as the latter are tightly constrained by nuclei. Therefore, ITMCs with small occupation of interstitial states gain stability compared to electrides. Last, a handful of theoretical descriptors have been developed to diagnose electrides \cite{zhao2016electride}, which heavily rely on the ground-state excess electron at vacant space. Consequently, they may not be directly applied to search for ITMCs, while extension to excited-state calculations is not trivial. Of course, one may regard our candidate ITMCs as a generalized form of electrides, but meanwhile keep in mind their distinction from the typical electride materials.

\section{Identifying K$_4$Ba$_2$[SnBi$_4$] as a molecular crystal}
\label{sec-chembonds}

There is no unified definition on molecular crystals. As indicated at the very beginning of this article, we adopt an inclusive description, that is, the primary components of the crystal are necessarily atomic clusters that are not extended along any dimension (i.e., 0D); the intracluster interactions ought to be stronger than intercluster ones (e.g., covalent bonding within the cluster and electrostatic forces among clusters) so that each cluster behaves as an entity and is therewith regarded as a molecule. The identity as a molecular crystal, or the 0D character, of K$_4$Ba$_2$[SnBi$_4$], has been uncovered several times in the main text, through the multiple surfaces free from dangling bonds (Section \ref{subsec-cleavage}), the strong response to moderate pressure (Section \ref{subsec-pressure}), and the phononic structure (isotropically low thermal conductivity of the lattice; Section \ref{subsec-thermoelectricity}). We provide more direct clarifications in this section, firstly analyzing the chemical bonding types of K$_4$Ba$_2$[SnBi$_4$].

Since K and Ba are much more electropositive (with the electronegativity 0.82 and 0.89, respectively) than Sn and Bi (electronegativity: 1.96 and 2.02), the valence electrons in the outermost $s$ shell of the former transfer to the latter, so that K and Ba become isolated cations altogether, and [SnBi$_4$]$^{8-}$ as a whole show the oxidation number $-8$. Further, within the substructure of [SnBi$_4$]$^{8-}$, covalent bonds are established between Sn and Bi, resulting from their close electronegativities. Indeed, the orbitals of Sn ($5s^25p^2$) are hybridized into $sp^3$, filled with four electrons. Each $sp^3$ electron is then paired with one $p$ electron of Bi ($6s^26p^3$), and four Bi atoms are thus connected with the central Sn in a tetrahedral shape through covalent bonding. The remaining two $p$ orbitals of each Bi that are half-occupied receive two extra electrons captured from cations (four Bi atoms are then assigned with eight electrons exactly), forming two lone pairs, whereby the [SnBi$_4$]$^{8-}$ polyanion reaches a full valence abiding by the octet rule. Note that the electron transfer is not complete, with minor electrons (much less than unity) remaining at the hollow site, which is key to the topological state (see Section \ref{subsec-bulk}), but inconsequential to the nominal valence of each cation or polyanion. In this way, [SnBi$_4$]$^{8-}$ is a molecular unit glued by the Sn-Bi covalent bonding, and interacts with K$^+$ and Ba$^{2+}$ through ionic bonding. In fact, the spacings between neighboring [SnBi$_4$]$^{8-}$ tetrahedra (5.3 - 5.6 {\AA}) are considerably greater than the intracluster Sn-Bi bond length (2.9 {\AA}), also suggesting the hallmark of molecular crystals \cite{kitaigorodsky2012molecular}. Similar analysis can be addressed to other ITMC candidates as well.

\begin{figure}[t]
	\centering
	\includegraphics[width=\linewidth]{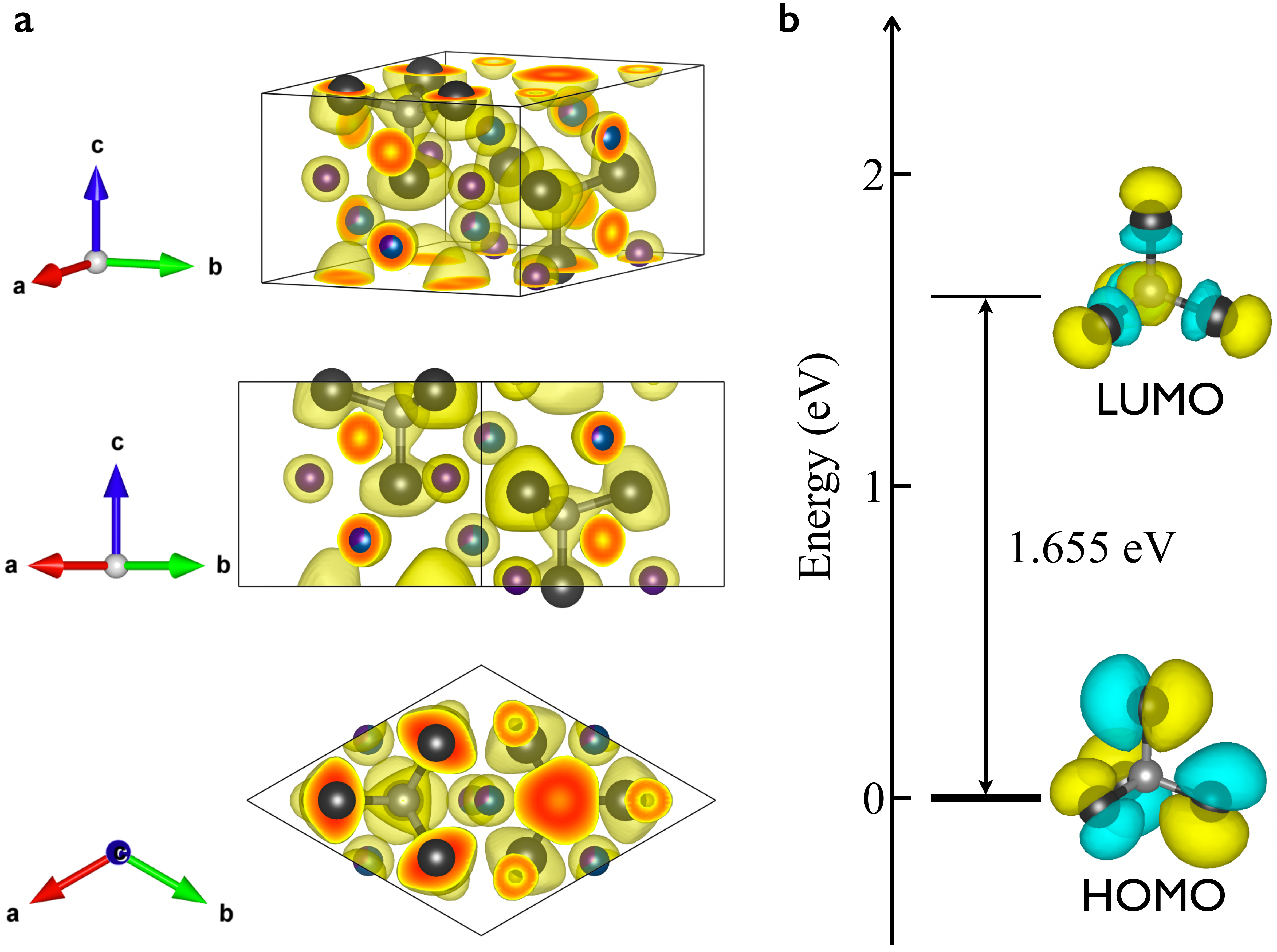}
    \caption{\sf Molecular character of K$_4$Ba$_2$[SnBi$_4$]. a) ELF of K$_4$Ba$_2$[SnBi$_4$]. The yellow bubble represents the ELF. Views from several different directions are shown. b) LUMO and one of HOMOs of [SnBi$_4$]$^{8-}$. Yellow and light blue indicate positive and negative components of the wavefunction, respectively.}
	\label{molecule}
\end{figure}

The arguments above are supported by our first-principles calculation. Electron localization function (ELF) is a variant of electron density, aiming to describe the degree of electron localization in real space \cite{becke1990simple, savin2005electron}. The local minimum of ELF is called the bonding attractor, with its occurrence between atomic shells indicative of shared-electron interactions (e.g., covalent bond), and its absence on the bond path reflecting unshared-electron interactions (e.g., ionic bond) \cite{silvi1994classification, zhao2016electride}. Figure \ref{molecule}a gives the ELF of K$_4$Ba$_2$[SnBi$_4$]. While bonding attractors appear in the middle region of Sn and Bi, suggesting the Sn-Bi covalent bond, the [SnBi$_4$]$^{8-}$ tetrahedra are completely separated from K$^+$ and Ba$^{2+}$ cations, as the evidence of ionic bonding in-between. This fact confirms the 0D nature of K$_4$Ba$_2$[SnBi$_4$].

We can acquire further insight by estimating the single [SnBi$_4$]$^{8-}$ molecule extracted from the bulk. For a quick view, we calculate the energy level of the neutral SnBi$_4$ cluster with DFT, and take the fourth (fifth) unoccupied orbital as the HOMO (LUMO) of the charged [SnBi$_4$]$^{8-}$ (since eight extra electrons would fill in four orbitals). Despite the rough approximation in this calculation, we may gain useful information consistent with early discussions. The highest energy for occupied states of [SnBi$_4$]$^{8-}$ is triply degenerate, with three HOMOs, originating from Bi $6p$ orbitals. Figure \ref{molecule}b depicts one of the HOMOs; they may respect the symmetry of the tetrahedron (or of the crystal if in the bulk) via linear combination. In the bulk, the degeneracy is removed by the crystal field, and the HOMO is inverted to the conduction band (Figures \ref{bulk}c,e). We also notice a wide HOMO-LUMO gap (1.655 eV) in [SnBi$_4$]$^{8-}$, in line with the closed shell of the molecule, indicating the possible stability of the individual [SnBi$_4$]$^{8-}$ cluster (e.g., via dissolving in appropriate solvents).
\begin{figure}[b]
	\centering
	\includegraphics[width=\linewidth]{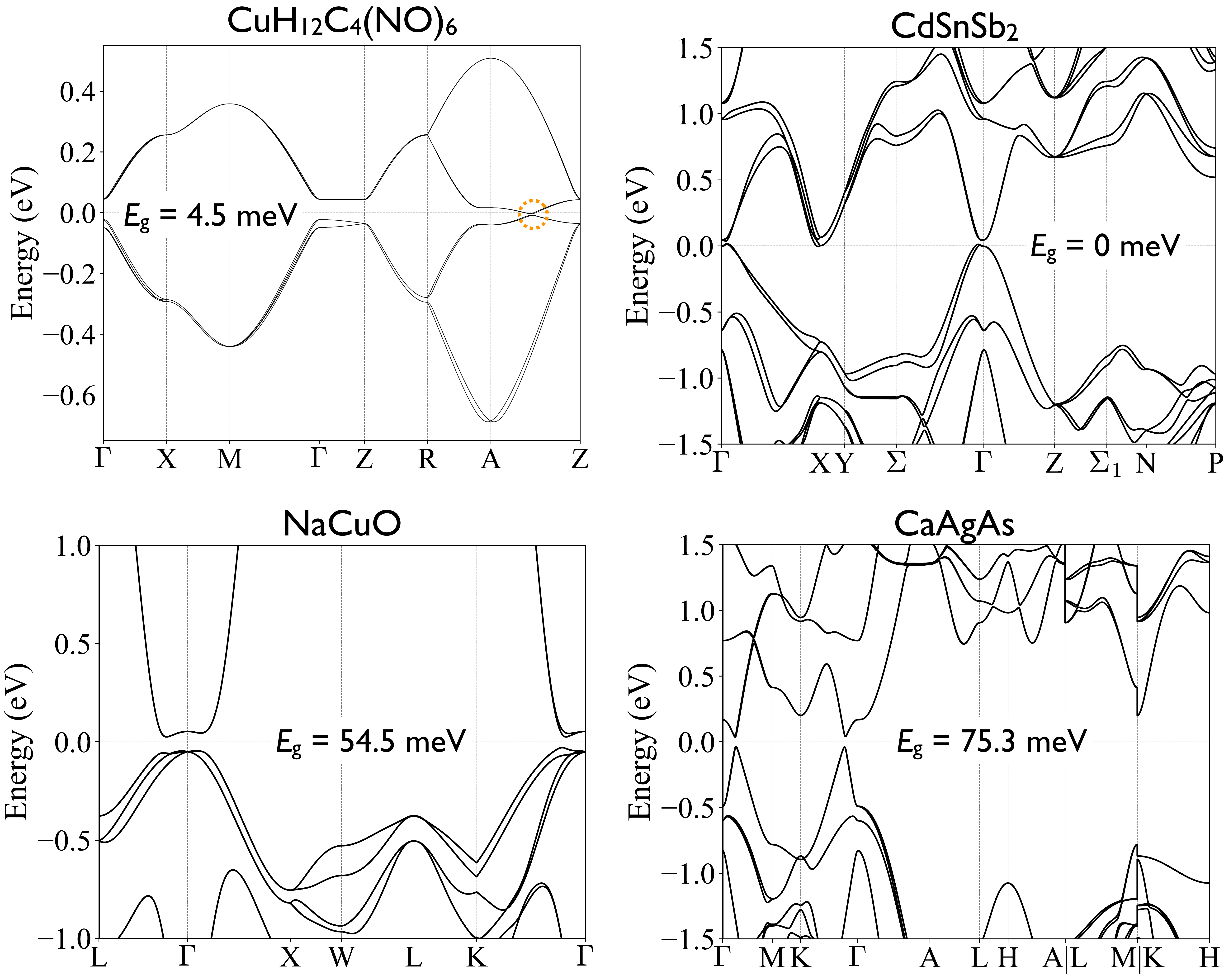}
	\caption{\sf Band structures of noncentrosymmetric TIs. Calculations are performed in the framework of DFT with the PBE parametrization. For CdSnSb$_2$, the indirect gap is evaluated to be zero, despite the finite direct gap.}
	\label{TIs_gap}
\end{figure}

One consequence of low structural dimensionality is the weak electronic dispersions, except for those caused by interstitial states (Figures \ref{bulk}c,d). Similar weak dispersions are observed in other charge-transfer-type low-dimensional systems \cite{PhysRevX.7.021019, noguchi2019weak, PhysRevX.11.031042, shi2021charge}, albeit not as flat as in van der Waals materials.

\section{Band gaps of noncentrosymmetric TIs}
\label{sec-TI_gaps}
We obtain large-gap noncentrosymmetric TIs from the topological materials database \texttt{Materiae} \cite{zhang2019catalogue}. Only those with the estimated band gap ($E_{\mathrm{g}}$) larger than 50 meV are selected. Note that TIs with broken inversion symmetry generally have a smaller band gap than inversion-symmetric ones; while for the latter $E_{\mathrm{g}}$ can reach hundreds of meV (e.g., tetradymites), for the former $E_g > 100$ meV is hardly found. Since the band structures in the materials database are estimated via a high-throughput flow and numerical inadequacy would be unavoidable, we recalculate the selected TIs using more adaptive settings mentioned in Section \ref{sec-methods} Methods. Figure \ref{TIs_gap} presents the relativistic band structures at the level of PBE functional, with narrower or comparable gaps relative to K$_4$Ba$_2$[SnBi$_4$] (75.4 meV). Although CuH$_{12}$C$_4$(NO)$_6$ and CdSnSb$_2$ (and its isostructural partner ZnSnSb$_2$) are predicted to have large $E_{\mathrm{g}}$ ($> 50$ meV) by the database, their gaps almost vanish according to our calculations.

\section{DFT band structure of the K$_4$Ba$_2$[SnBi$_4$] slab}
\label{sec-DFTslabband}

\begin{figure}[hb]
	\centering
	\includegraphics[width=0.55\linewidth]{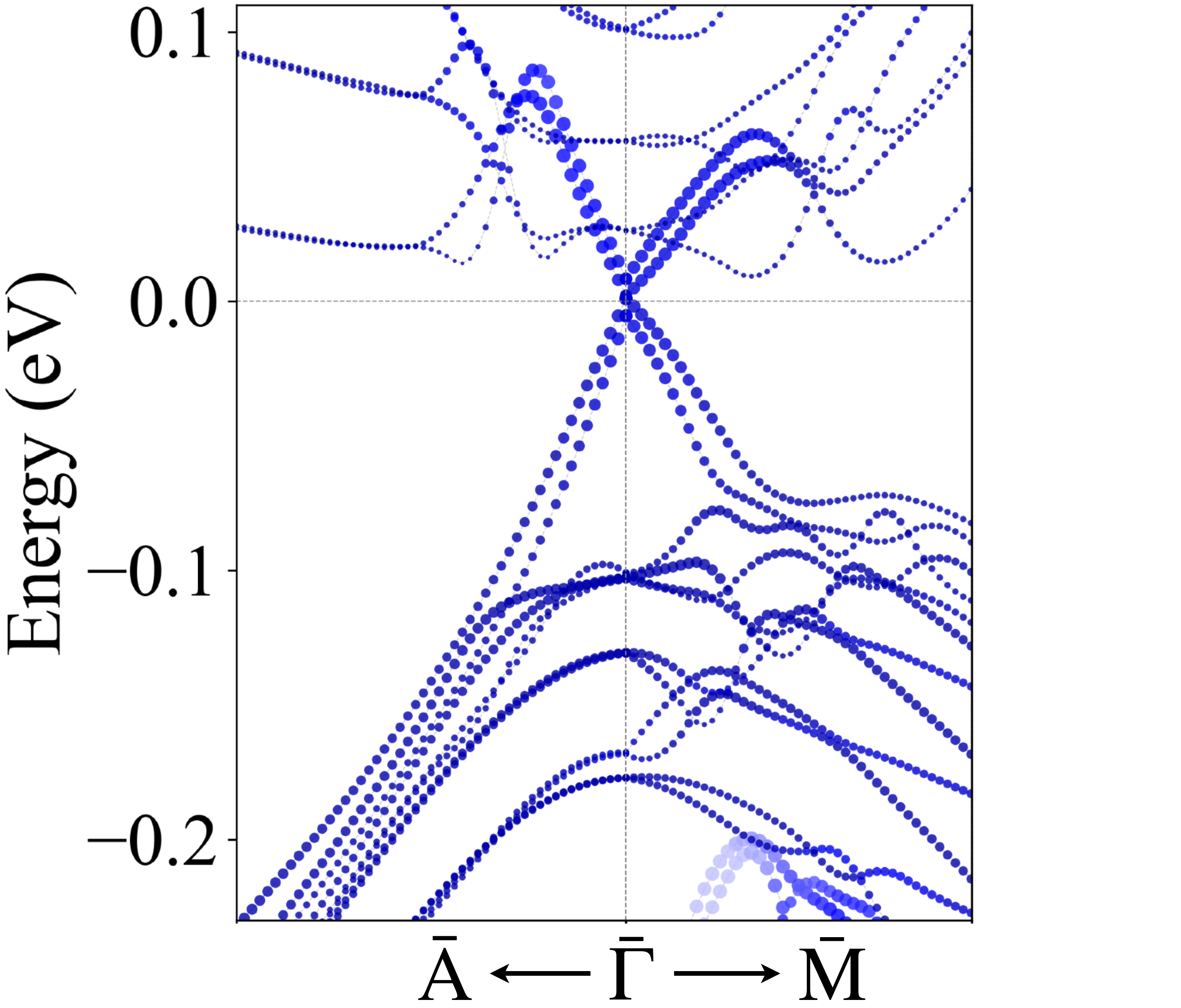}
	\caption{\sf DFT band structure of an 8-layer $(10\bar{1}0)$ K$_4$Ba$_2$[SnBi$_4$] slab. High-symmetry $\boldsymbol{k}$ points are given in Figure \ref{bulk}b. Energy is measured from the Fermi level. The size of dots indicates contribution of outermost layers.}
	\label{dft_slab_band}
\end{figure}

We calculate the band structure of an 8-layer $(10\bar10)$ K$_4$Ba$_2$[SnBi$_4$] slab (thickness $\approx$ 79 {\AA}) with SOC under DFT, as illustrated in Figure \ref{dft_slab_band}. The linear gapless TSS are consistent with the prediction by Wannier functions (Figure \ref{cleavage}e), despite the slight difference of the projected bulk bands due to the boundary effect. The asymmetric (0001) slab is skipped because complicated setup of surface defects or adatoms is needed to cure the energy divergence caused by the polar surface \cite{tasker1979stability}. Still, we can expect a surface spectrum akin to the tight-binding model after neutralizing the surface, just like the correspondence in the $(10\bar10)$ slab.

\section{Work function of K$_4$Ba$_2$[SnBi$_4$]}
\label{sec-workfunc}
\begin{figure}[ht]
	\centering
	\includegraphics[width=\linewidth]{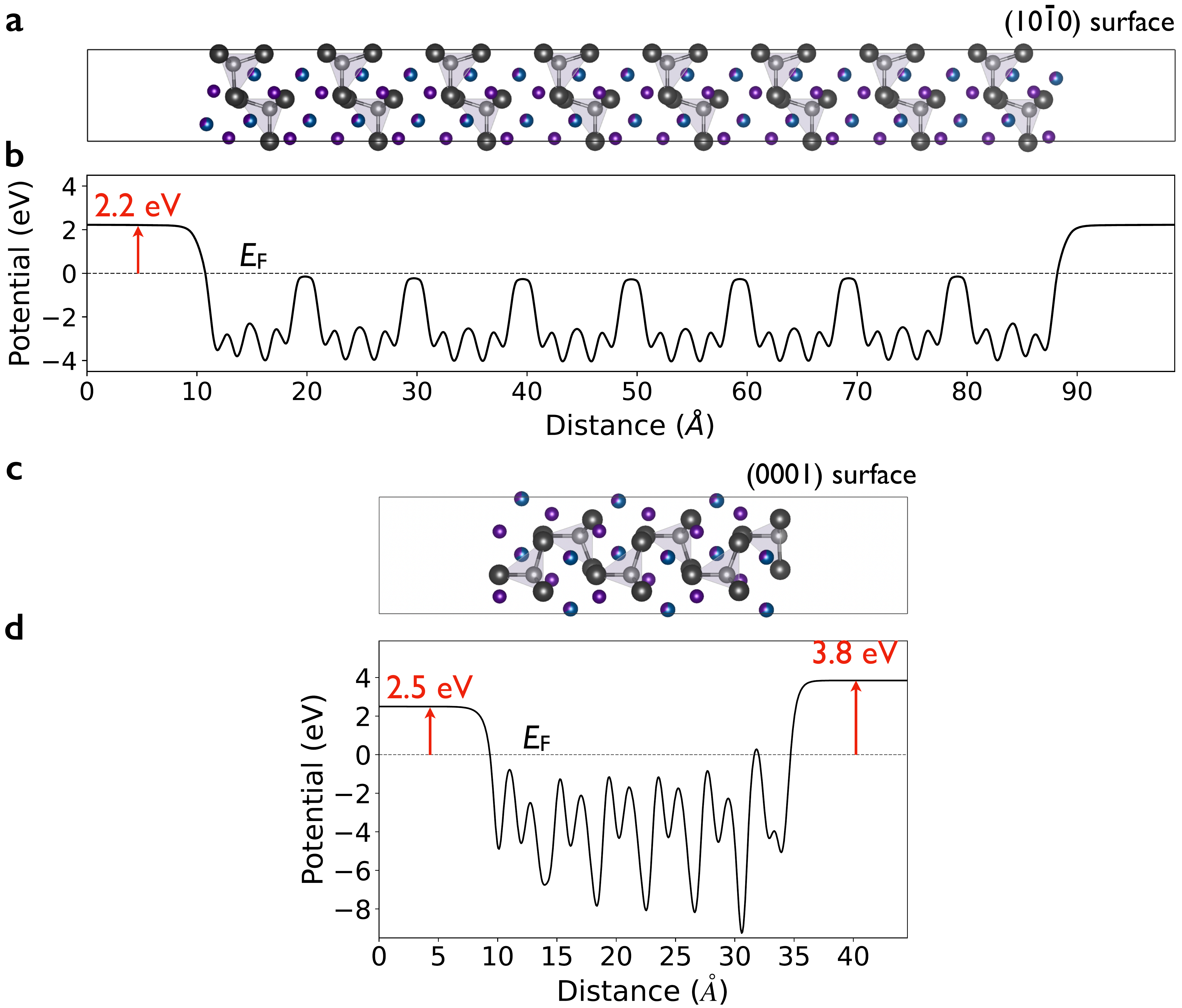}
	\caption{\sf Work function of K$_4$Ba$_2$[SnBi$_4$]. a) 8-layer $(10\bar10)$ K$_4$Ba$_2$[SnBi$_4$] slab. b) Averaged Hartree potential along the $[10\bar10]$ direction. c) 3-layer $(0001)$ K$_4$Ba$_2$[SnBi$_4$] slab. d) Averaged Hartree potential along the $(0001)$ direction. The Fermi level has been set as the reference level of energy.}
	\label{workfunc}
\end{figure}

The work function of K$_4$Ba$_2$[SnBi$_4$] $(10\bar10)$ is estimated through an 8-layer slab (Figure \ref{workfunc}a) with the inclusion of SOC. The averaged Hartree potential energy along the $[10\bar10]$ direction is presented in Figure \ref{workfunc}b, showing that the vacuum level is 2.2 eV above $E_{\mathrm{F}}$, indicative of the work function.

We additionally characterize the work function for the polar (0001) slab. To circumvent the intricate setting of defects or adatoms that is required to simulate the significant reconstruction of the charged surface, we restrict our scope to a thin film (3 layers, thickness $\approx$ 25 {\AA}) as a compromise, as shown in Figure \ref{workfunc}c. The effective screening medium method \cite{PhysRevB.73.115407} is implemented to eliminate the spurious dipole interactions between periodic images. Atoms in the outermost layers are relaxed. Figure \ref{workfunc}d depicts the unbalanced work function for each end [3.8 eV for $(0001)$, and 2.5 eV for $(000\bar1)$], as routinely seen in polar slabs. Here the $(0001)$ vacuum potential is greatly lifted by the electric field of the negative charge on this surface. A low work function can be retrieved through surface neutralization.

\section{Interfaces of K$_4$Ba$_2$[SnBi$_4$] with molecules}
\label{sec-interfaces}
\begin{figure}[ht]
	\centering
	\includegraphics[width=\linewidth]{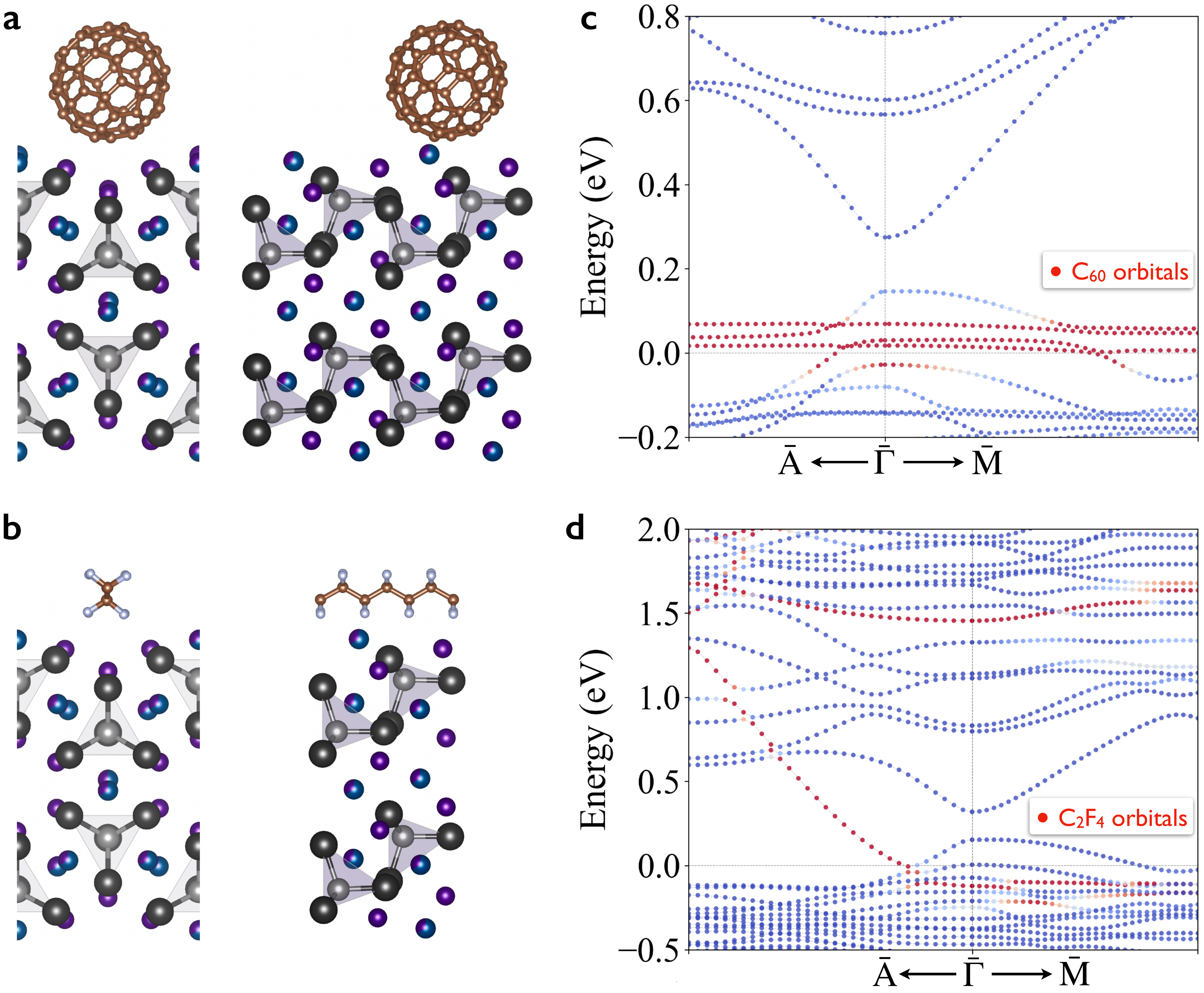}
	\caption{\sf Interface systems of K$_4$Ba$_2$[SnBi$_4$] with other molecules. a,b) A $(10\bar10)$ K$_4$Ba$_2$[SnBi$_4$] slab with the adsorption of fullerene C$_{60}$ (a) and PTFE (C$_2$F$_4$)$_n$ (b). Both front and side views are provided. c,d) Band structures of the interface system with C$_{60}$ (c) and (C$_2$F$_4$)$_n$ (d). SOC is not included. States arising from the adsorbed molecules are highlighted in red.}
	\label{interfaces}
\end{figure}

We investigate the interface of K$_4$Ba$_2$[SnBi$_4$] with the fullerene C$_{60}$. The C$_{60}$ molecule occupies the hollow on the $(10\bar10)$ surface of K$_4$Ba$_2$[SnBi$_4$], as shown in the left panel of Figure \ref{interfaces}a. As the diameter of C$_{60}$ is quite large ($\approx 7.1$ {\AA}) and close to the lattice constant $c$ ($\approx 8.3$ {\AA}) of K$_4$Ba$_2$[SnBi$_4$], we double the unit cell along the $\boldsymbol{c}$ direction (i.e., [0001]). To reduce the computational expense, only a 2-layer slab of K$_4$Ba$_2$[SnBi$_4$] is used, and SOC is switched off; we aim at a qualitatively heuristic result rather than quantitative rigor. The outermost cations and polyanions, together with the distance of C$_{60}$ to the surface are fully relaxed. The position of C$_{60}$ along [0001] is optimized (as given in the right panel of Figure \ref{interfaces}a) so that the total energy of the interface system is at the minimum. Figure \ref{interfaces}c displays the band structure of this interface system. The LUMOs of C$_{60}$ are successfully doped, owing to the low work function of K$_4$Ba$_2$[SnBi$_4$]. Importantly, the doped C$_{60}$ orbitals are next to the surface Dirac cone of K$_4$Ba$_2$[SnBi$_4$] [not manifested in Figure \ref{interfaces}c as SOC is absent], revealing the intriguing possibility that the massless Dirac Fermions might interact with the fullerene molecule or fulleride crystal with further fine tuning.

A $(10\bar10)$ K$_4$Ba$_2$[SnBi$_4$] slab interfaced with a polytetrafluoroethylene (PTFE) (C$_2$F$_4$)$_n$ chain is studied as well, as shown in Figure \ref{interfaces}b. The PTFE is a stable fluorocarbon polymer with a small electron affinity. When contacted with K$_4$Ba$_2$[SnBi$_4$], the LUMO of (C$_2$F$_4$)$_n$ is filled and adjacent to the Dirac cone (Figure \ref{interfaces}d), just like C$_{60}$, as a reflection of doping TSS to molecules or non-topological insulators even with a large gap.

\section{Influence of the \textit{k}-grid on the band gap}
\label{sec-kgrid}

\begin{table}[b]
    \centering
    \begin{tabular}{c|c|c}\hline
        $\boldsymbol{k}$-grid & Resolution ($\times 2\pi$ {\AA$^{-1}$}) & Band gap (eV) \\ \hline
        $2\times2\times2$ & 0.050 & 0.0201 \\
        $3\times3\times4$ & 0.030 & 0.0206 \\
        $4\times4\times5$ & 0.024 & 0.0207 \\
        $5\times5\times6$ & 0.020 & 0.0208 \\
        $6\times6\times7$ & 0.017 & 0.0208 \\ \hline
    \end{tabular}
    \caption{\sf Band gap of K$_4$Ba$_2$[SnBi$_4$] under GGA without SOC.}
    \label{gap-cvg}
\end{table}
We calculate the band gap of K$_4$Ba$_2$[SnBi$_4$] with GGA and no SOC, under a series of $\boldsymbol{k}$-grids with increasing resolutions (Table \ref{gap-cvg}). The gap turns out to be quite insensitive to the $\boldsymbol{k}$-grid, the variation of which is only at a sub-meV level. Convergence has been almost reached with the resolution of $2\pi\times0.030$ {\AA$^{-1}$} ($3\times3\times4$ grid), as indicated in Table \ref{gap-cvg}. Further increasing the resolution improves the gap of merely 0.1-0.2 meV. Therefore, the setting $2\pi\times0.030$ {\AA$^{-1}$} ($4\times4\times5$ grid) as given in Section \ref{sec-methods} Methods can be claimed reasonably sufficient. One can then expect the predicted gap in the presence of SOC (i.e., 75.4 meV under GGA+SOC), or with a different exchange functional (i.e., 85.7 meV under HSE06+SOC) to reach a sub-meV precision, since the dependence on $\boldsymbol{k}$-grid for the same bulk system is transferable.

\section{Justification of the VCA}
\label{sec-vca}
\begin{figure}
	\centering
	\includegraphics[width=\linewidth]{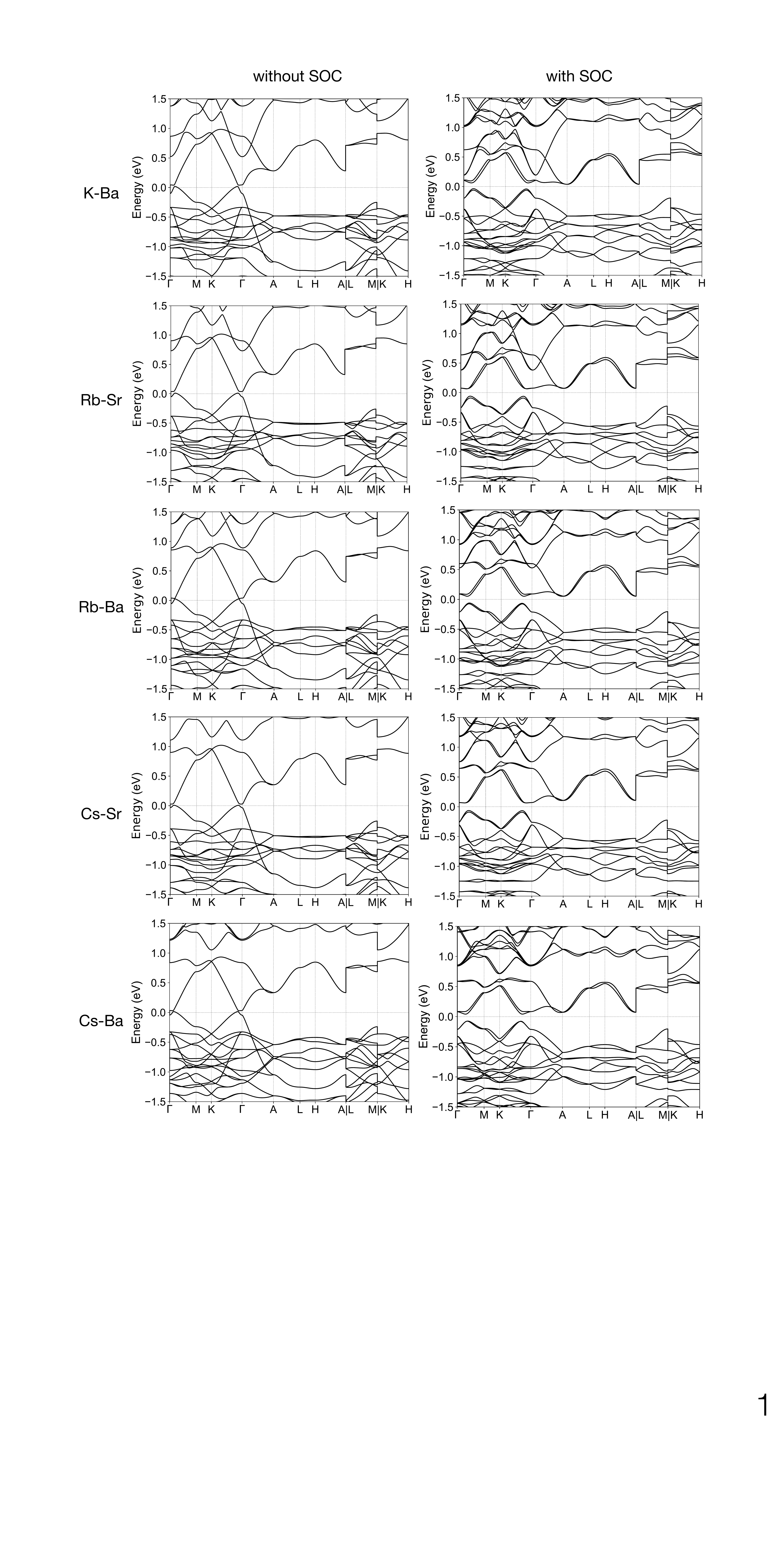}
	\caption{\sf Band structures of K$_4$Ba$_2$[SnBi$_4$] under different settings of the VCA. The first row (K-Ba) means that the pseudopotential is obtained by the interpolation of 1/3 K and 2/3 Ba; analogously for the remaining rows.}
	\label{vca_validation}
\end{figure}

The fractional occupation sites in K$_4$Ba$_2$[SnBi$_4$] (Figure \ref{bulk}a) are described with the pseudopotential interpolated by those of K and Ba (i.e., VCA). To justify this approximation, the band structure is evaluated under the pseudopotentials interpolated also by other pairs of alkali and alkaline earth elements. Here we take Rb-Sr, Rb-Ba, Cs-Sr, and Cs-Ba. Figure \ref{vca_validation} shows that the band structures in the vicinity of $E_{\mathrm{F}}$ keep almost unchanged regardless of the mixing type, no matter whether SOC is included or not. We can, therefore, safely harness the VCA, with no need of implementing more sophisticated approximations, to which the electronic properties of K$_4$Ba$_2$[SnBi$_4$] are insensitive.
\vfill
\newpage


\begin{thebibliography}{91}%
\makeatletter
\providecommand \@ifxundefined [1]{%
 \@ifx{#1\undefined}
}%
\providecommand \@ifnum [1]{%
 \ifnum #1\expandafter \@firstoftwo
 \else \expandafter \@secondoftwo
 \fi
}%
\providecommand \@ifx [1]{%
 \ifx #1\expandafter \@firstoftwo
 \else \expandafter \@secondoftwo
 \fi
}%
\providecommand \natexlab [1]{#1}%
\providecommand \enquote  [1]{``#1''}%
\providecommand \bibnamefont  [1]{#1}%
\providecommand \bibfnamefont [1]{#1}%
\providecommand \citenamefont [1]{#1}%
\providecommand \href@noop [0]{\@secondoftwo}%
\providecommand \href [0]{\begingroup \@sanitize@url \@href}%
\providecommand \@href[1]{\@@startlink{#1}\@@href}%
\providecommand \@@href[1]{\endgroup#1\@@endlink}%
\providecommand \@sanitize@url [0]{\catcode `\\12\catcode `\$12\catcode
  `\&12\catcode `\#12\catcode `\^12\catcode `\_12\catcode `\%12\relax}%
\providecommand \@@startlink[1]{}%
\providecommand \@@endlink[0]{}%
\providecommand \url  [0]{\begingroup\@sanitize@url \@url }%
\providecommand \@url [1]{\endgroup\@href {#1}{\urlprefix }}%
\providecommand \urlprefix  [0]{URL }%
\providecommand \Eprint [0]{\href }%
\providecommand \doibase [0]{http://dx.doi.org/}%
\providecommand \selectlanguage [0]{\@gobble}%
\providecommand \bibinfo  [0]{\@secondoftwo}%
\providecommand \bibfield  [0]{\@secondoftwo}%
\providecommand \translation [1]{[#1]}%
\providecommand \BibitemOpen [0]{}%
\providecommand \bibitemStop [0]{}%
\providecommand \bibitemNoStop [0]{.\EOS\space}%
\providecommand \EOS [0]{\spacefactor3000\relax}%
\providecommand \BibitemShut  [1]{\csname bibitem#1\endcsname}%
\let\auto@bib@innerbib\@empty
\bibitem [{\citenamefont {Kitaigorodsky}(2012)}]{kitaigorodsky2012molecular}%
  \BibitemOpen
  \bibfield  {author} {\bibinfo {author} {\bibfnamefont {Alexander I.}~\bibnamefont
  {Kitaigorodsky}},\ }\href@noop {} {\enquote {\bibinfo {title} {Molecular
  crystals and molecules}}},\ Vol.~\bibinfo {volume} {29}\ (\bibinfo
  {publisher} {Elsevier},\ \bibinfo {year} {2012})\BibitemShut {NoStop}%
\bibitem [{\citenamefont {Kronik}\ and\ \citenamefont
  {Neaton}(2016)}]{kronik2016excited}%
  \BibitemOpen
  \bibfield  {author} {\bibinfo {author} {\bibfnamefont {Leeor}\ \bibnamefont
  {Kronik}}\ and\ \bibinfo {author} {\bibfnamefont {Jeffrey~B.}\ \bibnamefont
  {Neaton}},\ }\bibfield  {title} {\enquote {\bibinfo {title} {Excited-state
  properties of molecular solids from first principles},}\ }\href {\doibase
  10.1146/annurev-physchem-040214-121351} {\bibfield  {journal} {\bibinfo
  {journal} {Annual Review of Physical Chemistry}\ }\textbf {\bibinfo {volume}
  {67}},\ \bibinfo {pages} {587--616} (\bibinfo {year} {2016})}\BibitemShut
  {NoStop}%
\bibitem [{\citenamefont {Claridge}\ \emph {et~al.}(2009)\citenamefont
  {Claridge}, \citenamefont {Castleman~Jr}, \citenamefont {Khanna},
  \citenamefont {Murray}, \citenamefont {Sen},\ and\ \citenamefont
  {Weiss}}]{claridge2009cluster}%
  \BibitemOpen
  \bibfield  {author} {\bibinfo {author} {\bibfnamefont {Shelley~A}\
  \bibnamefont {Claridge}}, \bibinfo {author} {\bibfnamefont {AW}~\bibnamefont
  {Castleman~Jr}}, \bibinfo {author} {\bibfnamefont {Shiv~N}\ \bibnamefont
  {Khanna}}, \bibinfo {author} {\bibfnamefont {Christopher~B}\ \bibnamefont
  {Murray}}, \bibinfo {author} {\bibfnamefont {Ayusman}\ \bibnamefont {Sen}}, \
  and\ \bibinfo {author} {\bibfnamefont {Paul~S}\ \bibnamefont {Weiss}},\
  }\bibfield  {title} {\enquote {\bibinfo {title} {Cluster-assembled
  materials},}\ }\href {\doibase 10.1021/nn800820e} {\bibfield  {journal}
  {\bibinfo  {journal} {ACS nano}\ }\textbf {\bibinfo {volume} {3}},\ \bibinfo
  {pages} {244--255} (\bibinfo {year} {2009})}\BibitemShut {NoStop}%
\bibitem [{\citenamefont {Yin}\ \emph {et~al.}(2017)\citenamefont {Yin},
  \citenamefont {Maity}, \citenamefont {De~Bastiani}, \citenamefont {Dursun},
  \citenamefont {Bakr}, \citenamefont {Br{\'e}das},\ and\ \citenamefont
  {Mohammed}}]{yin2017molecular}%
  \BibitemOpen
  \bibfield  {author} {\bibinfo {author} {\bibfnamefont {Jun}\ \bibnamefont
  {Yin}}, \bibinfo {author} {\bibfnamefont {Partha}\ \bibnamefont {Maity}},
  \bibinfo {author} {\bibfnamefont {Michele}\ \bibnamefont {De~Bastiani}},
  \bibinfo {author} {\bibfnamefont {Ibrahim}\ \bibnamefont {Dursun}}, \bibinfo
  {author} {\bibfnamefont {Osman~M}\ \bibnamefont {Bakr}}, \bibinfo {author}
  {\bibfnamefont {Jean-Luc}\ \bibnamefont {Br{\'e}das}}, \ and\ \bibinfo
  {author} {\bibfnamefont {Omar~F}\ \bibnamefont {Mohammed}},\ }\bibfield
  {title} {\enquote {\bibinfo {title} {Molecular behavior of zero-dimensional
  perovskites},}\ }\href {\doibase 10.1126/sciadv.1701793} {\bibfield
  {journal} {\bibinfo  {journal} {Science advances}\ }\textbf {\bibinfo
  {volume} {3}},\ \bibinfo {pages} {e1701793} (\bibinfo {year}
  {2017})}\BibitemShut {NoStop}%
\bibitem [{\citenamefont {Ju}\ \emph {et~al.}(2018)\citenamefont {Ju},
  \citenamefont {Dai}, \citenamefont {Ma}, \citenamefont {Zhou},\ and\
  \citenamefont {Zeng}}]{ju2018zero}%
  \BibitemOpen
  \bibfield  {author} {\bibinfo {author} {\bibfnamefont {Ming-Gang}\
  \bibnamefont {Ju}}, \bibinfo {author} {\bibfnamefont {Jun}\ \bibnamefont
  {Dai}}, \bibinfo {author} {\bibfnamefont {Liang}\ \bibnamefont {Ma}},
  \bibinfo {author} {\bibfnamefont {Yuanyuan}\ \bibnamefont {Zhou}}, \ and\
  \bibinfo {author} {\bibfnamefont {Xiao~Cheng}\ \bibnamefont {Zeng}},\
  }\bibfield  {title} {\enquote {\bibinfo {title} {Zero-dimensional
  organic--inorganic perovskite variant: transition between molecular and solid
  crystal},}\ }\href {\doibase 10.1021/jacs.8b03917} {\bibfield  {journal}
  {\bibinfo  {journal} {Journal of the American Chemical Society}\ }\textbf
  {\bibinfo {volume} {140}},\ \bibinfo {pages} {10456--10463} (\bibinfo {year}
  {2018})}\BibitemShut {NoStop}%
\bibitem [{\citenamefont {Smits}\ \emph {et~al.}(2008)\citenamefont {Smits},
  \citenamefont {Mathijssen}, \citenamefont {Van~Hal}, \citenamefont
  {Setayesh}, \citenamefont {Geuns}, \citenamefont {Mutsaers}, \citenamefont
  {Cantatore}, \citenamefont {Wondergem}, \citenamefont {Werzer}, \citenamefont
  {Resel} \emph {et~al.}}]{smits2008bottom}%
  \BibitemOpen
  \bibfield  {author} {\bibinfo {author} {\bibfnamefont {Edsger~CP}\
  \bibnamefont {Smits}}, \bibinfo {author} {\bibfnamefont {Simon~GJ}\
  \bibnamefont {Mathijssen}}, \bibinfo {author} {\bibfnamefont {Paul~A}\
  \bibnamefont {Van~Hal}}, \bibinfo {author} {\bibfnamefont {Sepas}\
  \bibnamefont {Setayesh}}, \bibinfo {author} {\bibfnamefont {Thomas~CT}\
  \bibnamefont {Geuns}}, \bibinfo {author} {\bibfnamefont {Kees~AHA}\
  \bibnamefont {Mutsaers}}, \bibinfo {author} {\bibfnamefont {Eugenio}\
  \bibnamefont {Cantatore}}, \bibinfo {author} {\bibfnamefont {Harry~J}\
  \bibnamefont {Wondergem}}, \bibinfo {author} {\bibfnamefont {Oliver}\
  \bibnamefont {Werzer}}, \bibinfo {author} {\bibfnamefont {Roland}\
  \bibnamefont {Resel}},  \emph {et~al.},\ }\bibfield  {title} {\enquote
  {\bibinfo {title} {Bottom-up organic integrated circuits},}\ }\href {\doibase
  10.1038/nature07320} {\bibfield  {journal} {\bibinfo  {journal} {Nature}\
  }\textbf {\bibinfo {volume} {455}},\ \bibinfo {pages} {956--959} (\bibinfo
  {year} {2008})}\BibitemShut {NoStop}%
\bibitem [{\citenamefont {Jena}\ and\ \citenamefont
  {Sun}(2018)}]{jena2018super}%
  \BibitemOpen
  \bibfield  {author} {\bibinfo {author} {\bibfnamefont {Puru}\ \bibnamefont
  {Jena}}\ and\ \bibinfo {author} {\bibfnamefont {Qiang}\ \bibnamefont {Sun}},\
  }\bibfield  {title} {\enquote {\bibinfo {title} {Super atomic clusters:
  design rules and potential for building blocks of materials},}\ }\href
  {\doibase 10.1021/acs.chemrev.7b00524} {\bibfield  {journal} {\bibinfo
  {journal} {Chemical reviews}\ }\textbf {\bibinfo {volume} {118}},\ \bibinfo
  {pages} {5755--5870} (\bibinfo {year} {2018})}\BibitemShut {NoStop}%
\bibitem [{\citenamefont {Han}\ \emph {et~al.}(2019)\citenamefont {Han},
  \citenamefont {Huang}, \citenamefont {Li}, \citenamefont {Wang},
  \citenamefont {Luo}, \citenamefont {Liu}, \citenamefont {Zhou}, \citenamefont
  {Li}, \citenamefont {Zhang}, \citenamefont {Cui} \emph
  {et~al.}}]{han2019two}%
  \BibitemOpen
  \bibfield  {author} {\bibinfo {author} {\bibfnamefont {Wei}\ \bibnamefont
  {Han}}, \bibinfo {author} {\bibfnamefont {Pu}~\bibnamefont {Huang}}, \bibinfo
  {author} {\bibfnamefont {Liang}\ \bibnamefont {Li}}, \bibinfo {author}
  {\bibfnamefont {Fakun}\ \bibnamefont {Wang}}, \bibinfo {author}
  {\bibfnamefont {Peng}\ \bibnamefont {Luo}}, \bibinfo {author} {\bibfnamefont
  {Kailang}\ \bibnamefont {Liu}}, \bibinfo {author} {\bibfnamefont {Xing}\
  \bibnamefont {Zhou}}, \bibinfo {author} {\bibfnamefont {Huiqiao}\
  \bibnamefont {Li}}, \bibinfo {author} {\bibfnamefont {Xiuwen}\ \bibnamefont
  {Zhang}}, \bibinfo {author} {\bibfnamefont {Yi}~\bibnamefont {Cui}},  \emph
  {et~al.},\ }\bibfield  {title} {\enquote {\bibinfo {title} {Two-dimensional
  inorganic molecular crystals},}\ }\href {\doibase 10.1038/s41467-019-12569-9}
  {\bibfield  {journal} {\bibinfo  {journal} {Nature communications}\ }\textbf
  {\bibinfo {volume} {10}},\ \bibinfo {pages} {1--10} (\bibinfo {year}
  {2019})}\BibitemShut {NoStop}%
\bibitem [{\citenamefont {Hasan}\ and\ \citenamefont
  {Kane}(2010)}]{RevModPhys.82.3045}%
  \BibitemOpen
  \bibfield  {author} {\bibinfo {author} {\bibfnamefont {M.~Z.}\ \bibnamefont
  {Hasan}}\ and\ \bibinfo {author} {\bibfnamefont {C.~L.}\ \bibnamefont
  {Kane}},\ }\bibfield  {title} {\enquote {\bibinfo {title} {Colloquium:
  Topological insulators},}\ }\href {\doibase 10.1103/RevModPhys.82.3045}
  {\bibfield  {journal} {\bibinfo  {journal} {Rev. Mod. Phys.}\ }\textbf
  {\bibinfo {volume} {82}},\ \bibinfo {pages} {3045--3067} (\bibinfo {year}
  {2010})}\BibitemShut {NoStop}%
\bibitem [{\citenamefont {Ando}(2013)}]{ando2013topological}%
  \BibitemOpen
  \bibfield  {author} {\bibinfo {author} {\bibfnamefont {Yoichi}\ \bibnamefont
  {Ando}},\ }\bibfield  {title} {\enquote {\bibinfo {title} {Topological
  insulator materials},}\ }\href {\doibase 10.7566/JPSJ.82.102001} {\bibfield
  {journal} {\bibinfo  {journal} {Journal of the Physical Society of Japan}\
  }\textbf {\bibinfo {volume} {82}},\ \bibinfo {pages} {102001} (\bibinfo
  {year} {2013})}\BibitemShut {NoStop}%
\bibitem [{\citenamefont {Armitage}\ \emph {et~al.}(2018)\citenamefont
  {Armitage}, \citenamefont {Mele},\ and\ \citenamefont
  {Vishwanath}}]{RevModPhys.90.015001}%
  \BibitemOpen
  \bibfield  {author} {\bibinfo {author} {\bibfnamefont {N.~P.}\ \bibnamefont
  {Armitage}}, \bibinfo {author} {\bibfnamefont {E.~J.}\ \bibnamefont {Mele}},
  \ and\ \bibinfo {author} {\bibfnamefont {Ashvin}\ \bibnamefont
  {Vishwanath}},\ }\bibfield  {title} {\enquote {\bibinfo {title} {Weyl and
  dirac semimetals in three-dimensional solids},}\ }\href {\doibase
  10.1103/RevModPhys.90.015001} {\bibfield  {journal} {\bibinfo  {journal}
  {Rev. Mod. Phys.}\ }\textbf {\bibinfo {volume} {90}},\ \bibinfo {pages}
  {015001} (\bibinfo {year} {2018})}\BibitemShut {NoStop}%
\bibitem [{\citenamefont {Hirayama}\ \emph
  {et~al.}(2018{\natexlab{a}})\citenamefont {Hirayama}, \citenamefont
  {Okugawa},\ and\ \citenamefont {Murakami}}]{hirayama2018topological}%
  \BibitemOpen
  \bibfield  {author} {\bibinfo {author} {\bibfnamefont {Motoaki}\ \bibnamefont
  {Hirayama}}, \bibinfo {author} {\bibfnamefont {Ryo}\ \bibnamefont {Okugawa}},
  \ and\ \bibinfo {author} {\bibfnamefont {Shuichi}\ \bibnamefont {Murakami}},\
  }\bibfield  {title} {\enquote {\bibinfo {title} {Topological semimetals
  studied by ab initio calculations},}\ }\href {\doibase
  10.7566/JPSJ.87.041002} {\bibfield  {journal} {\bibinfo  {journal} {Journal
  of the Physical Society of Japan}\ }\textbf {\bibinfo {volume} {87}},\
  \bibinfo {pages} {041002} (\bibinfo {year} {2018}{\natexlab{a}})}\BibitemShut
  {NoStop}%
\bibitem [{\citenamefont {Nayak}\ \emph {et~al.}(2008)\citenamefont {Nayak},
  \citenamefont {Simon}, \citenamefont {Stern}, \citenamefont {Freedman},\ and\
  \citenamefont {Das~Sarma}}]{RevModPhys.80.1083}%
  \BibitemOpen
  \bibfield  {author} {\bibinfo {author} {\bibfnamefont {Chetan}\ \bibnamefont
  {Nayak}}, \bibinfo {author} {\bibfnamefont {Steven~H.}\ \bibnamefont
  {Simon}}, \bibinfo {author} {\bibfnamefont {Ady}\ \bibnamefont {Stern}},
  \bibinfo {author} {\bibfnamefont {Michael}\ \bibnamefont {Freedman}}, \ and\
  \bibinfo {author} {\bibfnamefont {Sankar}\ \bibnamefont {Das~Sarma}},\
  }\bibfield  {title} {\enquote {\bibinfo {title} {Non-abelian anyons and
  topological quantum computation},}\ }\href {\doibase
  10.1103/RevModPhys.80.1083} {\bibfield  {journal} {\bibinfo  {journal} {Rev.
  Mod. Phys.}\ }\textbf {\bibinfo {volume} {80}},\ \bibinfo {pages}
  {1083--1159} (\bibinfo {year} {2008})}\BibitemShut {NoStop}%
\bibitem [{\citenamefont {Ozawa}\ \emph {et~al.}(2019)\citenamefont {Ozawa},
  \citenamefont {Price}, \citenamefont {Amo}, \citenamefont {Goldman},
  \citenamefont {Hafezi}, \citenamefont {Lu}, \citenamefont {Rechtsman},
  \citenamefont {Schuster}, \citenamefont {Simon}, \citenamefont {Zilberberg},\
  and\ \citenamefont {Carusotto}}]{RevModPhys.91.015006}%
  \BibitemOpen
  \bibfield  {author} {\bibinfo {author} {\bibfnamefont {Tomoki}\ \bibnamefont
  {Ozawa}}, \bibinfo {author} {\bibfnamefont {Hannah~M.}\ \bibnamefont
  {Price}}, \bibinfo {author} {\bibfnamefont {Alberto}\ \bibnamefont {Amo}},
  \bibinfo {author} {\bibfnamefont {Nathan}\ \bibnamefont {Goldman}}, \bibinfo
  {author} {\bibfnamefont {Mohammad}\ \bibnamefont {Hafezi}}, \bibinfo {author}
  {\bibfnamefont {Ling}\ \bibnamefont {Lu}}, \bibinfo {author} {\bibfnamefont
  {Mikael~C.}\ \bibnamefont {Rechtsman}}, \bibinfo {author} {\bibfnamefont
  {David}\ \bibnamefont {Schuster}}, \bibinfo {author} {\bibfnamefont
  {Jonathan}\ \bibnamefont {Simon}}, \bibinfo {author} {\bibfnamefont {Oded}\
  \bibnamefont {Zilberberg}}, \ and\ \bibinfo {author} {\bibfnamefont {Iacopo}\
  \bibnamefont {Carusotto}},\ }\bibfield  {title} {\enquote {\bibinfo {title}
  {Topological photonics},}\ }\href {\doibase 10.1103/RevModPhys.91.015006}
  {\bibfield  {journal} {\bibinfo  {journal} {Rev. Mod. Phys.}\ }\textbf
  {\bibinfo {volume} {91}},\ \bibinfo {pages} {015006} (\bibinfo {year}
  {2019})}\BibitemShut {NoStop}%
\bibitem [{\citenamefont {Bergholtz}\ \emph {et~al.}(2021)\citenamefont
  {Bergholtz}, \citenamefont {Budich},\ and\ \citenamefont
  {Kunst}}]{RevModPhys.93.015005}%
  \BibitemOpen
  \bibfield  {author} {\bibinfo {author} {\bibfnamefont {Emil~J.}\ \bibnamefont
  {Bergholtz}}, \bibinfo {author} {\bibfnamefont {Jan~Carl}\ \bibnamefont
  {Budich}}, \ and\ \bibinfo {author} {\bibfnamefont {Flore~K.}\ \bibnamefont
  {Kunst}},\ }\bibfield  {title} {\enquote {\bibinfo {title} {Exceptional
  topology of non-Hermitian systems},}\ }\href {\doibase
  10.1103/RevModPhys.93.015005} {\bibfield  {journal} {\bibinfo  {journal}
  {Rev. Mod. Phys.}\ }\textbf {\bibinfo {volume} {93}},\ \bibinfo {pages}
  {015005} (\bibinfo {year} {2021})}\BibitemShut {NoStop}%
\bibitem [{Note1()}]{Note1}%
  \BibitemOpen
  \bibinfo {note} {In this article, we concentrate on topological materials
  within the scope of topological semimetals \cite {RevModPhys.90.015001,
  hirayama2018topological} or first-order topological insulators \cite
  {RevModPhys.82.3045, ando2013topological}, with lately proposed phases such
  as higher-order topological insulating states characterized by protected
  gapless hinge or corner modes \cite {schindler2018higher} out of the
  discussions and awaiting for future exploration.}\BibitemShut {Stop}%
\bibitem [{\citenamefont {Kajita}\ \emph {et~al.}(2014)\citenamefont {Kajita},
  \citenamefont {Nishio}, \citenamefont {Tajima}, \citenamefont {Suzumura},\
  and\ \citenamefont {Kobayashi}}]{kajita2014molecular}%
  \BibitemOpen
  \bibfield  {author} {\bibinfo {author} {\bibfnamefont {Koji}\ \bibnamefont
  {Kajita}}, \bibinfo {author} {\bibfnamefont {Yutaka}\ \bibnamefont {Nishio}},
  \bibinfo {author} {\bibfnamefont {Naoya}\ \bibnamefont {Tajima}}, \bibinfo
  {author} {\bibfnamefont {Yoshikazu}\ \bibnamefont {Suzumura}}, \ and\
  \bibinfo {author} {\bibfnamefont {Akito}\ \bibnamefont {Kobayashi}},\
  }\bibfield  {title} {\enquote {\bibinfo {title} {Molecular dirac fermion
  systems — theoretical and experimental approaches —},}\ }\href {\doibase
  10.7566/JPSJ.83.072002} {\bibfield  {journal} {\bibinfo  {journal} {Journal
  of the Physical Society of Japan}\ }\textbf {\bibinfo {volume} {83}},\
  \bibinfo {pages} {072002} (\bibinfo {year} {2014})}\BibitemShut {NoStop}%
\bibitem [{\citenamefont {Kato}\ \emph {et~al.}(2017)\citenamefont {Kato},
  \citenamefont {Cui}, \citenamefont {Tsumuraya}, \citenamefont {Miyazaki},\
  and\ \citenamefont {Suzumura}}]{kato2017emergence}%
  \BibitemOpen
  \bibfield  {author} {\bibinfo {author} {\bibfnamefont {Reizo}\ \bibnamefont
  {Kato}}, \bibinfo {author} {\bibfnamefont {HengBo}\ \bibnamefont {Cui}},
  \bibinfo {author} {\bibfnamefont {Takao}\ \bibnamefont {Tsumuraya}}, \bibinfo
  {author} {\bibfnamefont {Tsuyoshi}\ \bibnamefont {Miyazaki}}, \ and\ \bibinfo
  {author} {\bibfnamefont {Yoshikazu}\ \bibnamefont {Suzumura}},\ }\bibfield
  {title} {\enquote {\bibinfo {title} {Emergence of the dirac electron system
  in a single-component molecular conductor under high pressure},}\ }\href
  {\doibase 10.1021/jacs.6b12187} {\bibfield  {journal} {\bibinfo  {journal}
  {Journal of the American Chemical Society}\ }\textbf {\bibinfo {volume}
  {139}},\ \bibinfo {pages} {1770--1773} (\bibinfo {year} {2017})}\BibitemShut
  {NoStop}%
\bibitem [{\citenamefont {Zhou}\ \emph {et~al.}(2019)\citenamefont {Zhou},
  \citenamefont {Ishibashi}, \citenamefont {Ishii}, \citenamefont {Sekine},
  \citenamefont {Takehara}, \citenamefont {Miyagawa}, \citenamefont {Kanoda},
  \citenamefont {Nishibori},\ and\ \citenamefont {Kobayashi}}]{zhou2019single}%
  \BibitemOpen
  \bibfield  {author} {\bibinfo {author} {\bibfnamefont {Biao}\ \bibnamefont
  {Zhou}}, \bibinfo {author} {\bibfnamefont {Shoji}\ \bibnamefont {Ishibashi}},
  \bibinfo {author} {\bibfnamefont {Tatsuru}\ \bibnamefont {Ishii}}, \bibinfo
  {author} {\bibfnamefont {Takahiko}\ \bibnamefont {Sekine}}, \bibinfo {author}
  {\bibfnamefont {Ryosuke}\ \bibnamefont {Takehara}}, \bibinfo {author}
  {\bibfnamefont {Kazuya}\ \bibnamefont {Miyagawa}}, \bibinfo {author}
  {\bibfnamefont {Kazushi}\ \bibnamefont {Kanoda}}, \bibinfo {author}
  {\bibfnamefont {Eiji}\ \bibnamefont {Nishibori}}, \ and\ \bibinfo {author}
  {\bibfnamefont {Akiko}\ \bibnamefont {Kobayashi}},\ }\bibfield  {title}
  {\enquote {\bibinfo {title} {Single-component molecular conductor [Pt(dmdt)$_2$] — a three-dimensional ambient-pressure molecular dirac electron system},}\
  }\href {\doibase 10.1039/c9cc00218a} {\bibfield  {journal} {\bibinfo
  {journal} {Chemical Communications}\ }\textbf {\bibinfo {volume} {55}},\
  \bibinfo {pages} {3327--3330} (\bibinfo {year} {2019})}\BibitemShut {NoStop}%
\bibitem [{\citenamefont {Slack}(1995)}]{slack1995handbook}%
  \BibitemOpen
  \bibfield  {author} {\bibinfo {author} {\bibfnamefont {Glen~A.}\ \bibnamefont
  {Slack}},\ }in\ \href {\doibase 10.1201/9781420049718} {\enquote {\bibinfo
  {booktitle} {CRC Handbook of Thermoelectrics}}}\ (\bibinfo  {publisher} {CRC
  Press},\ \bibinfo {address} {Boca Raton, FL},\ \bibinfo {year} {1995})\ pp.\
  \bibinfo {pages} {407--440}\BibitemShut {NoStop}%
\bibitem [{\citenamefont {Dye}(2003)}]{dye2003electrons}%
  \BibitemOpen
  \bibfield  {author} {\bibinfo {author} {\bibfnamefont {James~L}\ \bibnamefont
  {Dye}},\ }\bibfield  {title} {\enquote {\bibinfo {title} {Electrons as
  anions},}\ }\href {https://www.science.org/doi/10.1126/science.1088103}
  {\bibfield  {journal} {\bibinfo  {journal} {Science}\ }\textbf {\bibinfo
  {volume} {301}},\ \bibinfo {pages} {607--608} (\bibinfo {year}
  {2003})}\BibitemShut {NoStop}%
\bibitem [{\citenamefont {Hosono}\ and\ \citenamefont
  {Kitano}(2021)}]{hosono2021advances}%
  \BibitemOpen
  \bibfield  {author} {\bibinfo {author} {\bibfnamefont {Hideo}\ \bibnamefont
  {Hosono}}\ and\ \bibinfo {author} {\bibfnamefont {Masaaki}\ \bibnamefont
  {Kitano}},\ }\bibfield  {title} {\enquote {\bibinfo {title} {Advances in
  materials and applications of inorganic electrides},}\ }\href
  {https://pubs.acs.org/doi/10.1021/acs.chemrev.0c01071} {\bibfield  {journal}
  {\bibinfo  {journal} {Chemical Reviews}\ }\textbf {\bibinfo {volume} {121}},\
  \bibinfo {pages} {3121--3185} (\bibinfo {year} {2021})}\BibitemShut {NoStop}%
\bibitem [{\citenamefont {Yu}\ \emph {et~al.}(2021)\citenamefont {Yu},
  \citenamefont {Hirayama}, \citenamefont {Flores-Livas}, \citenamefont
  {Huebsch}, \citenamefont {Nomoto},\ and\ \citenamefont
  {Arita}}]{PhysRevMaterials.5.044203}%
  \BibitemOpen
  \bibfield  {author} {\bibinfo {author} {\bibfnamefont {Tonghua}\ \bibnamefont
  {Yu}}, \bibinfo {author} {\bibfnamefont {Motoaki}\ \bibnamefont {Hirayama}},
  \bibinfo {author} {\bibfnamefont {Jos\'e~A.}\ \bibnamefont {Flores-Livas}},
  \bibinfo {author} {\bibfnamefont {Marie-Therese}\ \bibnamefont {Huebsch}},
  \bibinfo {author} {\bibfnamefont {Takuya}\ \bibnamefont {Nomoto}}, \ and\
  \bibinfo {author} {\bibfnamefont {Ryotaro}\ \bibnamefont {Arita}},\
  }\bibfield  {title} {\enquote {\bibinfo {title} {First-principles design of
  halide-reduced electrides: Magnetism and topological phases},}\ }\href
  {\doibase 10.1103/PhysRevMaterials.5.044203} {\bibfield  {journal} {\bibinfo
  {journal} {Phys. Rev. Materials}\ }\textbf {\bibinfo {volume} {5}},\ \bibinfo
  {pages} {044203} (\bibinfo {year} {2021})}\BibitemShut {NoStop}%
\bibitem [{\citenamefont {Matsuishi}\ \emph {et~al.}(2003)\citenamefont
  {Matsuishi}, \citenamefont {Toda}, \citenamefont {Miyakawa}, \citenamefont
  {Hayashi}, \citenamefont {Kamiya}, \citenamefont {Hirano}, \citenamefont
  {Tanaka},\ and\ \citenamefont {Hosono}}]{matsuishi2003high}%
  \BibitemOpen
  \bibfield  {author} {\bibinfo {author} {\bibfnamefont {Satoru}\ \bibnamefont
  {Matsuishi}}, \bibinfo {author} {\bibfnamefont {Yoshitake}\ \bibnamefont
  {Toda}}, \bibinfo {author} {\bibfnamefont {Masashi}\ \bibnamefont
  {Miyakawa}}, \bibinfo {author} {\bibfnamefont {Katsuro}\ \bibnamefont
  {Hayashi}}, \bibinfo {author} {\bibfnamefont {Toshio}\ \bibnamefont
  {Kamiya}}, \bibinfo {author} {\bibfnamefont {Masahiro}\ \bibnamefont
  {Hirano}}, \bibinfo {author} {\bibfnamefont {Isao}\ \bibnamefont {Tanaka}}, \
  and\ \bibinfo {author} {\bibfnamefont {Hideo}\ \bibnamefont {Hosono}},\
  }\bibfield  {title} {\enquote {\bibinfo {title} {High-density electron anions
  in a nanoporous single crystal: [Ca$_{24}$Al$_{28}$O$_{64}$]$^{4+}$(4e$^-$)},}\ }\href {\doibase
  10.1126/science.1083842} {\bibfield  {journal} {\bibinfo  {journal}
  {Science}\ }\textbf {\bibinfo {volume} {301}},\ \bibinfo {pages} {626--629}
  (\bibinfo {year} {2003})}\BibitemShut {NoStop}%
\bibitem [{\citenamefont {Hirayama}\ \emph
  {et~al.}(2018{\natexlab{b}})\citenamefont {Hirayama}, \citenamefont
  {Matsuishi}, \citenamefont {Hosono},\ and\ \citenamefont
  {Murakami}}]{PhysRevX.8.031067}%
  \BibitemOpen
  \bibfield  {author} {\bibinfo {author} {\bibfnamefont {Motoaki}\ \bibnamefont
  {Hirayama}}, \bibinfo {author} {\bibfnamefont {Satoru}\ \bibnamefont
  {Matsuishi}}, \bibinfo {author} {\bibfnamefont {Hideo}\ \bibnamefont
  {Hosono}}, \ and\ \bibinfo {author} {\bibfnamefont {Shuichi}\ \bibnamefont
  {Murakami}},\ }\bibfield  {title} {\enquote {\bibinfo {title} {Electrides as
  a new platform of topological materials},}\ }\href {\doibase
  10.1103/PhysRevX.8.031067} {\bibfield  {journal} {\bibinfo  {journal} {Phys.
  Rev. X}\ }\textbf {\bibinfo {volume} {8}},\ \bibinfo {pages} {031067}
  (\bibinfo {year} {2018}{\natexlab{b}})}\BibitemShut {NoStop}%
\bibitem [{\citenamefont {Nie}\ \emph {et~al.}(2021)\citenamefont {Nie},
  \citenamefont {Qian}, \citenamefont {Gao}, \citenamefont {Fang},
  \citenamefont {Weng},\ and\ \citenamefont {Wang}}]{PhysRevB.103.205133}%
  \BibitemOpen
  \bibfield  {author} {\bibinfo {author} {\bibfnamefont {Simin}\ \bibnamefont
  {Nie}}, \bibinfo {author} {\bibfnamefont {Yuting}\ \bibnamefont {Qian}},
  \bibinfo {author} {\bibfnamefont {Jiacheng}\ \bibnamefont {Gao}}, \bibinfo
  {author} {\bibfnamefont {Zhong}\ \bibnamefont {Fang}}, \bibinfo {author}
  {\bibfnamefont {Hongming}\ \bibnamefont {Weng}}, \ and\ \bibinfo {author}
  {\bibfnamefont {Zhijun}\ \bibnamefont {Wang}},\ }\bibfield  {title} {\enquote
  {\bibinfo {title} {Application of topological quantum chemistry in
  electrides},}\ }\href {\doibase 10.1103/PhysRevB.103.205133} {\bibfield
  {journal} {\bibinfo  {journal} {Phys. Rev. B}\ }\textbf {\bibinfo {volume}
  {103}},\ \bibinfo {pages} {205133} (\bibinfo {year} {2021})}\BibitemShut
  {NoStop}%
\bibitem [{\citenamefont {Kauzlarich}\ \emph {et~al.}(2007)\citenamefont
  {Kauzlarich}, \citenamefont {Brown},\ and\ \citenamefont
  {Snyder}}]{kauzlarich2007zintl}%
  \BibitemOpen
  \bibfield  {author} {\bibinfo {author} {\bibfnamefont {Susan~M}\ \bibnamefont
  {Kauzlarich}}, \bibinfo {author} {\bibfnamefont {Shawna~R}\ \bibnamefont
  {Brown}}, \ and\ \bibinfo {author} {\bibfnamefont {G~Jeffrey}\ \bibnamefont
  {Snyder}},\ }\bibfield  {title} {\enquote {\bibinfo {title} {Zintl phases for
  thermoelectric devices},}\ }\href {\doibase 10.1039/B702266B} {\bibfield
  {journal} {\bibinfo  {journal} {Dalton Transactions}\ ,\ \bibinfo {pages}
  {2099--2107}} (\bibinfo {year} {2007})}\BibitemShut {NoStop}%
\bibitem [{\citenamefont {Kauzlarich}\ \emph {et~al.}(2017)\citenamefont
  {Kauzlarich}, \citenamefont {Zevalkink}, \citenamefont {Toberer},\ and\
  \citenamefont {Snyder}}]{kauzlarich2017zintl}%
  \BibitemOpen
  \bibfield  {author} {\bibinfo {author} {\bibfnamefont {Susan~M.}\
  \bibnamefont {Kauzlarich}}, \bibinfo {author} {\bibfnamefont {Alex}\
  \bibnamefont {Zevalkink}}, \bibinfo {author} {\bibfnamefont {Eric}\
  \bibnamefont {Toberer}}, \ and\ \bibinfo {author} {\bibfnamefont {G.~Jeff}\
  \bibnamefont {Snyder}},\ }\enquote {\bibinfo {title} {Zintl phases: Recent
  developments in thermoelectrics and future outlook},}\ in\ \href {\doibase
  10.1039/9781782624042-00001} {\bibinfo {booktitle} {Thermoelectric
  Materials and Devices}},\ Vol.\ \bibinfo {volume} {2017-January}\ (\bibinfo
  {publisher} {Royal Society of Chemistry},\ \bibinfo {address} {United
  Kingdom},\ \bibinfo {year} {2017})\ pp.\ \bibinfo {pages} {1--26}\BibitemShut
  {NoStop}%
\bibitem [{\citenamefont {Eisenmann}\ and\ \citenamefont
  {R{\"o}{\ss}ler}(2000)}]{eisenmann2000pniktogenidostannate}%
  \BibitemOpen
  \bibfield  {author} {\bibinfo {author} {\bibfnamefont {B.}~\bibnamefont
  {Eisenmann}}\ and\ \bibinfo {author} {\bibfnamefont {U.}~\bibnamefont
  {R{\"o}{\ss}ler}},\ }\bibfield  {title} {\bibinfo {title}
  {Pniktogenidostannate(IV) mit isolierten Tetraeder-Anionen: Neue Vertreter
  (E1)$_4$(E2)$_2$[Sn(E15)$_4$] (mit E1= Na, K; E2= Ca, Sr, Ba; E15= P, As, Sb, Bi)
  vom Na$_6$[ZnO$_4$]-Typ und die {\"U}berstrukturvariante von K$_4$Sr$_2$[SnAs$_4$]},\
  }\href
  {https://doi.org/https://doi.org/10.1002/(SICI)1521-3749(200006)626:6<1373::AID-ZAAC1373>3.0.CO;2-J}
  {\bibfield  {journal} {\bibinfo  {journal} {Zeitschrift f{\"u}r anorganische
  und allgemeine Chemie}\ }\textbf {\bibinfo {volume} {626}},\ \bibinfo {pages}
  {1373} (\bibinfo {year} {2000})}\BibitemShut {NoStop}%
\bibitem [{\citenamefont {Perdew}\ \emph {et~al.}(1996)\citenamefont {Perdew},
  \citenamefont {Burke},\ and\ \citenamefont
  {Ernzerhof}}]{PhysRevLett.77.3865}%
  \BibitemOpen
  \bibfield  {author} {\bibinfo {author} {\bibfnamefont {John~P.}\ \bibnamefont
  {Perdew}}, \bibinfo {author} {\bibfnamefont {Kieron}\ \bibnamefont {Burke}},
  \ and\ \bibinfo {author} {\bibfnamefont {Matthias}\ \bibnamefont
  {Ernzerhof}},\ }\bibfield  {title} {\enquote {\bibinfo {title} {Generalized
  gradient approximation made simple},}\ }\href {\doibase
  10.1103/PhysRevLett.77.3865} {\bibfield  {journal} {\bibinfo  {journal}
  {Phys. Rev. Lett.}\ }\textbf {\bibinfo {volume} {77}},\ \bibinfo {pages}
  {3865--3868} (\bibinfo {year} {1996})}\BibitemShut {NoStop}%
\bibitem [{\citenamefont {Marzari}\ and\ \citenamefont
  {Vanderbilt}(1997)}]{PhysRevB.56.12847}%
  \BibitemOpen
  \bibfield  {author} {\bibinfo {author} {\bibfnamefont {Nicola}\ \bibnamefont
  {Marzari}}\ and\ \bibinfo {author} {\bibfnamefont {David}\ \bibnamefont
  {Vanderbilt}},\ }\bibfield  {title} {\enquote {\bibinfo {title} {Maximally
  localized generalized wannier functions for composite energy bands},}\ }\href
  {\doibase 10.1103/PhysRevB.56.12847} {\bibfield  {journal} {\bibinfo
  {journal} {Phys. Rev. B}\ }\textbf {\bibinfo {volume} {56}},\ \bibinfo
  {pages} {12847--12865} (\bibinfo {year} {1997})}\BibitemShut {NoStop}%
\bibitem [{\citenamefont {Souza}\ \emph {et~al.}(2001)\citenamefont {Souza},
  \citenamefont {Marzari},\ and\ \citenamefont
  {Vanderbilt}}]{PhysRevB.65.035109}%
  \BibitemOpen
  \bibfield  {author} {\bibinfo {author} {\bibfnamefont {Ivo}\ \bibnamefont
  {Souza}}, \bibinfo {author} {\bibfnamefont {Nicola}\ \bibnamefont {Marzari}},
  \ and\ \bibinfo {author} {\bibfnamefont {David}\ \bibnamefont {Vanderbilt}},\
  }\bibfield  {title} {\enquote {\bibinfo {title} {Maximally localized wannier
  functions for entangled energy bands},}\ }\href {\doibase
  10.1103/PhysRevB.65.035109} {\bibfield  {journal} {\bibinfo  {journal} {Phys.
  Rev. B}\ }\textbf {\bibinfo {volume} {65}},\ \bibinfo {pages} {035109}
  (\bibinfo {year} {2001})}\BibitemShut {NoStop}%
\bibitem [{\citenamefont {Vidal}\ \emph {et~al.}(2011)\citenamefont {Vidal},
  \citenamefont {Zhang}, \citenamefont {Yu}, \citenamefont {Luo},\ and\
  \citenamefont {Zunger}}]{PhysRevB.84.041109}%
  \BibitemOpen
  \bibfield  {author} {\bibinfo {author} {\bibfnamefont {J.}~\bibnamefont
  {Vidal}}, \bibinfo {author} {\bibfnamefont {X.}~\bibnamefont {Zhang}},
  \bibinfo {author} {\bibfnamefont {L.}~\bibnamefont {Yu}}, \bibinfo {author}
  {\bibfnamefont {J.-W.}\ \bibnamefont {Luo}}, \ and\ \bibinfo {author}
  {\bibfnamefont {A.}~\bibnamefont {Zunger}},\ }\bibfield  {title} {\enquote
  {\bibinfo {title} {False-positive and false-negative assignments of
  topological insulators in density functional theory and hybrids},}\ }\href
  {\doibase 10.1103/PhysRevB.84.041109} {\bibfield  {journal} {\bibinfo
  {journal} {Phys. Rev. B}\ }\textbf {\bibinfo {volume} {84}},\ \bibinfo
  {pages} {041109} (\bibinfo {year} {2011})}\BibitemShut {NoStop}%
\bibitem [{\citenamefont {Heyd}\ \emph {et~al.}(2003)\citenamefont {Heyd},
  \citenamefont {Scuseria},\ and\ \citenamefont {Ernzerhof}}]{heyd2003hybrid}%
  \BibitemOpen
  \bibfield  {author} {\bibinfo {author} {\bibfnamefont {Jochen}\ \bibnamefont
  {Heyd}}, \bibinfo {author} {\bibfnamefont {Gustavo~E}\ \bibnamefont
  {Scuseria}}, \ and\ \bibinfo {author} {\bibfnamefont {Matthias}\ \bibnamefont
  {Ernzerhof}},\ }\bibfield  {title} {\enquote {\bibinfo {title} {Hybrid
  functionals based on a screened coulomb potential},}\ }\href {\doibase
  10.1063/1.1564060} {\bibfield  {journal} {\bibinfo  {journal} {The Journal of
  chemical physics}\ }\textbf {\bibinfo {volume} {118}},\ \bibinfo {pages}
  {8207--8215} (\bibinfo {year} {2003})}\BibitemShut {NoStop}%
\bibitem [{\citenamefont {Chen}\ \emph {et~al.}(2009)\citenamefont {Chen},
  \citenamefont {Analytis}, \citenamefont {Chu}, \citenamefont {Liu},
  \citenamefont {Mo}, \citenamefont {Qi}, \citenamefont {Zhang}, \citenamefont
  {Lu}, \citenamefont {Dai}, \citenamefont {Fang} \emph
  {et~al.}}]{chen2009experimental}%
  \BibitemOpen
  \bibfield  {author} {\bibinfo {author} {\bibfnamefont {YL}~\bibnamefont
  {Chen}}, \bibinfo {author} {\bibfnamefont {James~G}\ \bibnamefont
  {Analytis}}, \bibinfo {author} {\bibfnamefont {J-H}\ \bibnamefont {Chu}},
  \bibinfo {author} {\bibfnamefont {ZK}~\bibnamefont {Liu}}, \bibinfo {author}
  {\bibfnamefont {S-K}\ \bibnamefont {Mo}}, \bibinfo {author} {\bibfnamefont
  {Xiao-Liang}\ \bibnamefont {Qi}}, \bibinfo {author} {\bibfnamefont
  {HJ}~\bibnamefont {Zhang}}, \bibinfo {author} {\bibfnamefont
  {DH}~\bibnamefont {Lu}}, \bibinfo {author} {\bibfnamefont {Xi}~\bibnamefont
  {Dai}}, \bibinfo {author} {\bibfnamefont {Zhong}\ \bibnamefont {Fang}},
  \emph {et~al.},\ }\bibfield  {title} {\enquote {\bibinfo {title}
  {Experimental realization of a three-dimensional topological insulator,
  Bi$_2$Te$_3$},}\ }\href {\doibase 10.1126/science.1173034} {\bibfield  {journal}
  {\bibinfo  {journal} {science}\ }\textbf {\bibinfo {volume} {325}},\ \bibinfo
  {pages} {178--181} (\bibinfo {year} {2009})}\BibitemShut {NoStop}%
\bibitem [{\citenamefont {Heremans}\ \emph {et~al.}(2017)\citenamefont
  {Heremans}, \citenamefont {Cava},\ and\ \citenamefont
  {Samarth}}]{heremans2017tetradymites}%
  \BibitemOpen
  \bibfield  {author} {\bibinfo {author} {\bibfnamefont {Joseph~P}\
  \bibnamefont {Heremans}}, \bibinfo {author} {\bibfnamefont {Robert~J}\
  \bibnamefont {Cava}}, \ and\ \bibinfo {author} {\bibfnamefont {Nitin}\
  \bibnamefont {Samarth}},\ }\bibfield  {title} {\enquote {\bibinfo {title}
  {Tetradymites as thermoelectrics and topological insulators},}\ }\href
  {\doibase doi.org/10.1038/natrevmats.2017.49} {\bibfield  {journal} {\bibinfo
   {journal} {Nature Reviews Materials}\ }\textbf {\bibinfo {volume} {2}},\
  \bibinfo {pages} {1--21} (\bibinfo {year} {2017})}\BibitemShut {NoStop}%
\bibitem [{\citenamefont {Huang}\ \emph {et~al.}(2021)\citenamefont {Huang},
  \citenamefont {Li}, \citenamefont {Yoon}, \citenamefont {Oh}, \citenamefont
  {Wu}, \citenamefont {Liu}, \citenamefont {Dhale}, \citenamefont {Zhou},
  \citenamefont {Guo}, \citenamefont {Zhang}, \citenamefont {Hashimoto},
  \citenamefont {Lu}, \citenamefont {Denlinger}, \citenamefont {Wang},
  \citenamefont {Lau}, \citenamefont {Birgeneau}, \citenamefont {Zhang},
  \citenamefont {Lv},\ and\ \citenamefont {Yi}}]{PhysRevX.11.031042}%
  \BibitemOpen
  \bibfield  {author} {\bibinfo {author} {\bibfnamefont {Jianwei}\ \bibnamefont
  {Huang}}, \bibinfo {author} {\bibfnamefont {Sheng}\ \bibnamefont {Li}},
  \bibinfo {author} {\bibfnamefont {Chiho}\ \bibnamefont {Yoon}}, \bibinfo
  {author} {\bibfnamefont {Ji~Seop}\ \bibnamefont {Oh}}, \bibinfo {author}
  {\bibfnamefont {Han}\ \bibnamefont {Wu}}, \bibinfo {author} {\bibfnamefont
  {Xiaoyuan}\ \bibnamefont {Liu}}, \bibinfo {author} {\bibfnamefont {Nikhil}\
  \bibnamefont {Dhale}}, \bibinfo {author} {\bibfnamefont {Yan-Feng}\
  \bibnamefont {Zhou}}, \bibinfo {author} {\bibfnamefont {Yucheng}\
  \bibnamefont {Guo}}, \bibinfo {author} {\bibfnamefont {Yichen}\ \bibnamefont
  {Zhang}}, \bibinfo {author} {\bibfnamefont {Makoto}\ \bibnamefont
  {Hashimoto}}, \bibinfo {author} {\bibfnamefont {Donghui}\ \bibnamefont {Lu}},
  \bibinfo {author} {\bibfnamefont {Jonathan}\ \bibnamefont {Denlinger}},
  \bibinfo {author} {\bibfnamefont {Xiqu}\ \bibnamefont {Wang}}, \bibinfo
  {author} {\bibfnamefont {Chun~Ning}\ \bibnamefont {Lau}}, \bibinfo {author}
  {\bibfnamefont {Robert~J.}\ \bibnamefont {Birgeneau}}, \bibinfo {author}
  {\bibfnamefont {Fan}\ \bibnamefont {Zhang}}, \bibinfo {author} {\bibfnamefont
  {Bing}\ \bibnamefont {Lv}}, \ and\ \bibinfo {author} {\bibfnamefont {Ming}\
  \bibnamefont {Yi}},\ }\bibfield  {title} {\enquote {\bibinfo {title}
  {Room-temperature topological phase transition in quasi-one-dimensional
  material ${\mathrm{Bi}}_{\mathrm{4}}{\mathrm{I}}_{\mathrm{4}}$},}\
  }\href {\doibase 10.1103/PhysRevX.11.031042} {\bibfield  {journal} {\bibinfo
  {journal} {Phys. Rev. X}\ }\textbf {\bibinfo {volume} {11}},\ \bibinfo
  {pages} {031042} (\bibinfo {year} {2021})}\BibitemShut {NoStop}%
\bibitem [{\citenamefont {Liu}\ and\ \citenamefont
  {Vanderbilt}(2014)}]{PhysRevB.90.155316}%
  \BibitemOpen
  \bibfield  {author} {\bibinfo {author} {\bibfnamefont {Jianpeng}\
  \bibnamefont {Liu}}\ and\ \bibinfo {author} {\bibfnamefont {David}\
  \bibnamefont {Vanderbilt}},\ }\bibfield  {title} {\enquote {\bibinfo {title}
  {Weyl semimetals from noncentrosymmetric topological insulators},}\ }\href
  {\doibase 10.1103/PhysRevB.90.155316} {\bibfield  {journal} {\bibinfo
  {journal} {Phys. Rev. B}\ }\textbf {\bibinfo {volume} {90}},\ \bibinfo
  {pages} {155316} (\bibinfo {year} {2014})}\BibitemShut {NoStop}%
\bibitem [{\citenamefont {Ochi}\ \emph {et~al.}(2016)\citenamefont {Ochi},
  \citenamefont {Arita}, \citenamefont {Trivedi},\ and\ \citenamefont
  {Okamoto}}]{PhysRevB.93.195149}%
  \BibitemOpen
  \bibfield  {author} {\bibinfo {author} {\bibfnamefont {Masayuki}\
  \bibnamefont {Ochi}}, \bibinfo {author} {\bibfnamefont {Ryotaro}\
  \bibnamefont {Arita}}, \bibinfo {author} {\bibfnamefont {Nandini}\
  \bibnamefont {Trivedi}}, \ and\ \bibinfo {author} {\bibfnamefont {Satoshi}\
  \bibnamefont {Okamoto}},\ }\bibfield  {title} {\enquote {\bibinfo {title}
  {Strain-induced topological transition in
  ${\mathrm{SrRu}}_{2}{\mathrm{O}}_{6}$ and
  ${\mathrm{CaOs}}_{2}{\mathrm{O}}_{6}$},}\ }\href {\doibase
  10.1103/PhysRevB.93.195149} {\bibfield  {journal} {\bibinfo  {journal} {Phys.
  Rev. B}\ }\textbf {\bibinfo {volume} {93}},\ \bibinfo {pages} {195149}
  (\bibinfo {year} {2016})}\BibitemShut {NoStop}%
\bibitem [{\citenamefont {Hsieh}\ \emph {et~al.}(2008)\citenamefont {Hsieh},
  \citenamefont {Qian}, \citenamefont {Wray}, \citenamefont {Xia},
  \citenamefont {Hor}, \citenamefont {Cava},\ and\ \citenamefont
  {Hasan}}]{hsieh2008topological}%
  \BibitemOpen
  \bibfield  {author} {\bibinfo {author} {\bibfnamefont {David}\ \bibnamefont
  {Hsieh}}, \bibinfo {author} {\bibfnamefont {Dong}\ \bibnamefont {Qian}},
  \bibinfo {author} {\bibfnamefont {Lewis}\ \bibnamefont {Wray}}, \bibinfo
  {author} {\bibfnamefont {Yiman}\ \bibnamefont {Xia}}, \bibinfo {author}
  {\bibfnamefont {Yew~San}\ \bibnamefont {Hor}}, \bibinfo {author}
  {\bibfnamefont {Robert~Joseph}\ \bibnamefont {Cava}}, \ and\ \bibinfo
  {author} {\bibfnamefont {M~Zahid}\ \bibnamefont {Hasan}},\ }\bibfield
  {title} {\enquote {\bibinfo {title} {A topological Dirac insulator in a
  quantum spin hall phase},}\ }\href {\doibase 10.1038/nature06843} {\bibfield
  {journal} {\bibinfo  {journal} {Nature}\ }\textbf {\bibinfo {volume} {452}},\
  \bibinfo {pages} {970--974} (\bibinfo {year} {2008})}\BibitemShut {NoStop}%
\bibitem [{\citenamefont {Goldsmid}(2010)}]{goldsmid2010thermoelectricity}%
  \BibitemOpen
  \bibfield  {author} {\bibinfo {author} {\bibfnamefont {H~Julian}\
  \bibnamefont {Goldsmid}},\ }\href {\doibase 10.1007/978-3-642-00716-3_6}
  {\enquote {\bibinfo {title} {Introduction to Thermoelectricity}}},\ Vol.\
  \bibinfo {volume} {121}\ (\bibinfo  {publisher} {Springer},\ \bibinfo
  {address} {Berlin, Heidelberg},\ \bibinfo {year} {2010})\BibitemShut
  {NoStop}%
\bibitem [{\citenamefont {Xu}\ \emph {et~al.}(2017)\citenamefont {Xu},
  \citenamefont {Xu},\ and\ \citenamefont {Zhu}}]{xu2017topological}%
  \BibitemOpen
  \bibfield  {author} {\bibinfo {author} {\bibfnamefont {Ning}\ \bibnamefont
  {Xu}}, \bibinfo {author} {\bibfnamefont {Yong}\ \bibnamefont {Xu}}, \ and\
  \bibinfo {author} {\bibfnamefont {Jia}\ \bibnamefont {Zhu}},\ }\bibfield
  {title} {\enquote {\bibinfo {title} {Topological insulators for
  thermoelectrics},}\ }\href {\doibase 10.1038/s41535-017-0054-3} {\bibfield
  {journal} {\bibinfo  {journal} {npj Quantum Materials}\ }\textbf {\bibinfo
  {volume} {2}},\ \bibinfo {pages} {1--9} (\bibinfo {year} {2017})}\BibitemShut
  {NoStop}%
\bibitem [{\citenamefont {Shi}\ \emph {et~al.}(2015)\citenamefont {Shi},
  \citenamefont {Parker}, \citenamefont {Du},\ and\ \citenamefont
  {Singh}}]{PhysRevApplied.3.014004}%
  \BibitemOpen
  \bibfield  {author} {\bibinfo {author} {\bibfnamefont {Hongliang}\
  \bibnamefont {Shi}}, \bibinfo {author} {\bibfnamefont {David}\ \bibnamefont
  {Parker}}, \bibinfo {author} {\bibfnamefont {Mao-Hua}\ \bibnamefont {Du}}, \
  and\ \bibinfo {author} {\bibfnamefont {David~J.}\ \bibnamefont {Singh}},\
  }\bibfield  {title} {\enquote {\bibinfo {title} {Connecting thermoelectric
  performance and topological-insulator behavior:
  ${\mathrm{Bi}}_{2}{\mathrm{Te}}_{3}$ and
  ${\mathrm{Bi}}_{2}{\mathrm{Te}}_{2}\mathrm{Se}$ from first principles},}\
  }\href {\doibase 10.1103/PhysRevApplied.3.014004} {\bibfield  {journal}
  {\bibinfo  {journal} {Phys. Rev. Applied}\ }\textbf {\bibinfo {volume} {3}},\
  \bibinfo {pages} {014004} (\bibinfo {year} {2015})}\BibitemShut {NoStop}%
\bibitem [{\citenamefont {Kuroki}\ and\ \citenamefont
  {Arita}(2007)}]{kuroki2007pudding}%
  \BibitemOpen
  \bibfield  {author} {\bibinfo {author} {\bibfnamefont {Kazuhiko}\
  \bibnamefont {Kuroki}}\ and\ \bibinfo {author} {\bibfnamefont {Ryotaro}\
  \bibnamefont {Arita}},\ }\bibfield  {title} {\enquote {\bibinfo {title}
  {“pudding mold” band drives large thermopower in naxcoo2},}\ }\href
  {\doibase doi.org/10.1143/JPSJ.76.083707} {\bibfield  {journal} {\bibinfo
  {journal} {Journal of the Physical Society of Japan}\ }\textbf {\bibinfo
  {volume} {76}},\ \bibinfo {pages} {083707--083707} (\bibinfo {year}
  {2007})}\BibitemShut {NoStop}%
\bibitem [{\citenamefont {Zhang}\ \emph {et~al.}(2010)\citenamefont {Zhang},
  \citenamefont {Du},\ and\ \citenamefont {Singh}}]{PhysRevB.81.075117}%
  \BibitemOpen
  \bibfield  {author} {\bibinfo {author} {\bibfnamefont {Lijun}\ \bibnamefont
  {Zhang}}, \bibinfo {author} {\bibfnamefont {Mao-Hua}\ \bibnamefont {Du}}, \
  and\ \bibinfo {author} {\bibfnamefont {D.~J.}\ \bibnamefont {Singh}},\
  }\bibfield  {title} {\enquote {\bibinfo {title} {Zintl-phase compounds with
  ${\text{snsb}}_{4}$ tetrahedral anions: Electronic structure and
  thermoelectric properties},}\ }\href {\doibase 10.1103/PhysRevB.81.075117}
  {\bibfield  {journal} {\bibinfo  {journal} {Phys. Rev. B}\ }\textbf {\bibinfo
  {volume} {81}},\ \bibinfo {pages} {075117} (\bibinfo {year}
  {2010})}\BibitemShut {NoStop}%
\bibitem [{\citenamefont {Hellman}\ and\ \citenamefont
  {Broido}(2014)}]{PhysRevB.90.134309}%
  \BibitemOpen
  \bibfield  {author} {\bibinfo {author} {\bibfnamefont {Olle}\ \bibnamefont
  {Hellman}}\ and\ \bibinfo {author} {\bibfnamefont {David~A.}\ \bibnamefont
  {Broido}},\ }\bibfield  {title} {\enquote {\bibinfo {title} {Phonon thermal
  transport in ${\mathrm{Bi}}_{2}{\mathrm{Te}}_{3}$ from first principles},}\
  }\href {\doibase 10.1103/PhysRevB.90.134309} {\bibfield  {journal} {\bibinfo
  {journal} {Phys. Rev. B}\ }\textbf {\bibinfo {volume} {90}},\ \bibinfo
  {pages} {134309} (\bibinfo {year} {2014})}\BibitemShut {NoStop}%
\bibitem [{\citenamefont {Takane}\ \emph {et~al.}(2016)\citenamefont {Takane},
  \citenamefont {Souma}, \citenamefont {Sato}, \citenamefont {Takahashi},
  \citenamefont {Segawa},\ and\ \citenamefont {Ando}}]{takane2016work}%
  \BibitemOpen
  \bibfield  {author} {\bibinfo {author} {\bibfnamefont {Daichi}\ \bibnamefont
  {Takane}}, \bibinfo {author} {\bibfnamefont {Seigo}\ \bibnamefont {Souma}},
  \bibinfo {author} {\bibfnamefont {Takafumi}\ \bibnamefont {Sato}}, \bibinfo
  {author} {\bibfnamefont {Takashi}\ \bibnamefont {Takahashi}}, \bibinfo
  {author} {\bibfnamefont {Kouji}\ \bibnamefont {Segawa}}, \ and\ \bibinfo
  {author} {\bibfnamefont {Yoichi}\ \bibnamefont {Ando}},\ }\bibfield  {title}
  {\enquote {\bibinfo {title} {Work function of bulk-insulating topological
  insulator Bi$_{2–x}$Sb$_x$Te$_{3–y}$Se$_y$},}\ }\href {\doibase 10.1063/1.4961987}
  {\bibfield  {journal} {\bibinfo  {journal} {Applied Physics Letters}\
  }\textbf {\bibinfo {volume} {109}},\ \bibinfo {pages} {091601} (\bibinfo
  {year} {2016})}\BibitemShut {NoStop}%
\bibitem [{\citenamefont {Chen}\ \emph {et~al.}(2011)\citenamefont {Chen},
  \citenamefont {Zhu}, \citenamefont {Xiao},\ and\ \citenamefont
  {Zhang}}]{PhysRevLett.107.056804}%
  \BibitemOpen
  \bibfield  {author} {\bibinfo {author} {\bibfnamefont {Hua}\ \bibnamefont
  {Chen}}, \bibinfo {author} {\bibfnamefont {Wenguang}\ \bibnamefont {Zhu}},
  \bibinfo {author} {\bibfnamefont {Di}~\bibnamefont {Xiao}}, \ and\ \bibinfo
  {author} {\bibfnamefont {Zhenyu}\ \bibnamefont {Zhang}},\ }\bibfield  {title}
  {\enquote {\bibinfo {title} {Co oxidation facilitated by robust surface
  states on au-covered topological insulators},}\ }\href {\doibase
  10.1103/PhysRevLett.107.056804} {\bibfield  {journal} {\bibinfo  {journal}
  {Phys. Rev. Lett.}\ }\textbf {\bibinfo {volume} {107}},\ \bibinfo {pages}
  {056804} (\bibinfo {year} {2011})}\BibitemShut {NoStop}%
\bibitem [{\citenamefont {Asbrand}\ and\ \citenamefont
  {Eisenmann}(1993)}]{asbrand1993crystal}%
  \BibitemOpen
  \bibfield  {author} {\bibinfo {author} {\bibfnamefont {M}~\bibnamefont
  {Asbrand}}\ and\ \bibinfo {author} {\bibfnamefont {B}~\bibnamefont
  {Eisenmann}},\ }\bibfield  {title} {\enquote {\bibinfo {title} {Crystal
  structure of decapotassium hexabismutidodistannate, K$_{10}$ [Sn$_2$Bi$_6$]},}\ }\href
  {\doibase 10.1524/zkri.1993.205.Part-2.323} {\bibfield  {journal} {\bibinfo
  {journal} {Zeitschrift f{\"u}r Kristallographie-Crystalline Materials}\
  }\textbf {\bibinfo {volume} {205}},\ \bibinfo {pages} {323--324} (\bibinfo
  {year} {1993})}\BibitemShut {NoStop}%
\bibitem [{\citenamefont {Asbrand}\ \emph {et~al.}(1998)\citenamefont
  {Asbrand}, \citenamefont {Eisenmann}, \citenamefont {Engelhardt},\ and\
  \citenamefont {R{\"o}{\ss}ler}}]{asbrand1998dimere}%
  \BibitemOpen
  \bibfield  {author} {\bibinfo {author} {\bibfnamefont {Matthias}\
  \bibnamefont {Asbrand}}, \bibinfo {author} {\bibfnamefont {Brigitte}\
  \bibnamefont {Eisenmann}}, \bibinfo {author} {\bibfnamefont {Holger}\
  \bibnamefont {Engelhardt}}, \ and\ \bibinfo {author} {\bibfnamefont {Ute}\
  \bibnamefont {R{\"o}{\ss}ler}},\ }\bibfield  {title} {\enquote {\bibinfo
  {title} {Dimeric and Polymeric Pnictidostannate(IV) Anions - Preparation and Crystal Structures of Na$_2$K$_3$[SnP$_3$], Na$_2$Cs$_3$[SnP$_3$] and Na$_2$K$_3$[SnBi$_3$]},}\ }\href {\doibase
  10.1515/znb-1998-0402} {\bibfield  {journal} {\bibinfo  {journal}
  {Zeitschrift f{\"u}r Naturforschung B}\ }\textbf {\bibinfo {volume} {53}},\
  \bibinfo {pages} {405--410} (\bibinfo {year} {1998})}\BibitemShut {NoStop}%
\bibitem [{\citenamefont {Behera}\ \emph {et~al.}(2013)\citenamefont {Behera},
  \citenamefont {Samanta},\ and\ \citenamefont {Jena}}]{behera2013nitrate}%
  \BibitemOpen
  \bibfield  {author} {\bibinfo {author} {\bibfnamefont {Swayamprabha}\
  \bibnamefont {Behera}}, \bibinfo {author} {\bibfnamefont {Devleena}\
  \bibnamefont {Samanta}}, \ and\ \bibinfo {author} {\bibfnamefont {Puru}\
  \bibnamefont {Jena}},\ }\bibfield  {title} {\enquote {\bibinfo {title}
  {Nitrate superhalogens as building blocks of hypersalts},}\ }\href {\doibase
  10.1021/jp405201r} {\bibfield  {journal} {\bibinfo  {journal} {The Journal of
  Physical Chemistry A}\ }\textbf {\bibinfo {volume} {117}},\ \bibinfo {pages}
  {5428--5434} (\bibinfo {year} {2013})}\BibitemShut {NoStop}%
\bibitem [{\citenamefont {Bradlyn}\ \emph {et~al.}(2017)\citenamefont
  {Bradlyn}, \citenamefont {Elcoro}, \citenamefont {Cano}, \citenamefont
  {Vergniory}, \citenamefont {Wang}, \citenamefont {Felser}, \citenamefont
  {Aroyo},\ and\ \citenamefont {Bernevig}}]{bradlyn2017topological}%
  \BibitemOpen
  \bibfield  {author} {\bibinfo {author} {\bibfnamefont {Barry}\ \bibnamefont
  {Bradlyn}}, \bibinfo {author} {\bibfnamefont {L}~\bibnamefont {Elcoro}},
  \bibinfo {author} {\bibfnamefont {Jennifer}\ \bibnamefont {Cano}}, \bibinfo
  {author} {\bibfnamefont {MG}~\bibnamefont {Vergniory}}, \bibinfo {author}
  {\bibfnamefont {Zhijun}\ \bibnamefont {Wang}}, \bibinfo {author}
  {\bibfnamefont {C}~\bibnamefont {Felser}}, \bibinfo {author} {\bibfnamefont
  {Mois~I}\ \bibnamefont {Aroyo}}, \ and\ \bibinfo {author} {\bibfnamefont
  {B~Andrei}\ \bibnamefont {Bernevig}},\ }\bibfield  {title} {\enquote
  {\bibinfo {title} {Topological quantum chemistry},}\ }\href {\doibase
  10.1038/nature23268} {\bibfield  {journal} {\bibinfo  {journal} {Nature}\
  }\textbf {\bibinfo {volume} {547}},\ \bibinfo {pages} {298--305} (\bibinfo
  {year} {2017})}\BibitemShut {NoStop}%
\bibitem [{\citenamefont {Liu}\ and\ \citenamefont
  {Zunger}(2017)}]{PhysRevX.7.021019}%
  \BibitemOpen
  \bibfield  {author} {\bibinfo {author} {\bibfnamefont {Qihang}\ \bibnamefont
  {Liu}}\ and\ \bibinfo {author} {\bibfnamefont {Alex}\ \bibnamefont
  {Zunger}},\ }\bibfield  {title} {\enquote {\bibinfo {title} {Predicted
  realization of cubic dirac fermion in quasi-one-dimensional transition-metal
  monochalcogenides},}\ }\href {\doibase 10.1103/PhysRevX.7.021019} {\bibfield
  {journal} {\bibinfo  {journal} {Phys. Rev. X}\ }\textbf {\bibinfo {volume}
  {7}},\ \bibinfo {pages} {021019} (\bibinfo {year} {2017})}\BibitemShut
  {NoStop}%
\bibitem [{\citenamefont {Noguchi}\ \emph {et~al.}(2019)\citenamefont
  {Noguchi}, \citenamefont {Takahashi}, \citenamefont {Kuroda}, \citenamefont
  {Ochi}, \citenamefont {Shirasawa}, \citenamefont {Sakano}, \citenamefont
  {Bareille}, \citenamefont {Nakayama}, \citenamefont {Watson}, \citenamefont
  {Yaji} \emph {et~al.}}]{noguchi2019weak}%
  \BibitemOpen
  \bibfield  {author} {\bibinfo {author} {\bibfnamefont {Ryo}\ \bibnamefont
  {Noguchi}}, \bibinfo {author} {\bibfnamefont {T}~\bibnamefont {Takahashi}},
  \bibinfo {author} {\bibfnamefont {K}~\bibnamefont {Kuroda}}, \bibinfo
  {author} {\bibfnamefont {M}~\bibnamefont {Ochi}}, \bibinfo {author}
  {\bibfnamefont {T}~\bibnamefont {Shirasawa}}, \bibinfo {author}
  {\bibfnamefont {M}~\bibnamefont {Sakano}}, \bibinfo {author} {\bibfnamefont
  {C}~\bibnamefont {Bareille}}, \bibinfo {author} {\bibfnamefont
  {M}~\bibnamefont {Nakayama}}, \bibinfo {author} {\bibfnamefont
  {MD}~\bibnamefont {Watson}}, \bibinfo {author} {\bibfnamefont
  {K}~\bibnamefont {Yaji}},  \emph {et~al.},\ }\bibfield  {title} {\enquote
  {\bibinfo {title} {A weak topological insulator state in
  quasi-one-dimensional bismuth iodide},}\ }\href {\doibase
  10.1038/s41586-019-0927-7} {\bibfield  {journal} {\bibinfo  {journal}
  {Nature}\ }\textbf {\bibinfo {volume} {566}},\ \bibinfo {pages} {518--522}
  (\bibinfo {year} {2019})}\BibitemShut {NoStop}%
\bibitem [{\citenamefont {Shi}\ \emph {et~al.}(2021)\citenamefont {Shi},
  \citenamefont {Wieder}, \citenamefont {Meyerheim}, \citenamefont {Sun},
  \citenamefont {Zhang}, \citenamefont {Li}, \citenamefont {Shen},
  \citenamefont {Qi}, \citenamefont {Yang}, \citenamefont {Jena} \emph
  {et~al.}}]{shi2021charge}%
  \BibitemOpen
  \bibfield  {author} {\bibinfo {author} {\bibfnamefont {Wujun}\ \bibnamefont
  {Shi}}, \bibinfo {author} {\bibfnamefont {Benjamin~J}\ \bibnamefont
  {Wieder}}, \bibinfo {author} {\bibfnamefont {Holger~L}\ \bibnamefont
  {Meyerheim}}, \bibinfo {author} {\bibfnamefont {Yan}\ \bibnamefont {Sun}},
  \bibinfo {author} {\bibfnamefont {Yang}\ \bibnamefont {Zhang}}, \bibinfo
  {author} {\bibfnamefont {Yiwei}\ \bibnamefont {Li}}, \bibinfo {author}
  {\bibfnamefont {Lei}\ \bibnamefont {Shen}}, \bibinfo {author} {\bibfnamefont
  {Yanpeng}\ \bibnamefont {Qi}}, \bibinfo {author} {\bibfnamefont {Lexian}\
  \bibnamefont {Yang}}, \bibinfo {author} {\bibfnamefont {Jagannath}\
  \bibnamefont {Jena}},  \emph {et~al.},\ }\bibfield  {title} {\enquote
  {\bibinfo {title} {A charge-density-wave topological semimetal},}\ }\href
  {\doibase 10.1038/s41567-020-01104-z} {\bibfield  {journal} {\bibinfo
  {journal} {Nature Physics}\ }\textbf {\bibinfo {volume} {17}},\ \bibinfo
  {pages} {381--387} (\bibinfo {year} {2021})}\BibitemShut {NoStop}%
\bibitem [{\citenamefont {Papoian}\ and\ \citenamefont
  {Hoffmann}(2000)}]{papoian2000hypervalent}%
  \BibitemOpen
  \bibfield  {author} {\bibinfo {author} {\bibfnamefont {Garegin~A.}\
  \bibnamefont {Papoian}}\ and\ \bibinfo {author} {\bibfnamefont {Roald}\
  \bibnamefont {Hoffmann}},\ }\bibfield  {title} {\enquote {\bibinfo {title}
  {Hypervalent bonding in one, two, and three dimensions: Extending the
  zintl--klemm concept to nonclassical electron-rich networks},}\ }\href
  {\doibase 10.1002/1521-3773(20000717)39:14<2408::AID-ANIE2408>3.0.CO;2-U}
  {\bibfield  {journal} {\bibinfo  {journal} {Angewandte Chemie International
  Edition}\ }\textbf {\bibinfo {volume} {39}},\ \bibinfo {pages} {2408--2448}
  (\bibinfo {year} {2000})}\BibitemShut {NoStop}%
\bibitem [{\citenamefont {Khoury}\ and\ \citenamefont
  {Schoop}(2021)}]{khoury2021chemical}%
  \BibitemOpen
  \bibfield  {author} {\bibinfo {author} {\bibfnamefont {Jason~F}\ \bibnamefont
  {Khoury}}\ and\ \bibinfo {author} {\bibfnamefont {Leslie~M}\ \bibnamefont
  {Schoop}},\ }\bibfield  {title} {\enquote {\bibinfo {title} {Chemical bonds
  in topological materials},}\ }\href {\doibase 10.1016/j.trechm.2021.04.011}
  {\bibfield  {journal} {\bibinfo  {journal} {Trends in Chemistry}\ }\textbf
  {\bibinfo {volume} {3}},\ \bibinfo {pages} {700--715} (\bibinfo {year}
  {2021})}\BibitemShut {NoStop}%
\bibitem [{\citenamefont {Tan}\ \emph {et~al.}(2017)\citenamefont {Tan},
  \citenamefont {Wu}, \citenamefont {Zhu}, \citenamefont {Shen}, \citenamefont
  {Zhu}, \citenamefont {Zhao}, \citenamefont {Huang}, \citenamefont {Tao},\
  and\ \citenamefont {Xia}}]{tan201714mgbi11}%
  \BibitemOpen
  \bibfield  {author} {\bibinfo {author} {\bibfnamefont {Wenjie}\ \bibnamefont
  {Tan}}, \bibinfo {author} {\bibfnamefont {Zhen}\ \bibnamefont {Wu}}, \bibinfo
  {author} {\bibfnamefont {Min}\ \bibnamefont {Zhu}}, \bibinfo {author}
  {\bibfnamefont {Jiajun}\ \bibnamefont {Shen}}, \bibinfo {author}
  {\bibfnamefont {Tiejun}\ \bibnamefont {Zhu}}, \bibinfo {author}
  {\bibfnamefont {Xinbing}\ \bibnamefont {Zhao}}, \bibinfo {author}
  {\bibfnamefont {Baibiao}\ \bibnamefont {Huang}}, \bibinfo {author}
  {\bibfnamefont {Xu-tang}\ \bibnamefont {Tao}}, \ and\ \bibinfo {author}
  {\bibfnamefont {Sheng-qing}\ \bibnamefont {Xia}},\ }\bibfield  {title}
  {\enquote {\bibinfo {title} {$A_{14}$MgBi$_{11}$ ($A$ = Ca, Sr, Eu): Magnesium bismuth
  based zintl phases as potential thermoelectric materials},}\ }\href {\doibase
  10.1021/acs.inorgchem.7b01548} {\bibfield  {journal} {\bibinfo  {journal}
  {Inorganic chemistry}\ }\textbf {\bibinfo {volume} {56}},\ \bibinfo {pages}
  {10576--10583} (\bibinfo {year} {2017})}\BibitemShut {NoStop}%
\bibitem [{\citenamefont {Varnava}\ \emph {et~al.}(2022)\citenamefont
  {Varnava}, \citenamefont {Berry}, \citenamefont {McQueen},\ and\
  \citenamefont {Vanderbilt}}]{PhysRevB.105.235128}%
  \BibitemOpen
  \bibfield  {author} {\bibinfo {author} {\bibfnamefont {Nicodemos}\
  \bibnamefont {Varnava}}, \bibinfo {author} {\bibfnamefont {Tanya}\
  \bibnamefont {Berry}}, \bibinfo {author} {\bibfnamefont {Tyrel~M.}\
  \bibnamefont {McQueen}}, \ and\ \bibinfo {author} {\bibfnamefont {David}\
  \bibnamefont {Vanderbilt}},\ }\bibfield  {title} {\enquote {\bibinfo {title}
  {Engineering magnetic topological insulators in
  ${\mathrm{Eu}}_{5}{M}_{2}{X}_{6}$ zintl compounds},}\ }\href {\doibase
  10.1103/PhysRevB.105.235128} {\bibfield  {journal} {\bibinfo  {journal}
  {Phys. Rev. B}\ }\textbf {\bibinfo {volume} {105}},\ \bibinfo {pages}
  {235128} (\bibinfo {year} {2022})}\BibitemShut {NoStop}%
\bibitem [{\citenamefont {Setyawan}\ and\ \citenamefont
  {Curtarolo}(2010)}]{setyawan2010high-throughput}%
  \BibitemOpen
  \bibfield  {author} {\bibinfo {author} {\bibfnamefont {Wahyu}\ \bibnamefont
  {Setyawan}}\ and\ \bibinfo {author} {\bibfnamefont {Stefano}\ \bibnamefont
  {Curtarolo}},\ }\bibfield  {title} {\enquote {\bibinfo {title}
  {High-throughput electronic band structure calculations: Challenges and
  tools},}\ }\href {\doibase https://doi.org/10.1016/j.commatsci.2010.05.010}
  {\bibfield  {journal} {\bibinfo  {journal} {Computational Materials Science}\
  }\textbf {\bibinfo {volume} {49}},\ \bibinfo {pages} {299 -- 312} (\bibinfo
  {year} {2010})}\BibitemShut {NoStop}%
\bibitem [{\citenamefont {Ruck}(2015)}]{ruck2015between}%
  \BibitemOpen
  \bibfield  {author} {\bibinfo {author} {\bibfnamefont {Michael}\ \bibnamefont
  {Ruck}},\ }\bibfield  {title} {\enquote {\bibinfo {title} {Between covalent
  and metallic bonding: From clusters to intermetallics of bismuth},}\ }in\
  \href {\doibase 10.1016/B978-0-12-409547-2.11445-3} {\bibinfo
  {booktitle} {Reference Module in Chemistry, Molecular Sciences and Chemical
  Engineering}}\ (\bibinfo  {publisher} {Elsevier},\ \bibinfo {year}
  {2015})\BibitemShut {NoStop}%
\bibitem [{\citenamefont {Reber}\ \emph {et~al.}(2007)\citenamefont {Reber},
  \citenamefont {Khanna},\ and\ \citenamefont
  {Castleman}}]{reber2007superatom}%
  \BibitemOpen
  \bibfield  {author} {\bibinfo {author} {\bibfnamefont {Arthur~C}\
  \bibnamefont {Reber}}, \bibinfo {author} {\bibfnamefont {Shiv~N}\
  \bibnamefont {Khanna}}, \ and\ \bibinfo {author} {\bibfnamefont {A~Welford}\
  \bibnamefont {Castleman}},\ }\bibfield  {title} {\enquote {\bibinfo {title}
  {Superatom compounds, clusters, and assemblies: Ultra alkali motifs and
  architectures},}\ }\href {\doibase 10.1021/ja071647n} {\bibfield  {journal}
  {\bibinfo  {journal} {Journal of the American Chemical Society}\ }\textbf
  {\bibinfo {volume} {129}},\ \bibinfo {pages} {10189--10194} (\bibinfo {year}
  {2007})}\BibitemShut {NoStop}%
\bibitem [{\citenamefont {Kresse}\ and\ \citenamefont
  {Joubert}(1999)}]{PhysRevB.59.1758}%
  \BibitemOpen
  \bibfield  {author} {\bibinfo {author} {\bibfnamefont {G.}~\bibnamefont
  {Kresse}}\ and\ \bibinfo {author} {\bibfnamefont {D.}~\bibnamefont
  {Joubert}},\ }\bibfield  {title} {\enquote {\bibinfo {title} {From ultrasoft
  pseudopotentials to the projector augmented-wave method},}\ }\href {\doibase
  10.1103/PhysRevB.59.1758} {\bibfield  {journal} {\bibinfo  {journal} {Phys.
  Rev. B}\ }\textbf {\bibinfo {volume} {59}},\ \bibinfo {pages} {1758--1775}
  (\bibinfo {year} {1999})}\BibitemShut {NoStop}%
\bibitem [{\citenamefont {Kresse}\ and\ \citenamefont
  {Furthm\"uller}(1996)}]{PhysRevB.54.11169}%
  \BibitemOpen
  \bibfield  {author} {\bibinfo {author} {\bibfnamefont {G.}~\bibnamefont
  {Kresse}}\ and\ \bibinfo {author} {\bibfnamefont {J.}~\bibnamefont
  {Furthm\"uller}},\ }\bibfield  {title} {\enquote {\bibinfo {title} {Efficient
  iterative schemes for ab initio total-energy calculations using a plane-wave
  basis set},}\ }\href {\doibase 10.1103/PhysRevB.54.11169} {\bibfield
  {journal} {\bibinfo  {journal} {Phys. Rev. B}\ }\textbf {\bibinfo {volume}
  {54}},\ \bibinfo {pages} {11169--11186} (\bibinfo {year} {1996})}\BibitemShut
  {NoStop}%
\bibitem [{\citenamefont {Csonka}\ \emph {et~al.}(2009)\citenamefont {Csonka},
  \citenamefont {Perdew}, \citenamefont {Ruzsinszky}, \citenamefont
  {Philipsen}, \citenamefont {Leb\`egue}, \citenamefont {Paier}, \citenamefont
  {Vydrov},\ and\ \citenamefont {\'Angy\'an}}]{PhysRevB.79.155107}%
  \BibitemOpen
  \bibfield  {author} {\bibinfo {author} {\bibfnamefont {G\'abor~I.}\
  \bibnamefont {Csonka}}, \bibinfo {author} {\bibfnamefont {John~P.}\
  \bibnamefont {Perdew}}, \bibinfo {author} {\bibfnamefont {Adrienn}\
  \bibnamefont {Ruzsinszky}}, \bibinfo {author} {\bibfnamefont {Pier H.~T.}\
  \bibnamefont {Philipsen}}, \bibinfo {author} {\bibfnamefont {S\'ebastien}\
  \bibnamefont {Leb\`egue}}, \bibinfo {author} {\bibfnamefont {Joachim}\
  \bibnamefont {Paier}}, \bibinfo {author} {\bibfnamefont {Oleg~A.}\
  \bibnamefont {Vydrov}}, \ and\ \bibinfo {author} {\bibfnamefont {J\'anos~G.}\
  \bibnamefont {\'Angy\'an}},\ }\bibfield  {title} {\enquote {\bibinfo {title}
  {Assessing the performance of recent density functionals for bulk solids},}\
  }\href {\doibase 10.1103/PhysRevB.79.155107} {\bibfield  {journal} {\bibinfo
  {journal} {Phys. Rev. B}\ }\textbf {\bibinfo {volume} {79}},\ \bibinfo
  {pages} {155107} (\bibinfo {year} {2009})}\BibitemShut {NoStop}%
\bibitem [{\citenamefont {Grimme}\ \emph {et~al.}(2010)\citenamefont {Grimme},
  \citenamefont {Antony}, \citenamefont {Ehrlich},\ and\ \citenamefont
  {Krieg}}]{grimme2010consistent}%
  \BibitemOpen
  \bibfield  {author} {\bibinfo {author} {\bibfnamefont {Stefan}\ \bibnamefont
  {Grimme}}, \bibinfo {author} {\bibfnamefont {Jens}\ \bibnamefont {Antony}},
  \bibinfo {author} {\bibfnamefont {Stephan}\ \bibnamefont {Ehrlich}}, \ and\
  \bibinfo {author} {\bibfnamefont {Helge}\ \bibnamefont {Krieg}},\ }\bibfield
  {title} {\enquote {\bibinfo {title} {A consistent and accurate ab initio
  parametrization of density functional dispersion correction (dft-d) for the
  94 elements h-pu},}\ }\href {\doibase 10.1063/1.3382344} {\bibfield
  {journal} {\bibinfo  {journal} {The Journal of Chemical Physics}\ }\textbf
  {\bibinfo {volume} {132}},\ \bibinfo {pages} {154104} (\bibinfo {year}
  {2010})}\BibitemShut {NoStop}%
\bibitem [{\citenamefont {Grimme}\ \emph {et~al.}(2011)\citenamefont {Grimme},
  \citenamefont {Ehrlich},\ and\ \citenamefont {Goerigk}}]{grimme2011effect}%
  \BibitemOpen
  \bibfield  {author} {\bibinfo {author} {\bibfnamefont {Stefan}\ \bibnamefont
  {Grimme}}, \bibinfo {author} {\bibfnamefont {Stephan}\ \bibnamefont
  {Ehrlich}}, \ and\ \bibinfo {author} {\bibfnamefont {Lars}\ \bibnamefont
  {Goerigk}},\ }\bibfield  {title} {\enquote {\bibinfo {title} {Effect of the
  damping function in dispersion corrected density functional theory},}\ }\href
  {\doibase https://doi.org/10.1002/jcc.21759} {\bibfield  {journal} {\bibinfo
  {journal} {Journal of Computational Chemistry}\ }\textbf {\bibinfo {volume}
  {32}},\ \bibinfo {pages} {1456--1465} (\bibinfo {year} {2011})}\BibitemShut
  {NoStop}%
\bibitem [{\citenamefont {Wang}\ \emph {et~al.}(2021)\citenamefont {Wang},
  \citenamefont {Xu}, \citenamefont {Liu}, \citenamefont {Tang},\ and\
  \citenamefont {Geng}}]{wang2021vaspkit}%
  \BibitemOpen
  \bibfield  {author} {\bibinfo {author} {\bibfnamefont {Vei}\ \bibnamefont
  {Wang}}, \bibinfo {author} {\bibfnamefont {Nan}\ \bibnamefont {Xu}}, \bibinfo
  {author} {\bibfnamefont {Jin-Cheng}\ \bibnamefont {Liu}}, \bibinfo {author}
  {\bibfnamefont {Gang}\ \bibnamefont {Tang}}, \ and\ \bibinfo {author}
  {\bibfnamefont {Wen-Tong}\ \bibnamefont {Geng}},\ }\bibfield  {title}
  {\enquote {\bibinfo {title} {Vaspkit: a user-friendly interface facilitating
  high-throughput computing and analysis using vasp code},}\ }\href
  {https://www.sciencedirect.com/science/article/pii/S0010465521001454?via%3Dihub}
  {\bibfield  {journal} {\bibinfo  {journal} {Computer Physics Communications}\
  ,\ \bibinfo {pages} {108033}} (\bibinfo {year} {2021})}\BibitemShut {NoStop}%
\bibitem [{\citenamefont {Curtarolo}\ \emph {et~al.}(2012)\citenamefont
  {Curtarolo}, \citenamefont {Setyawan}, \citenamefont {Hart}, \citenamefont
  {Jahnatek}, \citenamefont {Chepulskii}, \citenamefont {Taylor}, \citenamefont
  {Wang}, \citenamefont {Xue}, \citenamefont {Yang}, \citenamefont {Levy} \emph
  {et~al.}}]{curtarolo2012aflow}%
  \BibitemOpen
  \bibfield  {author} {\bibinfo {author} {\bibfnamefont {Stefano}\ \bibnamefont
  {Curtarolo}}, \bibinfo {author} {\bibfnamefont {Wahyu}\ \bibnamefont
  {Setyawan}}, \bibinfo {author} {\bibfnamefont {Gus~LW}\ \bibnamefont {Hart}},
  \bibinfo {author} {\bibfnamefont {Michal}\ \bibnamefont {Jahnatek}}, \bibinfo
  {author} {\bibfnamefont {Roman~V}\ \bibnamefont {Chepulskii}}, \bibinfo
  {author} {\bibfnamefont {Richard~H}\ \bibnamefont {Taylor}}, \bibinfo
  {author} {\bibfnamefont {Shidong}\ \bibnamefont {Wang}}, \bibinfo {author}
  {\bibfnamefont {Junkai}\ \bibnamefont {Xue}}, \bibinfo {author}
  {\bibfnamefont {Kesong}\ \bibnamefont {Yang}}, \bibinfo {author}
  {\bibfnamefont {Ohad}\ \bibnamefont {Levy}},  \emph {et~al.},\ }\bibfield
  {title} {\enquote {\bibinfo {title} {Aflow: An automatic framework for
  high-throughput materials discovery},}\ }\href {\doibase
  10.1016/j.commatsci.2012.02.005} {\bibfield  {journal} {\bibinfo  {journal}
  {Computational Materials Science}\ }\textbf {\bibinfo {volume} {58}},\
  \bibinfo {pages} {218--226} (\bibinfo {year} {2012})}\BibitemShut {NoStop}%
\bibitem [{\citenamefont {Herath}\ \emph {et~al.}(2020)\citenamefont {Herath},
  \citenamefont {Tavadze}, \citenamefont {He}, \citenamefont {Bousquet},
  \citenamefont {Singh}, \citenamefont {Muñoz},\ and\ \citenamefont
  {Romero}}]{herath2020pyprocar}%
  \BibitemOpen
  \bibfield  {author} {\bibinfo {author} {\bibfnamefont {Uthpala}\ \bibnamefont
  {Herath}}, \bibinfo {author} {\bibfnamefont {Pedram}\ \bibnamefont
  {Tavadze}}, \bibinfo {author} {\bibfnamefont {Xu}~\bibnamefont {He}},
  \bibinfo {author} {\bibfnamefont {Eric}\ \bibnamefont {Bousquet}}, \bibinfo
  {author} {\bibfnamefont {Sobhit}\ \bibnamefont {Singh}}, \bibinfo {author}
  {\bibfnamefont {Francisco}\ \bibnamefont {Muñoz}}, \ and\ \bibinfo {author}
  {\bibfnamefont {Aldo~H.}\ \bibnamefont {Romero}},\ }\bibfield  {title}
  {\enquote {\bibinfo {title} {Pyprocar: A python library for electronic
  structure pre/post-processing},}\ }\href {\doibase
  https://doi.org/10.1016/j.cpc.2019.107080} {\bibfield  {journal} {\bibinfo
  {journal} {Computer Physics Communications}\ }\textbf {\bibinfo {volume}
  {251}},\ \bibinfo {pages} {107080} (\bibinfo {year} {2020})}\BibitemShut
  {NoStop}%
\bibitem [{\citenamefont {Togo}\ and\ \citenamefont
  {Tanaka}(2018)}]{togo2018texttt}%
  \BibitemOpen
  \bibfield  {author} {\bibinfo {author} {\bibfnamefont {Atsushi}\ \bibnamefont
  {Togo}}\ and\ \bibinfo {author} {\bibfnamefont {Isao}\ \bibnamefont
  {Tanaka}},\ }\bibfield  {title} {\enquote {\bibinfo {title} {Spglib: a
  software library for crystal symmetry search},}\ }\href
  {https://arxiv.org/abs/1808.01590} {\bibfield  {journal} {\bibinfo  {journal}
  {arXiv preprint arXiv:1808.01590}\ } (\bibinfo {year} {2018})}\BibitemShut
  {NoStop}%
\bibitem [{\citenamefont {Momma}\ and\ \citenamefont
  {Izumi}(2008)}]{momma2008vesta}%
  \BibitemOpen
  \bibfield  {author} {\bibinfo {author} {\bibfnamefont {Koichi}\ \bibnamefont
  {Momma}}\ and\ \bibinfo {author} {\bibfnamefont {Fujio}\ \bibnamefont
  {Izumi}},\ }\bibfield  {title} {\enquote {\bibinfo {title} {Vesta: a
  three-dimensional visualization system for electronic and structural
  analysis},}\ }\href {\doibase 10.1107/S0021889808012016} {\bibfield
  {journal} {\bibinfo  {journal} {Journal of Applied Crystallography}\ }\textbf
  {\bibinfo {volume} {41}},\ \bibinfo {pages} {653--658} (\bibinfo {year}
  {2008})}\BibitemShut {NoStop}%
\bibitem [{\citenamefont {Kawamura}(2019)}]{kawamura2019fermisurfer}%
  \BibitemOpen
  \bibfield  {author} {\bibinfo {author} {\bibfnamefont {Mitsuaki}\
  \bibnamefont {Kawamura}},\ }\bibfield  {title} {\enquote {\bibinfo {title}
  {Fermisurfer: Fermi-surface viewer providing multiple representation
  schemes},}\ }\href {\doibase doi.org/10.1016/j.cpc.2019.01.017} {\bibfield
  {journal} {\bibinfo  {journal} {Computer Physics Communications}\ }\textbf
  {\bibinfo {volume} {239}},\ \bibinfo {pages} {197--203} (\bibinfo {year}
  {2019})}\BibitemShut {NoStop}%
\bibitem [{\citenamefont {Nordheim}(1931)}]{nordheim1931elektronentheorie}%
  \BibitemOpen
  \bibfield  {author} {\bibinfo {author} {\bibfnamefont {Lothar}\ \bibnamefont
  {Nordheim}},\ }\bibfield  {title} {\enquote {\bibinfo {title} {Zur Elektronentheorie der Metalle. I},}\ }\href {\doibase
  10.1002/andp.19314010507} {\bibfield  {journal} {\bibinfo  {journal} {Annalen
  der Physik}\ }\textbf {\bibinfo {volume} {401}},\ \bibinfo {pages} {607--640}
  (\bibinfo {year} {1931})}\BibitemShut {NoStop}%
\bibitem [{\citenamefont {Bellaiche}\ and\ \citenamefont
  {Vanderbilt}(2000)}]{PhysRevB.61.7877}%
  \BibitemOpen
  \bibfield  {author} {\bibinfo {author} {\bibfnamefont {L.}~\bibnamefont
  {Bellaiche}}\ and\ \bibinfo {author} {\bibfnamefont {David}\ \bibnamefont
  {Vanderbilt}},\ }\bibfield  {title} {\enquote {\bibinfo {title} {Virtual
  crystal approximation revisited: Application to dielectric and piezoelectric
  properties of perovskites},}\ }\href {\doibase 10.1103/PhysRevB.61.7877}
  {\bibfield  {journal} {\bibinfo  {journal} {Phys. Rev. B}\ }\textbf {\bibinfo
  {volume} {61}},\ \bibinfo {pages} {7877--7882} (\bibinfo {year}
  {2000})}\BibitemShut {NoStop}%
\bibitem [{\citenamefont {Zunger}\ \emph {et~al.}(1990)\citenamefont {Zunger},
  \citenamefont {Wei}, \citenamefont {Ferreira},\ and\ \citenamefont
  {Bernard}}]{PhysRevLett.65.353}%
  \BibitemOpen
  \bibfield  {author} {\bibinfo {author} {\bibfnamefont {Alex}\ \bibnamefont
  {Zunger}}, \bibinfo {author} {\bibfnamefont {S.-H.}\ \bibnamefont {Wei}},
  \bibinfo {author} {\bibfnamefont {L.~G.}\ \bibnamefont {Ferreira}}, \ and\
  \bibinfo {author} {\bibfnamefont {James~E.}\ \bibnamefont {Bernard}},\
  }\bibfield  {title} {\enquote {\bibinfo {title} {Special quasirandom
  structures},}\ }\href {\doibase 10.1103/PhysRevLett.65.353} {\bibfield
  {journal} {\bibinfo  {journal} {Phys. Rev. Lett.}\ }\textbf {\bibinfo
  {volume} {65}},\ \bibinfo {pages} {353--356} (\bibinfo {year}
  {1990})}\BibitemShut {NoStop}%
\bibitem [{\citenamefont {Soven}(1967)}]{PhysRev.156.809}%
  \BibitemOpen
  \bibfield  {author} {\bibinfo {author} {\bibfnamefont {Paul}\ \bibnamefont
  {Soven}},\ }\bibfield  {title} {\enquote {\bibinfo {title}
  {Coherent-potential model of substitutional disordered alloys},}\ }\href
  {\doibase 10.1103/PhysRev.156.809} {\bibfield  {journal} {\bibinfo  {journal}
  {Phys. Rev.}\ }\textbf {\bibinfo {volume} {156}},\ \bibinfo {pages}
  {809--813} (\bibinfo {year} {1967})}\BibitemShut {NoStop}%
\bibitem [{\citenamefont {Wimmer}\ \emph {et~al.}(2021)\citenamefont {Wimmer},
  \citenamefont {S{\'a}nchez-Barriga}, \citenamefont {K{\"u}ppers},
  \citenamefont {Ney}, \citenamefont {Schierle}, \citenamefont {Freyse},
  \citenamefont {Caha}, \citenamefont {Michali{\v{c}}ka}, \citenamefont
  {Liebmann}, \citenamefont {Primetzhofer} \emph {et~al.}}]{wimmer2021mn}%
  \BibitemOpen
  \bibfield  {author} {\bibinfo {author} {\bibfnamefont {Stefan}\ \bibnamefont
  {Wimmer}}, \bibinfo {author} {\bibfnamefont {Jaime}\ \bibnamefont
  {S{\'a}nchez-Barriga}}, \bibinfo {author} {\bibfnamefont {Philipp}\
  \bibnamefont {K{\"u}ppers}}, \bibinfo {author} {\bibfnamefont {Andreas}\
  \bibnamefont {Ney}}, \bibinfo {author} {\bibfnamefont {Enrico}\ \bibnamefont
  {Schierle}}, \bibinfo {author} {\bibfnamefont {Friedrich}\ \bibnamefont
  {Freyse}}, \bibinfo {author} {\bibfnamefont {Ondrej}\ \bibnamefont {Caha}},
  \bibinfo {author} {\bibfnamefont {Jan}\ \bibnamefont {Michali{\v{c}}ka}},
  \bibinfo {author} {\bibfnamefont {Marcus}\ \bibnamefont {Liebmann}}, \bibinfo
  {author} {\bibfnamefont {Daniel}\ \bibnamefont {Primetzhofer}},  et~al.,\ }\bibfield  {title} {\enquote {\bibinfo {title} {Mn-rich MnSb$_2$Te$_4$:
  A topological insulator with magnetic gap closing at high curie temperatures
  of 45-50 K},}\ }\href {\doibase 10.1002/adma.202102935} {\bibfield
  {journal} {\bibinfo  {journal} {Advanced Materials}\ }\textbf {\bibinfo
  {volume} {33}},\ \bibinfo {pages} {2102935} (\bibinfo {year}
  {2021})}\BibitemShut {NoStop}%
\bibitem [{\citenamefont {Pizzi}\ \emph {et~al.}(2020)\citenamefont {Pizzi},
  \citenamefont {Vitale}, \citenamefont {Arita}, \citenamefont {Blügel},
  \citenamefont {Freimuth}, \citenamefont {G{\'{e}}ranton}, \citenamefont
  {Gibertini}, \citenamefont {Gresch}, \citenamefont {Johnson}, \citenamefont
  {Koretsune}, \citenamefont {Iba{\~{n}}ez-Azpiroz}, \citenamefont {Lee},
  \citenamefont {Lihm}, \citenamefont {Marchand}, \citenamefont {Marrazzo},
  \citenamefont {Mokrousov}, \citenamefont {Mustafa}, \citenamefont {Nohara},
  \citenamefont {Nomura}, \citenamefont {Paulatto}, \citenamefont
  {Ponc{\'{e}}}, \citenamefont {Ponweiser}, \citenamefont {Qiao}, \citenamefont
  {Thöle}, \citenamefont {Tsirkin}, \citenamefont {Wierzbowska}, \citenamefont
  {Marzari}, \citenamefont {Vanderbilt}, \citenamefont {Souza}, \citenamefont
  {Mostofi},\ and\ \citenamefont {Yates}}]{pizzi2020wannier90}%
  \BibitemOpen
  \bibfield  {author} {\bibinfo {author} {\bibfnamefont {Giovanni}\
  \bibnamefont {Pizzi}}, \bibinfo {author} {\bibfnamefont {Valerio}\
  \bibnamefont {Vitale}}, \bibinfo {author} {\bibfnamefont {Ryotaro}\
  \bibnamefont {Arita}}, \bibinfo {author} {\bibfnamefont {Stefan}\
  \bibnamefont {Blügel}}, \bibinfo {author} {\bibfnamefont {Frank}\
  \bibnamefont {Freimuth}}, \bibinfo {author} {\bibfnamefont {Guillaume}\
  \bibnamefont {G{\'{e}}ranton}}, \bibinfo {author} {\bibfnamefont {Marco}\
  \bibnamefont {Gibertini}}, \bibinfo {author} {\bibfnamefont {Dominik}\
  \bibnamefont {Gresch}}, \bibinfo {author} {\bibfnamefont {Charles}\
  \bibnamefont {Johnson}}, \bibinfo {author} {\bibfnamefont {Takashi}\
  \bibnamefont {Koretsune}}, \bibinfo {author} {\bibfnamefont {Julen}\
  \bibnamefont {Iba{\~{n}}ez-Azpiroz}}, \bibinfo {author} {\bibfnamefont
  {Hyungjun}\ \bibnamefont {Lee}}, \bibinfo {author} {\bibfnamefont {Jae-Mo}\
  \bibnamefont {Lihm}}, \bibinfo {author} {\bibfnamefont {Daniel}\ \bibnamefont
  {Marchand}}, \bibinfo {author} {\bibfnamefont {Antimo}\ \bibnamefont
  {Marrazzo}}, \bibinfo {author} {\bibfnamefont {Yuriy}\ \bibnamefont
  {Mokrousov}}, \bibinfo {author} {\bibfnamefont {Jamal~I}\ \bibnamefont
  {Mustafa}}, \bibinfo {author} {\bibfnamefont {Yoshiro}\ \bibnamefont
  {Nohara}}, \bibinfo {author} {\bibfnamefont {Yusuke}\ \bibnamefont {Nomura}},
  \bibinfo {author} {\bibfnamefont {Lorenzo}\ \bibnamefont {Paulatto}},
  \bibinfo {author} {\bibfnamefont {Samuel}\ \bibnamefont {Ponc{\'{e}}}},
  \bibinfo {author} {\bibfnamefont {Thomas}\ \bibnamefont {Ponweiser}},
  \bibinfo {author} {\bibfnamefont {Junfeng}\ \bibnamefont {Qiao}}, \bibinfo
  {author} {\bibfnamefont {Florian}\ \bibnamefont {Thöle}}, \bibinfo {author}
  {\bibfnamefont {Stepan~S}\ \bibnamefont {Tsirkin}}, \bibinfo {author}
  {\bibfnamefont {Ma{\l}gorzata}\ \bibnamefont {Wierzbowska}}, \bibinfo
  {author} {\bibfnamefont {Nicola}\ \bibnamefont {Marzari}}, \bibinfo {author}
  {\bibfnamefont {David}\ \bibnamefont {Vanderbilt}}, \bibinfo {author}
  {\bibfnamefont {Ivo}\ \bibnamefont {Souza}}, \bibinfo {author} {\bibfnamefont
  {Arash~A}\ \bibnamefont {Mostofi}}, \ and\ \bibinfo {author} {\bibfnamefont
  {Jonathan~R}\ \bibnamefont {Yates}},\ }\bibfield  {title} {\enquote {\bibinfo
  {title} {Wannier90 as a community code: new features and applications},}\
  }\href {\doibase 10.1088/1361-648x/ab51ff} {\bibfield  {journal} {\bibinfo
  {journal} {Journal of Physics: Condensed Matter}\ }\textbf {\bibinfo {volume}
  {32}},\ \bibinfo {pages} {165902} (\bibinfo {year} {2020})}\BibitemShut
  {NoStop}%
\bibitem [{\citenamefont {Gresch}\ \emph {et~al.}(2018)\citenamefont {Gresch},
  \citenamefont {Wu}, \citenamefont {Winkler}, \citenamefont {H\"auselmann},
  \citenamefont {Troyer},\ and\ \citenamefont
  {Soluyanov}}]{PhysRevMaterials.2.103805}%
  \BibitemOpen
  \bibfield  {author} {\bibinfo {author} {\bibfnamefont {Dominik}\ \bibnamefont
  {Gresch}}, \bibinfo {author} {\bibfnamefont {QuanSheng}\ \bibnamefont {Wu}},
  \bibinfo {author} {\bibfnamefont {Georg~W.}\ \bibnamefont {Winkler}},
  \bibinfo {author} {\bibfnamefont {Rico}\ \bibnamefont {H\"auselmann}},
  \bibinfo {author} {\bibfnamefont {Matthias}\ \bibnamefont {Troyer}}, \ and\
  \bibinfo {author} {\bibfnamefont {Alexey~A.}\ \bibnamefont {Soluyanov}},\
  }\bibfield  {title} {\enquote {\bibinfo {title} {Automated construction of
  symmetrized wannier-like tight-binding models from ab initio calculations},}\
  }\href {\doibase 10.1103/PhysRevMaterials.2.103805} {\bibfield  {journal}
  {\bibinfo  {journal} {Phys. Rev. Materials}\ }\textbf {\bibinfo {volume}
  {2}},\ \bibinfo {pages} {103805} (\bibinfo {year} {2018})}\BibitemShut
  {NoStop}%
\bibitem [{\citenamefont {Soluyanov}\ and\ \citenamefont
  {Vanderbilt}(2011)}]{PhysRevB.83.235401}%
  \BibitemOpen
  \bibfield  {author} {\bibinfo {author} {\bibfnamefont {Alexey~A.}\
  \bibnamefont {Soluyanov}}\ and\ \bibinfo {author} {\bibfnamefont {David}\
  \bibnamefont {Vanderbilt}},\ }\bibfield  {title} {\enquote {\bibinfo {title}
  {Computing topological invariants without inversion symmetry},}\ }\href
  {\doibase 10.1103/PhysRevB.83.235401} {\bibfield  {journal} {\bibinfo
  {journal} {Phys. Rev. B}\ }\textbf {\bibinfo {volume} {83}},\ \bibinfo
  {pages} {235401} (\bibinfo {year} {2011})}\BibitemShut {NoStop}%
\bibitem [{\citenamefont {Wu}\ \emph {et~al.}(2018)\citenamefont {Wu},
  \citenamefont {Zhang}, \citenamefont {Song}, \citenamefont {Troyer},\ and\
  \citenamefont {Soluyanov}}]{wu2018wanniertools}%
  \BibitemOpen
  \bibfield  {author} {\bibinfo {author} {\bibfnamefont {QuanSheng}\
  \bibnamefont {Wu}}, \bibinfo {author} {\bibfnamefont {ShengNan}\ \bibnamefont
  {Zhang}}, \bibinfo {author} {\bibfnamefont {Hai-Feng}\ \bibnamefont {Song}},
  \bibinfo {author} {\bibfnamefont {Matthias}\ \bibnamefont {Troyer}}, \ and\
  \bibinfo {author} {\bibfnamefont {Alexey~A.}\ \bibnamefont {Soluyanov}},\
  }\bibfield  {title} {\enquote {\bibinfo {title} {Wanniertools: An open-source
  software package for novel topological materials},}\ }\href {\doibase
  https://doi.org/10.1016/j.cpc.2017.09.033} {\bibfield  {journal} {\bibinfo
  {journal} {Computer Physics Communications}\ }\textbf {\bibinfo {volume}
  {224}},\ \bibinfo {pages} {405 -- 416} (\bibinfo {year} {2018})}\BibitemShut
  {NoStop}%
\bibitem [{\citenamefont {Sancho}\ \emph {et~al.}(1985)\citenamefont {Sancho},
  \citenamefont {Sancho}, \citenamefont {Sancho},\ and\ \citenamefont
  {Rubio}}]{sancho1985highly}%
  \BibitemOpen
  \bibfield  {author} {\bibinfo {author} {\bibfnamefont {M~P~Lopez}\
  \bibnamefont {Sancho}}, \bibinfo {author} {\bibfnamefont {J~M~Lopez}\
  \bibnamefont {Sancho}}, \bibinfo {author} {\bibfnamefont {J~M~L}\
  \bibnamefont {Sancho}}, \ and\ \bibinfo {author} {\bibfnamefont
  {J}~\bibnamefont {Rubio}},\ }\bibfield  {title} {\enquote {\bibinfo {title}
  {Highly convergent schemes for the calculation of bulk and surface green
  functions},}\ }\href {\doibase 10.1088/0305-4608/15/4/009} {\bibfield
  {journal} {\bibinfo  {journal} {Journal of Physics F: Metal Physics}\
  }\textbf {\bibinfo {volume} {15}},\ \bibinfo {pages} {851--858} (\bibinfo
  {year} {1985})}\BibitemShut {NoStop}%
\bibitem [{\citenamefont {Pizzi}\ \emph {et~al.}(2014)\citenamefont {Pizzi},
  \citenamefont {Volja}, \citenamefont {Kozinsky}, \citenamefont {Fornari},\
  and\ \citenamefont {Marzari}}]{pizzi2014boltzwann}%
  \BibitemOpen
  \bibfield  {author} {\bibinfo {author} {\bibfnamefont {Giovanni}\
  \bibnamefont {Pizzi}}, \bibinfo {author} {\bibfnamefont {Dmitri}\
  \bibnamefont {Volja}}, \bibinfo {author} {\bibfnamefont {Boris}\ \bibnamefont
  {Kozinsky}}, \bibinfo {author} {\bibfnamefont {Marco}\ \bibnamefont
  {Fornari}}, \ and\ \bibinfo {author} {\bibfnamefont {Nicola}\ \bibnamefont
  {Marzari}},\ }\bibfield  {title} {\enquote {\bibinfo {title} {Boltzwann: A
  code for the evaluation of thermoelectric and electronic transport properties
  with a maximally-localized wannier functions basis},}\ }\href {\doibase
  10.1016/j.cpc.2013.09.015} {\bibfield  {journal} {\bibinfo  {journal}
  {Computer Physics Communications}\ }\textbf {\bibinfo {volume} {185}},\
  \bibinfo {pages} {422--429} (\bibinfo {year} {2014})}\BibitemShut {NoStop}%
\bibitem [{\citenamefont {Giannozzi}\ \emph {et~al.}(2009)\citenamefont
  {Giannozzi}, \citenamefont {Baroni}, \citenamefont {Bonini}, \citenamefont
  {Calandra}, \citenamefont {Car}, \citenamefont {Cavazzoni}, \citenamefont
  {Ceresoli}, \citenamefont {Chiarotti}, \citenamefont {Cococcioni},
  \citenamefont {Dabo}, \citenamefont {{Dal Corso}}, \citenamefont
  {de~Gironcoli}, \citenamefont {Fabris}, \citenamefont {Fratesi},
  \citenamefont {Gebauer}, \citenamefont {Gerstmann}, \citenamefont
  {Gougoussis}, \citenamefont {Kokalj}, \citenamefont {Lazzeri}, \citenamefont
  {Martin-Samos}, \citenamefont {Marzari}, \citenamefont {Mauri}, \citenamefont
  {Mazzarello}, \citenamefont {Paolini}, \citenamefont {Pasquarello},
  \citenamefont {Paulatto}, \citenamefont {Sbraccia}, \citenamefont {Scandolo},
  \citenamefont {Sclauzero}, \citenamefont {Seitsonen}, \citenamefont
  {Smogunov}, \citenamefont {Umari},\ and\ \citenamefont
  {Wentzcovitch}}]{giannozzi2009quantum}%
  \BibitemOpen
  \bibfield  {author} {\bibinfo {author} {\bibfnamefont {Paolo}\ \bibnamefont
  {Giannozzi}}, \bibinfo {author} {\bibfnamefont {Stefano}\ \bibnamefont
  {Baroni}}, \bibinfo {author} {\bibfnamefont {Nicola}\ \bibnamefont {Bonini}},
  \bibinfo {author} {\bibfnamefont {Matteo}\ \bibnamefont {Calandra}}, \bibinfo
  {author} {\bibfnamefont {Roberto}\ \bibnamefont {Car}}, \bibinfo {author}
  {\bibfnamefont {Carlo}\ \bibnamefont {Cavazzoni}}, \bibinfo {author}
  {\bibfnamefont {Davide}\ \bibnamefont {Ceresoli}}, \bibinfo {author}
  {\bibfnamefont {Guido~L}\ \bibnamefont {Chiarotti}}, \bibinfo {author}
  {\bibfnamefont {Matteo}\ \bibnamefont {Cococcioni}}, \bibinfo {author}
  {\bibfnamefont {Ismaila}\ \bibnamefont {Dabo}}, \bibinfo {author}
  {\bibfnamefont {Andrea}\ \bibnamefont {{Dal Corso}}}, \bibinfo {author}
  {\bibfnamefont {Stefano}\ \bibnamefont {de~Gironcoli}}, \bibinfo {author}
  {\bibfnamefont {Stefano}\ \bibnamefont {Fabris}}, \bibinfo {author}
  {\bibfnamefont {Guido}\ \bibnamefont {Fratesi}}, \bibinfo {author}
  {\bibfnamefont {Ralph}\ \bibnamefont {Gebauer}}, \bibinfo {author}
  {\bibfnamefont {Uwe}\ \bibnamefont {Gerstmann}}, \bibinfo {author}
  {\bibfnamefont {Christos}\ \bibnamefont {Gougoussis}}, \bibinfo {author}
  {\bibfnamefont {Anton}\ \bibnamefont {Kokalj}}, \bibinfo {author}
  {\bibfnamefont {Michele}\ \bibnamefont {Lazzeri}}, \bibinfo {author}
  {\bibfnamefont {Layla}\ \bibnamefont {Martin-Samos}}, \bibinfo {author}
  {\bibfnamefont {Nicola}\ \bibnamefont {Marzari}}, \bibinfo {author}
  {\bibfnamefont {Francesco}\ \bibnamefont {Mauri}}, \bibinfo {author}
  {\bibfnamefont {Riccardo}\ \bibnamefont {Mazzarello}}, \bibinfo {author}
  {\bibfnamefont {Stefano}\ \bibnamefont {Paolini}}, \bibinfo {author}
  {\bibfnamefont {Alfredo}\ \bibnamefont {Pasquarello}}, \bibinfo {author}
  {\bibfnamefont {Lorenzo}\ \bibnamefont {Paulatto}}, \bibinfo {author}
  {\bibfnamefont {Carlo}\ \bibnamefont {Sbraccia}}, \bibinfo {author}
  {\bibfnamefont {Sandro}\ \bibnamefont {Scandolo}}, \bibinfo {author}
  {\bibfnamefont {Gabriele}\ \bibnamefont {Sclauzero}}, \bibinfo {author}
  {\bibfnamefont {Ari~P}\ \bibnamefont {Seitsonen}}, \bibinfo {author}
  {\bibfnamefont {Alexander}\ \bibnamefont {Smogunov}}, \bibinfo {author}
  {\bibfnamefont {Paolo}\ \bibnamefont {Umari}}, \ and\ \bibinfo {author}
  {\bibfnamefont {Renata~M}\ \bibnamefont {Wentzcovitch}},\ }\bibfield  {title}
  {\enquote {\bibinfo {title} {Quantum espresso: a modular and open-source
  software project for quantum simulations of materials},}\ }\href
  {http://www.quantum-espresso.org} {\bibfield  {journal} {\bibinfo  {journal}
  {Journal of Physics: Condensed Matter}\ }\textbf {\bibinfo {volume} {21}},\
  \bibinfo {pages} {395502 (19pp)} (\bibinfo {year} {2009})}\BibitemShut
  {NoStop}%
\bibitem [{\citenamefont {Giannozzi}\ \emph {et~al.}(2017)\citenamefont
  {Giannozzi}, \citenamefont {Andreussi}, \citenamefont {Brumme}, \citenamefont
  {Bunau}, \citenamefont {Nardelli}, \citenamefont {Calandra}, \citenamefont
  {Car}, \citenamefont {Cavazzoni}, \citenamefont {Ceresoli}, \citenamefont
  {Cococcioni}, \citenamefont {Colonna}, \citenamefont {Carnimeo},
  \citenamefont {Corso}, \citenamefont {de~Gironcoli}, \citenamefont {Delugas},
  \citenamefont {Jr}, \citenamefont {Ferretti}, \citenamefont {Floris},
  \citenamefont {Fratesi}, \citenamefont {Fugallo}, \citenamefont {Gebauer},
  \citenamefont {Gerstmann}, \citenamefont {Giustino}, \citenamefont {Gorni},
  \citenamefont {Jia}, \citenamefont {Kawamura}, \citenamefont {Ko},
  \citenamefont {Kokalj}, \citenamefont {Küçükbenli}, \citenamefont
  {Lazzeri}, \citenamefont {Marsili}, \citenamefont {Marzari}, \citenamefont
  {Mauri}, \citenamefont {Nguyen}, \citenamefont {Nguyen}, \citenamefont {de-la
  Roza}, \citenamefont {Paulatto}, \citenamefont {Poncé}, \citenamefont
  {Rocca}, \citenamefont {Sabatini}, \citenamefont {Santra}, \citenamefont
  {Schlipf}, \citenamefont {Seitsonen}, \citenamefont {Smogunov}, \citenamefont
  {Timrov}, \citenamefont {Thonhauser}, \citenamefont {Umari}, \citenamefont
  {Vast}, \citenamefont {Wu},\ and\ \citenamefont
  {Baroni}}]{giannozzi2017advanced}%
  \BibitemOpen
  \bibfield  {author} {\bibinfo {author} {\bibfnamefont {P}~\bibnamefont
  {Giannozzi}}, \bibinfo {author} {\bibfnamefont {O}~\bibnamefont {Andreussi}},
  \bibinfo {author} {\bibfnamefont {T}~\bibnamefont {Brumme}}, \bibinfo
  {author} {\bibfnamefont {O}~\bibnamefont {Bunau}}, \bibinfo {author}
  {\bibfnamefont {M~Buongiorno}\ \bibnamefont {Nardelli}}, \bibinfo {author}
  {\bibfnamefont {M}~\bibnamefont {Calandra}}, \bibinfo {author} {\bibfnamefont
  {R}~\bibnamefont {Car}}, \bibinfo {author} {\bibfnamefont {C}~\bibnamefont
  {Cavazzoni}}, \bibinfo {author} {\bibfnamefont {D}~\bibnamefont {Ceresoli}},
  \bibinfo {author} {\bibfnamefont {M}~\bibnamefont {Cococcioni}}, \bibinfo
  {author} {\bibfnamefont {N}~\bibnamefont {Colonna}}, \bibinfo {author}
  {\bibfnamefont {I}~\bibnamefont {Carnimeo}}, \bibinfo {author} {\bibfnamefont
  {A~Dal}\ \bibnamefont {Corso}}, \bibinfo {author} {\bibfnamefont
  {S}~\bibnamefont {de~Gironcoli}}, \bibinfo {author} {\bibfnamefont
  {P}~\bibnamefont {Delugas}}, \bibinfo {author} {\bibfnamefont {R~A~DiStasio}\
  \bibnamefont {Jr}}, \bibinfo {author} {\bibfnamefont {A}~\bibnamefont
  {Ferretti}}, \bibinfo {author} {\bibfnamefont {A}~\bibnamefont {Floris}},
  \bibinfo {author} {\bibfnamefont {G}~\bibnamefont {Fratesi}}, \bibinfo
  {author} {\bibfnamefont {G}~\bibnamefont {Fugallo}}, \bibinfo {author}
  {\bibfnamefont {R}~\bibnamefont {Gebauer}}, \bibinfo {author} {\bibfnamefont
  {U}~\bibnamefont {Gerstmann}}, \bibinfo {author} {\bibfnamefont
  {F}~\bibnamefont {Giustino}}, \bibinfo {author} {\bibfnamefont
  {T}~\bibnamefont {Gorni}}, \bibinfo {author} {\bibfnamefont {J}~\bibnamefont
  {Jia}}, \bibinfo {author} {\bibfnamefont {M}~\bibnamefont {Kawamura}},
  \bibinfo {author} {\bibfnamefont {H-Y}\ \bibnamefont {Ko}}, \bibinfo {author}
  {\bibfnamefont {A}~\bibnamefont {Kokalj}}, \bibinfo {author} {\bibfnamefont
  {E}~\bibnamefont {Küçükbenli}}, \bibinfo {author} {\bibfnamefont
  {M}~\bibnamefont {Lazzeri}}, \bibinfo {author} {\bibfnamefont
  {M}~\bibnamefont {Marsili}}, \bibinfo {author} {\bibfnamefont
  {N}~\bibnamefont {Marzari}}, \bibinfo {author} {\bibfnamefont
  {F}~\bibnamefont {Mauri}}, \bibinfo {author} {\bibfnamefont {N~L}\
  \bibnamefont {Nguyen}}, \bibinfo {author} {\bibfnamefont {H-V}\ \bibnamefont
  {Nguyen}}, \bibinfo {author} {\bibfnamefont {A~Otero}\ \bibnamefont {de-la
  Roza}}, \bibinfo {author} {\bibfnamefont {L}~\bibnamefont {Paulatto}},
  \bibinfo {author} {\bibfnamefont {S}~\bibnamefont {Poncé}}, \bibinfo
  {author} {\bibfnamefont {D}~\bibnamefont {Rocca}}, \bibinfo {author}
  {\bibfnamefont {R}~\bibnamefont {Sabatini}}, \bibinfo {author} {\bibfnamefont
  {B}~\bibnamefont {Santra}}, \bibinfo {author} {\bibfnamefont {M}~\bibnamefont
  {Schlipf}}, \bibinfo {author} {\bibfnamefont {A~P}\ \bibnamefont
  {Seitsonen}}, \bibinfo {author} {\bibfnamefont {A}~\bibnamefont {Smogunov}},
  \bibinfo {author} {\bibfnamefont {I}~\bibnamefont {Timrov}}, \bibinfo
  {author} {\bibfnamefont {T}~\bibnamefont {Thonhauser}}, \bibinfo {author}
  {\bibfnamefont {P}~\bibnamefont {Umari}}, \bibinfo {author} {\bibfnamefont
  {N}~\bibnamefont {Vast}}, \bibinfo {author} {\bibfnamefont {X}~\bibnamefont
  {Wu}}, \ and\ \bibinfo {author} {\bibfnamefont {S}~\bibnamefont {Baroni}},\
  }\bibfield  {title} {\enquote {\bibinfo {title} {Advanced capabilities for
  materials modelling with quantum espresso},}\ }\href
  {http://stacks.iop.org/0953-8984/29/i=46/a=465901} {\bibfield  {journal}
  {\bibinfo  {journal} {Journal of Physics: Condensed Matter}\ }\textbf
  {\bibinfo {volume} {29}},\ \bibinfo {pages} {465901} (\bibinfo {year}
  {2017})}\BibitemShut {NoStop}%
\bibitem [{\citenamefont {Dal~Corso}(2014)}]{dal2014pseudopotentials}%
  \BibitemOpen
  \bibfield  {author} {\bibinfo {author} {\bibfnamefont {Andrea}\ \bibnamefont
  {Dal~Corso}},\ }\bibfield  {title} {\enquote {\bibinfo {title}
  {Pseudopotentials periodic table: From h to pu},}\ }\href {\doibase
  10.1016/j.commatsci.2014.07.043} {\bibfield  {journal} {\bibinfo  {journal}
  {Computational Materials Science}\ }\textbf {\bibinfo {volume} {95}},\
  \bibinfo {pages} {337--350} (\bibinfo {year} {2014})}\BibitemShut {NoStop}%
\bibitem [{\citenamefont {Tadano}\ \emph {et~al.}(2014)\citenamefont {Tadano},
  \citenamefont {Gohda},\ and\ \citenamefont
  {Tsuneyuki}}]{tadano2014anharmonic}%
  \BibitemOpen
  \bibfield  {author} {\bibinfo {author} {\bibfnamefont {T}~\bibnamefont
  {Tadano}}, \bibinfo {author} {\bibfnamefont {Y}~\bibnamefont {Gohda}}, \ and\
  \bibinfo {author} {\bibfnamefont {S}~\bibnamefont {Tsuneyuki}},\ }\bibfield
  {title} {\enquote {\bibinfo {title} {Anharmonic force constants extracted
  from first-principles molecular dynamics: applications to heat transfer
  simulations},}\ }\href {\doibase 10.1088/0953-8984/26/22/225402} {\bibfield
  {journal} {\bibinfo  {journal} {Journal of Physics: Condensed Matter}\
  }\textbf {\bibinfo {volume} {26}},\ \bibinfo {pages} {225402} (\bibinfo
  {year} {2014})}\BibitemShut {NoStop}%
\bibitem [{\citenamefont {Cahen}\ and\ \citenamefont
  {Kahn}(2003)}]{cahen2003electron}%
  \BibitemOpen
  \bibfield  {author} {\bibinfo {author} {\bibfnamefont {D.}~\bibnamefont
  {Cahen}}\ and\ \bibinfo {author} {\bibfnamefont {A.}~\bibnamefont {Kahn}},\
  }\bibfield  {title} {\enquote {\bibinfo {title} {Electron energetics at
  surfaces and interfaces: Concepts and experiments},}\ }\href {\doibase
  https://doi.org/10.1002/adma.200390065} {\bibfield  {journal} {\bibinfo
  {journal} {Advanced Materials}\ }\textbf {\bibinfo {volume} {15}},\ \bibinfo
  {pages} {271--277} (\bibinfo {year} {2003})}\BibitemShut {NoStop}%
\bibitem [{\citenamefont {Fall}\ \emph {et~al.}(1999)\citenamefont {Fall},
  \citenamefont {Binggeli},\ and\ \citenamefont
  {Baldereschi}}]{fall1999deriving}%
  \BibitemOpen
  \bibfield  {author} {\bibinfo {author} {\bibfnamefont {C~J}\ \bibnamefont
  {Fall}}, \bibinfo {author} {\bibfnamefont {N}~\bibnamefont {Binggeli}}, \
  and\ \bibinfo {author} {\bibfnamefont {A}~\bibnamefont {Baldereschi}},\
  }\bibfield  {title} {\enquote {\bibinfo {title} {Deriving accurate work
  functions from thin-slab calculations},}\ }\href {\doibase
  10.1088/0953-8984/11/13/006} {\bibfield  {journal} {\bibinfo  {journal}
  {Journal of Physics: Condensed Matter}\ }\textbf {\bibinfo {volume} {11}},\
  \bibinfo {pages} {2689--2696} (\bibinfo {year} {1999})}\BibitemShut {NoStop}%
\bibitem [{\citenamefont {Schindler}\ \emph {et~al.}(2018)\citenamefont
  {Schindler}, \citenamefont {Cook}, \citenamefont {Vergniory}, \citenamefont
  {Wang}, \citenamefont {Parkin}, \citenamefont {Bernevig},\ and\ \citenamefont
  {Neupert}}]{schindler2018higher}%
  \BibitemOpen
  \bibfield  {author} {\bibinfo {author} {\bibfnamefont {Frank}\ \bibnamefont
  {Schindler}}, \bibinfo {author} {\bibfnamefont {Ashley~M}\ \bibnamefont
  {Cook}}, \bibinfo {author} {\bibfnamefont {Maia~G}\ \bibnamefont
  {Vergniory}}, \bibinfo {author} {\bibfnamefont {Zhijun}\ \bibnamefont
  {Wang}}, \bibinfo {author} {\bibfnamefont {Stuart~SP}\ \bibnamefont
  {Parkin}}, \bibinfo {author} {\bibfnamefont {B~Andrei}\ \bibnamefont
  {Bernevig}}, \ and\ \bibinfo {author} {\bibfnamefont {Titus}\ \bibnamefont
  {Neupert}},\ }\bibfield  {title} {\enquote {\bibinfo {title} {Higher-order
  topological insulators},}\ }\href {\doibase 10.1126/sciadv.aat0346}
  {\bibfield  {journal} {\bibinfo  {journal} {Science advances}\ }\textbf
  {\bibinfo {volume} {4}},\ \bibinfo {pages} {eaat0346} (\bibinfo {year}
  {2018})}\BibitemShut {NoStop}%
\end{thebibliography}

\begin{thebibliography}{25}%
\makeatletter
\providecommand \@ifxundefined [1]{%
 \@ifx{#1\undefined}
}%
\providecommand \@ifnum [1]{%
 \ifnum #1\expandafter \@firstoftwo
 \else \expandafter \@secondoftwo
 \fi
}%
\providecommand \@ifx [1]{%
 \ifx #1\expandafter \@firstoftwo
 \else \expandafter \@secondoftwo
 \fi
}%
\providecommand \natexlab [1]{#1}%
\providecommand \enquote  [1]{``#1''}%
\providecommand \bibnamefont  [1]{#1}%
\providecommand \bibfnamefont [1]{#1}%
\providecommand \citenamefont [1]{#1}%
\providecommand \href@noop [0]{\@secondoftwo}%
\providecommand \href [0]{\begingroup \@sanitize@url \@href}%
\providecommand \@href[1]{\@@startlink{#1}\@@href}%
\providecommand \@@href[1]{\endgroup#1\@@endlink}%
\providecommand \@sanitize@url [0]{\catcode `\\12\catcode `\$12\catcode
  `\&12\catcode `\#12\catcode `\^12\catcode `\_12\catcode `\%12\relax}%
\providecommand \@@startlink[1]{}%
\providecommand \@@endlink[0]{}%
\providecommand \url  [0]{\begingroup\@sanitize@url \@url }%
\providecommand \@url [1]{\endgroup\@href {#1}{\urlprefix }}%
\providecommand \urlprefix  [0]{URL }%
\providecommand \Eprint [0]{\href }%
\providecommand \doibase [0]{http://dx.doi.org/}%
\providecommand \selectlanguage [0]{\@gobble}%
\providecommand \bibinfo  [0]{\@secondoftwo}%
\providecommand \bibfield  [0]{\@secondoftwo}%
\providecommand \translation [1]{[#1]}%
\providecommand \BibitemOpen [0]{}%
\providecommand \bibitemStop [0]{}%
\providecommand \bibitemNoStop [0]{.\EOS\space}%
\providecommand \EOS [0]{\spacefactor3000\relax}%
\providecommand \BibitemShut  [1]{\csname bibitem#1\endcsname}%
\let\auto@bib@innerbib\@empty
\bibitem [{\citenamefont {Dye}(2003)}]{dye2003electrons}%
  \BibitemOpen
  \bibfield  {author} {\bibinfo {author} {\bibfnamefont {James~L}\ \bibnamefont
  {Dye}},\ }\bibfield  {title} {\enquote {\bibinfo {title} {Electrons as
  anions},}\ }\href {https://www.science.org/doi/10.1126/science.1088103}
  {\bibfield  {journal} {\bibinfo  {journal} {Science}\ }\textbf {\bibinfo
  {volume} {301}},\ \bibinfo {pages} {607--608} (\bibinfo {year}
  {2003})}\BibitemShut {NoStop}%
\bibitem [{\citenamefont {Ma}\ \emph {et~al.}(2009)\citenamefont {Ma},
  \citenamefont {Eremets}, \citenamefont {Oganov}, \citenamefont {Xie},
  \citenamefont {Trojan}, \citenamefont {Medvedev}, \citenamefont {Lyakhov},
  \citenamefont {Valle},\ and\ \citenamefont {Prakapenka}}]{ma2009transparent}%
  \BibitemOpen
  \bibfield  {author} {\bibinfo {author} {\bibfnamefont {Yanming}\ \bibnamefont
  {Ma}}, \bibinfo {author} {\bibfnamefont {Mikhail}\ \bibnamefont {Eremets}},
  \bibinfo {author} {\bibfnamefont {Artem~R}\ \bibnamefont {Oganov}}, \bibinfo
  {author} {\bibfnamefont {Yu}~\bibnamefont {Xie}}, \bibinfo {author}
  {\bibfnamefont {Ivan}\ \bibnamefont {Trojan}}, \bibinfo {author}
  {\bibfnamefont {Sergey}\ \bibnamefont {Medvedev}}, \bibinfo {author}
  {\bibfnamefont {Andriy~O}\ \bibnamefont {Lyakhov}}, \bibinfo {author}
  {\bibfnamefont {Mario}\ \bibnamefont {Valle}}, \ and\ \bibinfo {author}
  {\bibfnamefont {Vitali}\ \bibnamefont {Prakapenka}},\ }\bibfield  {title}
  {\enquote {\bibinfo {title} {Transparent dense sodium},}\ }\href {\doibase
  10.1038/nature07786} {\bibfield  {journal} {\bibinfo  {journal} {Nature}\
  }\textbf {\bibinfo {volume} {458}},\ \bibinfo {pages} {182--185} (\bibinfo
  {year} {2009})}\BibitemShut {NoStop}%
\bibitem [{\citenamefont {Hosono}\ and\ \citenamefont
  {Kitano}(2021)}]{hosono2021advances}%
  \BibitemOpen
  \bibfield  {author} {\bibinfo {author} {\bibfnamefont {Hideo}\ \bibnamefont
  {Hosono}}\ and\ \bibinfo {author} {\bibfnamefont {Masaaki}\ \bibnamefont
  {Kitano}},\ }\bibfield  {title} {\enquote {\bibinfo {title} {Advances in
  materials and applications of inorganic electrides},}\ }\href
  {https://pubs.acs.org/doi/10.1021/acs.chemrev.0c01071} {\bibfield  {journal}
  {\bibinfo  {journal} {Chemical Reviews}\ }\textbf {\bibinfo {volume} {121}},\
  \bibinfo {pages} {3121--3185} (\bibinfo {year} {2021})}\BibitemShut {NoStop}%
\bibitem [{\citenamefont {Arita}\ \emph {et~al.}(2004)\citenamefont {Arita},
  \citenamefont {Miyake}, \citenamefont {Kotani}, \citenamefont {Schilfgaarde},
  \citenamefont {Oka}, \citenamefont {Kuroki}, \citenamefont {Nozue},\ and\
  \citenamefont {Aoki}}]{PhysRevB.69.195106}%
  \BibitemOpen
  \bibfield  {author} {\bibinfo {author} {\bibfnamefont {Ryotaro}\ \bibnamefont
  {Arita}}, \bibinfo {author} {\bibfnamefont {Takashi}\ \bibnamefont {Miyake}},
  \bibinfo {author} {\bibfnamefont {Takao}\ \bibnamefont {Kotani}}, \bibinfo
  {author} {\bibfnamefont {Mark~van}\ \bibnamefont {Schilfgaarde}}, \bibinfo
  {author} {\bibfnamefont {Takashi}\ \bibnamefont {Oka}}, \bibinfo {author}
  {\bibfnamefont {Kazuhiko}\ \bibnamefont {Kuroki}}, \bibinfo {author}
  {\bibfnamefont {Yasuo}\ \bibnamefont {Nozue}}, \ and\ \bibinfo {author}
  {\bibfnamefont {Hideo}\ \bibnamefont {Aoki}},\ }\bibfield  {title} {\enquote
  {\bibinfo {title} {Electronic properties of alkali-metal loaded zeolites:
  Supercrystal mott insulators},}\ }\href {\doibase 10.1103/PhysRevB.69.195106}
  {\bibfield  {journal} {\bibinfo  {journal} {Phys. Rev. B}\ }\textbf {\bibinfo
  {volume} {69}},\ \bibinfo {pages} {195106} (\bibinfo {year}
  {2004})}\BibitemShut {NoStop}%
\bibitem [{\citenamefont {Pickard}\ and\ \citenamefont
  {Needs}(2011)}]{PhysRevLett.107.087201}%
  \BibitemOpen
  \bibfield  {author} {\bibinfo {author} {\bibfnamefont {Chris~J.}\
  \bibnamefont {Pickard}}\ and\ \bibinfo {author} {\bibfnamefont {R.~J.}\
  \bibnamefont {Needs}},\ }\bibfield  {title} {\enquote {\bibinfo {title}
  {Predicted pressure-induced $s$-band ferromagnetism in alkali metals},}\
  }\href {\doibase 10.1103/PhysRevLett.107.087201} {\bibfield  {journal}
  {\bibinfo  {journal} {Phys. Rev. Lett.}\ }\textbf {\bibinfo {volume} {107}},\
  \bibinfo {pages} {087201} (\bibinfo {year} {2011})}\BibitemShut {NoStop}%
\bibitem [{\citenamefont {Lu}\ \emph {et~al.}(2018)\citenamefont {Lu},
  \citenamefont {Wang}, \citenamefont {Li}, \citenamefont {Wu}, \citenamefont
  {Kanno}, \citenamefont {Tada},\ and\ \citenamefont
  {Hosono}}]{PhysRevB.98.125128}%
  \BibitemOpen
  \bibfield  {author} {\bibinfo {author} {\bibfnamefont {Yangfan}\ \bibnamefont
  {Lu}}, \bibinfo {author} {\bibfnamefont {Junjie}\ \bibnamefont {Wang}},
  \bibinfo {author} {\bibfnamefont {Jiang}\ \bibnamefont {Li}}, \bibinfo
  {author} {\bibfnamefont {Jiazhen}\ \bibnamefont {Wu}}, \bibinfo {author}
  {\bibfnamefont {Shu}\ \bibnamefont {Kanno}}, \bibinfo {author} {\bibfnamefont
  {Tomofumi}\ \bibnamefont {Tada}}, \ and\ \bibinfo {author} {\bibfnamefont
  {Hideo}\ \bibnamefont {Hosono}},\ }\bibfield  {title} {\enquote {\bibinfo
  {title} {Realization of mott-insulating electrides in dimorphic
  $\mathrm{Y}{\mathrm{b}}_{5}\mathrm{S}{\mathrm{b}}_{3}$},}\ }\href {\doibase
  10.1103/PhysRevB.98.125128} {\bibfield  {journal} {\bibinfo  {journal} {Phys.
  Rev. B}\ }\textbf {\bibinfo {volume} {98}},\ \bibinfo {pages} {125128}
  (\bibinfo {year} {2018})}\BibitemShut {NoStop}%
\bibitem [{\citenamefont {Sui}\ \emph {et~al.}(2019)\citenamefont {Sui},
  \citenamefont {Wang},\ and\ \citenamefont {Duan}}]{sui2019prediction}%
  \BibitemOpen
  \bibfield  {author} {\bibinfo {author} {\bibfnamefont {Xuelei}\ \bibnamefont
  {Sui}}, \bibinfo {author} {\bibfnamefont {Jianfeng}\ \bibnamefont {Wang}}, \
  and\ \bibinfo {author} {\bibfnamefont {Wenhui}\ \bibnamefont {Duan}},\
  }\bibfield  {title} {\enquote {\bibinfo {title} {Prediction of stoner-type
  magnetism in low-dimensional electrides},}\ }\href {\doibase
  10.1021/acs.jpcc.8b10379} {\bibfield  {journal} {\bibinfo  {journal} {The
  Journal of Physical Chemistry C}\ }\textbf {\bibinfo {volume} {123}},\
  \bibinfo {pages} {5003--5009} (\bibinfo {year} {2019})}\BibitemShut {NoStop}%
\bibitem [{\citenamefont {Zhou}\ \emph {et~al.}(2020)\citenamefont {Zhou},
  \citenamefont {Feng},\ and\ \citenamefont {Shen}}]{PhysRevB.102.180407}%
  \BibitemOpen
  \bibfield  {author} {\bibinfo {author} {\bibfnamefont {Jun}\ \bibnamefont
  {Zhou}}, \bibinfo {author} {\bibfnamefont {Yuan~Ping}\ \bibnamefont {Feng}},
  \ and\ \bibinfo {author} {\bibfnamefont {Lei}\ \bibnamefont {Shen}},\
  }\bibfield  {title} {\enquote {\bibinfo {title} {Atomic-orbital-free
  intrinsic ferromagnetism in electrenes},}\ }\href {\doibase
  10.1103/PhysRevB.102.180407} {\bibfield  {journal} {\bibinfo  {journal}
  {Phys. Rev. B}\ }\textbf {\bibinfo {volume} {102}},\ \bibinfo {pages}
  {180407} (\bibinfo {year} {2020})}\BibitemShut {NoStop}%
\bibitem [{\citenamefont {Novoselov}\ \emph {et~al.}(2021)\citenamefont
  {Novoselov}, \citenamefont {Anisimov},\ and\ \citenamefont
  {Oganov}}]{PhysRevB.103.235126}%
  \BibitemOpen
  \bibfield  {author} {\bibinfo {author} {\bibfnamefont {Dmitry~Y.}\
  \bibnamefont {Novoselov}}, \bibinfo {author} {\bibfnamefont {Vladimir~I.}\
  \bibnamefont {Anisimov}}, \ and\ \bibinfo {author} {\bibfnamefont {Artem~R.}\
  \bibnamefont {Oganov}},\ }\bibfield  {title} {\enquote {\bibinfo {title}
  {Strong electronic correlations in interstitial magnetic centers of
  zero-dimensional electride
  $\ensuremath{\beta}\ensuremath{-}{\mathrm{Yb}}_{5}{\mathrm{Sb}}_{3}$},}\
  }\href {\doibase 10.1103/PhysRevB.103.235126} {\bibfield  {journal} {\bibinfo
   {journal} {Phys. Rev. B}\ }\textbf {\bibinfo {volume} {103}},\ \bibinfo
  {pages} {235126} (\bibinfo {year} {2021})}\BibitemShut {NoStop}%
\bibitem [{\citenamefont {Bradlyn}\ \emph {et~al.}(2017)\citenamefont
  {Bradlyn}, \citenamefont {Elcoro}, \citenamefont {Cano}, \citenamefont
  {Vergniory}, \citenamefont {Wang}, \citenamefont {Felser}, \citenamefont
  {Aroyo},\ and\ \citenamefont {Bernevig}}]{bradlyn2017topological}%
  \BibitemOpen
  \bibfield  {author} {\bibinfo {author} {\bibfnamefont {Barry}\ \bibnamefont
  {Bradlyn}}, \bibinfo {author} {\bibfnamefont {L}~\bibnamefont {Elcoro}},
  \bibinfo {author} {\bibfnamefont {Jennifer}\ \bibnamefont {Cano}}, \bibinfo
  {author} {\bibfnamefont {MG}~\bibnamefont {Vergniory}}, \bibinfo {author}
  {\bibfnamefont {Zhijun}\ \bibnamefont {Wang}}, \bibinfo {author}
  {\bibfnamefont {C}~\bibnamefont {Felser}}, \bibinfo {author} {\bibfnamefont
  {Mois~I}\ \bibnamefont {Aroyo}}, \ and\ \bibinfo {author} {\bibfnamefont
  {B~Andrei}\ \bibnamefont {Bernevig}},\ }\bibfield  {title} {\enquote
  {\bibinfo {title} {Topological quantum chemistry},}\ }\href {\doibase
  10.1038/nature23268} {\bibfield  {journal} {\bibinfo  {journal} {Nature}\
  }\textbf {\bibinfo {volume} {547}},\ \bibinfo {pages} {298--305} (\bibinfo
  {year} {2017})}\BibitemShut {NoStop}%
\bibitem [{\citenamefont {Nie}\ \emph {et~al.}(2021)\citenamefont {Nie},
  \citenamefont {Qian}, \citenamefont {Gao}, \citenamefont {Fang},
  \citenamefont {Weng},\ and\ \citenamefont {Wang}}]{PhysRevB.103.205133}%
  \BibitemOpen
  \bibfield  {author} {\bibinfo {author} {\bibfnamefont {Simin}\ \bibnamefont
  {Nie}}, \bibinfo {author} {\bibfnamefont {Yuting}\ \bibnamefont {Qian}},
  \bibinfo {author} {\bibfnamefont {Jiacheng}\ \bibnamefont {Gao}}, \bibinfo
  {author} {\bibfnamefont {Zhong}\ \bibnamefont {Fang}}, \bibinfo {author}
  {\bibfnamefont {Hongming}\ \bibnamefont {Weng}}, \ and\ \bibinfo {author}
  {\bibfnamefont {Zhijun}\ \bibnamefont {Wang}},\ }\bibfield  {title} {\enquote
  {\bibinfo {title} {Application of topological quantum chemistry in
  electrides},}\ }\href {\doibase 10.1103/PhysRevB.103.205133} {\bibfield
  {journal} {\bibinfo  {journal} {Phys. Rev. B}\ }\textbf {\bibinfo {volume}
  {103}},\ \bibinfo {pages} {205133} (\bibinfo {year} {2021})}\BibitemShut
  {NoStop}%
\bibitem [{\citenamefont {Matsuishi}\ \emph {et~al.}(2003)\citenamefont
  {Matsuishi}, \citenamefont {Toda}, \citenamefont {Miyakawa}, \citenamefont
  {Hayashi}, \citenamefont {Kamiya}, \citenamefont {Hirano}, \citenamefont
  {Tanaka},\ and\ \citenamefont {Hosono}}]{matsuishi2003high}%
  \BibitemOpen
  \bibfield  {author} {\bibinfo {author} {\bibfnamefont {Satoru}\ \bibnamefont
  {Matsuishi}}, \bibinfo {author} {\bibfnamefont {Yoshitake}\ \bibnamefont
  {Toda}}, \bibinfo {author} {\bibfnamefont {Masashi}\ \bibnamefont
  {Miyakawa}}, \bibinfo {author} {\bibfnamefont {Katsuro}\ \bibnamefont
  {Hayashi}}, \bibinfo {author} {\bibfnamefont {Toshio}\ \bibnamefont
  {Kamiya}}, \bibinfo {author} {\bibfnamefont {Masahiro}\ \bibnamefont
  {Hirano}}, \bibinfo {author} {\bibfnamefont {Isao}\ \bibnamefont {Tanaka}}, \
  and\ \bibinfo {author} {\bibfnamefont {Hideo}\ \bibnamefont {Hosono}},\
  }\bibfield  {title} {\enquote {\bibinfo {title} {High-density electron anions
  in a nanoporous single crystal:[Ca$_{24}$Al$_{28}$O$_{64}$]$^{4+}$(4e$^-$)},}\ }\href {\doibase
  10.1126/science.1083842} {\bibfield  {journal} {\bibinfo  {journal}
  {Science}\ }\textbf {\bibinfo {volume} {301}},\ \bibinfo {pages} {626--629}
  (\bibinfo {year} {2003})}\BibitemShut {NoStop}%
\bibitem [{\citenamefont {Lee}\ \emph {et~al.}(2013)\citenamefont {Lee},
  \citenamefont {Kim}, \citenamefont {Toda}, \citenamefont {Matsuishi},\ and\
  \citenamefont {Hosono}}]{lee2013dicalcium}%
  \BibitemOpen
  \bibfield  {author} {\bibinfo {author} {\bibfnamefont {Kimoon}\ \bibnamefont
  {Lee}}, \bibinfo {author} {\bibfnamefont {Sung~Wng}\ \bibnamefont {Kim}},
  \bibinfo {author} {\bibfnamefont {Yoshitake}\ \bibnamefont {Toda}}, \bibinfo
  {author} {\bibfnamefont {Satoru}\ \bibnamefont {Matsuishi}}, \ and\ \bibinfo
  {author} {\bibfnamefont {Hideo}\ \bibnamefont {Hosono}},\ }\bibfield  {title}
  {\enquote {\bibinfo {title} {Dicalcium nitride as a two-dimensional electride
  with an anionic electron layer},}\ }\href {\doibase 10.1038/nature11812}
  {\bibfield  {journal} {\bibinfo  {journal} {Nature}\ }\textbf {\bibinfo
  {volume} {494}},\ \bibinfo {pages} {336--340} (\bibinfo {year}
  {2013})}\BibitemShut {NoStop}%
\bibitem [{\citenamefont {Zhao}\ \emph {et~al.}(2016)\citenamefont {Zhao},
  \citenamefont {Kan},\ and\ \citenamefont {Li}}]{zhao2016electride}%
  \BibitemOpen
  \bibfield  {author} {\bibinfo {author} {\bibfnamefont {Songtao}\ \bibnamefont
  {Zhao}}, \bibinfo {author} {\bibfnamefont {Erjun}\ \bibnamefont {Kan}}, \
  and\ \bibinfo {author} {\bibfnamefont {Zhenyu}\ \bibnamefont {Li}},\
  }\bibfield  {title} {\enquote {\bibinfo {title} {Electride: from
  computational characterization to theoretical design},}\ }\href {\doibase
  10.1002/wcms.1258} {\bibfield  {journal} {\bibinfo  {journal} {WIREs
  Computational Molecular Science}\ }\textbf {\bibinfo {volume} {6}},\ \bibinfo
  {pages} {430--440} (\bibinfo {year} {2016})}\BibitemShut {NoStop}%
\bibitem [{\citenamefont {Kitaigorodsky}(2012)}]{kitaigorodsky2012molecular}%
  \BibitemOpen
  \bibfield  {author} {\bibinfo {author} {\bibfnamefont {Alexander I.}~\bibnamefont
  {Kitaigorodsky}},\ }\href@noop {} {\enquote {\bibinfo {title} {Molecular
  crystals and molecules}}},\ Vol.~\bibinfo {volume} {29}\ (\bibinfo
  {publisher} {Elsevier},\ \bibinfo {year} {2012})\BibitemShut {NoStop}%
\bibitem [{\citenamefont {Becke}\ and\ \citenamefont
  {Edgecombe}(1990)}]{becke1990simple}%
  \BibitemOpen
  \bibfield  {author} {\bibinfo {author} {\bibfnamefont {A.~D.}\ \bibnamefont
  {Becke}}\ and\ \bibinfo {author} {\bibfnamefont {K.~E.}\ \bibnamefont
  {Edgecombe}},\ }\bibfield  {title} {\enquote {\bibinfo {title} {A simple
  measure of electron localization in atomic and molecular systems},}\ }\href
  {\doibase 10.1063/1.458517} {\bibfield  {journal} {\bibinfo  {journal} {The
  Journal of Chemical Physics}\ }\textbf {\bibinfo {volume} {92}},\ \bibinfo
  {pages} {5397--5403} (\bibinfo {year} {1990})}\BibitemShut {NoStop}%
\bibitem [{\citenamefont {Savin}(2005)}]{savin2005electron}%
  \BibitemOpen
  \bibfield  {author} {\bibinfo {author} {\bibfnamefont {Andreas}\ \bibnamefont
  {Savin}},\ }\bibfield  {title} {\enquote {\bibinfo {title} {The electron
  localization function (elf) and its relatives: interpretations and
  difficulties},}\ }\href {\doibase
  https://doi.org/10.1016/j.theochem.2005.02.034} {\bibfield  {journal}
  {\bibinfo  {journal} {Journal of Molecular Structure: THEOCHEM}\ }\textbf
  {\bibinfo {volume} {727}},\ \bibinfo {pages} {127 -- 131} (\bibinfo {year}
  {2005})}\BibitemShut {NoStop}%
\bibitem [{\citenamefont {Silvi}\ and\ \citenamefont
  {Savin}(1994)}]{silvi1994classification}%
  \BibitemOpen
  \bibfield  {author} {\bibinfo {author} {\bibfnamefont {Bernard}\ \bibnamefont
  {Silvi}}\ and\ \bibinfo {author} {\bibfnamefont {Andreas}\ \bibnamefont
  {Savin}},\ }\bibfield  {title} {\enquote {\bibinfo {title} {Classification of
  chemical bonds based on topological analysis of electron localization
  functions},}\ }\href {\doibase 10.1038/371683a0} {\bibfield  {journal}
  {\bibinfo  {journal} {Nature}\ }\textbf {\bibinfo {volume} {371}},\ \bibinfo
  {pages} {683--686} (\bibinfo {year} {1994})}\BibitemShut {NoStop}%
\bibitem [{\citenamefont {Liu}\ and\ \citenamefont
  {Zunger}(2017)}]{PhysRevX.7.021019}%
  \BibitemOpen
  \bibfield  {author} {\bibinfo {author} {\bibfnamefont {Qihang}\ \bibnamefont
  {Liu}}\ and\ \bibinfo {author} {\bibfnamefont {Alex}\ \bibnamefont
  {Zunger}},\ }\bibfield  {title} {\enquote {\bibinfo {title} {Predicted
  realization of cubic dirac fermion in quasi-one-dimensional transition-metal
  monochalcogenides},}\ }\href {\doibase 10.1103/PhysRevX.7.021019} {\bibfield
  {journal} {\bibinfo  {journal} {Phys. Rev. X}\ }\textbf {\bibinfo {volume}
  {7}},\ \bibinfo {pages} {021019} (\bibinfo {year} {2017})}\BibitemShut
  {NoStop}%
\bibitem [{\citenamefont {Noguchi}\ \emph {et~al.}(2019)\citenamefont
  {Noguchi}, \citenamefont {Takahashi}, \citenamefont {Kuroda}, \citenamefont
  {Ochi}, \citenamefont {Shirasawa}, \citenamefont {Sakano}, \citenamefont
  {Bareille}, \citenamefont {Nakayama}, \citenamefont {Watson}, \citenamefont
  {Yaji} \emph {et~al.}}]{noguchi2019weak}%
  \BibitemOpen
  \bibfield  {author} {\bibinfo {author} {\bibfnamefont {Ryo}\ \bibnamefont
  {Noguchi}}, \bibinfo {author} {\bibfnamefont {T}~\bibnamefont {Takahashi}},
  \bibinfo {author} {\bibfnamefont {K}~\bibnamefont {Kuroda}}, \bibinfo
  {author} {\bibfnamefont {M}~\bibnamefont {Ochi}}, \bibinfo {author}
  {\bibfnamefont {T}~\bibnamefont {Shirasawa}}, \bibinfo {author}
  {\bibfnamefont {M}~\bibnamefont {Sakano}}, \bibinfo {author} {\bibfnamefont
  {C}~\bibnamefont {Bareille}}, \bibinfo {author} {\bibfnamefont
  {M}~\bibnamefont {Nakayama}}, \bibinfo {author} {\bibfnamefont
  {MD}~\bibnamefont {Watson}}, \bibinfo {author} {\bibfnamefont
  {K}~\bibnamefont {Yaji}},  \emph {et~al.},\ }\bibfield  {title} {\enquote
  {\bibinfo {title} {A weak topological insulator state in
  quasi-one-dimensional bismuth iodide},}\ }\href {\doibase
  10.1038/s41586-019-0927-7} {\bibfield  {journal} {\bibinfo  {journal}
  {Nature}\ }\textbf {\bibinfo {volume} {566}},\ \bibinfo {pages} {518--522}
  (\bibinfo {year} {2019})}\BibitemShut {NoStop}%
\bibitem [{\citenamefont {Huang}\ \emph {et~al.}(2021)\citenamefont {Huang},
  \citenamefont {Li}, \citenamefont {Yoon}, \citenamefont {Oh}, \citenamefont
  {Wu}, \citenamefont {Liu}, \citenamefont {Dhale}, \citenamefont {Zhou},
  \citenamefont {Guo}, \citenamefont {Zhang}, \citenamefont {Hashimoto},
  \citenamefont {Lu}, \citenamefont {Denlinger}, \citenamefont {Wang},
  \citenamefont {Lau}, \citenamefont {Birgeneau}, \citenamefont {Zhang},
  \citenamefont {Lv},\ and\ \citenamefont {Yi}}]{PhysRevX.11.031042}%
  \BibitemOpen
  \bibfield  {author} {\bibinfo {author} {\bibfnamefont {Jianwei}\ \bibnamefont
  {Huang}}, \bibinfo {author} {\bibfnamefont {Sheng}\ \bibnamefont {Li}},
  \bibinfo {author} {\bibfnamefont {Chiho}\ \bibnamefont {Yoon}}, \bibinfo
  {author} {\bibfnamefont {Ji~Seop}\ \bibnamefont {Oh}}, \bibinfo {author}
  {\bibfnamefont {Han}\ \bibnamefont {Wu}}, \bibinfo {author} {\bibfnamefont
  {Xiaoyuan}\ \bibnamefont {Liu}}, \bibinfo {author} {\bibfnamefont {Nikhil}\
  \bibnamefont {Dhale}}, \bibinfo {author} {\bibfnamefont {Yan-Feng}\
  \bibnamefont {Zhou}}, \bibinfo {author} {\bibfnamefont {Yucheng}\
  \bibnamefont {Guo}}, \bibinfo {author} {\bibfnamefont {Yichen}\ \bibnamefont
  {Zhang}}, \bibinfo {author} {\bibfnamefont {Makoto}\ \bibnamefont
  {Hashimoto}}, \bibinfo {author} {\bibfnamefont {Donghui}\ \bibnamefont {Lu}},
  \bibinfo {author} {\bibfnamefont {Jonathan}\ \bibnamefont {Denlinger}},
  \bibinfo {author} {\bibfnamefont {Xiqu}\ \bibnamefont {Wang}}, \bibinfo
  {author} {\bibfnamefont {Chun~Ning}\ \bibnamefont {Lau}}, \bibinfo {author}
  {\bibfnamefont {Robert~J.}\ \bibnamefont {Birgeneau}}, \bibinfo {author}
  {\bibfnamefont {Fan}\ \bibnamefont {Zhang}}, \bibinfo {author} {\bibfnamefont
  {Bing}\ \bibnamefont {Lv}}, \ and\ \bibinfo {author} {\bibfnamefont {Ming}\
  \bibnamefont {Yi}},\ }\bibfield  {title} {\enquote {\bibinfo {title}
  {Room-temperature topological phase transition in quasi-one-dimensional
  material ${\mathrm{b}\mathrm{i}}_{\mathrm{4}}{\mathrm{i}}_{\mathrm{4}}$},}\
  }\href {\doibase 10.1103/PhysRevX.11.031042} {\bibfield  {journal} {\bibinfo
  {journal} {Phys. Rev. X}\ }\textbf {\bibinfo {volume} {11}},\ \bibinfo
  {pages} {031042} (\bibinfo {year} {2021})}\BibitemShut {NoStop}%
\bibitem [{\citenamefont {Shi}\ \emph {et~al.}(2021)\citenamefont {Shi},
  \citenamefont {Wieder}, \citenamefont {Meyerheim}, \citenamefont {Sun},
  \citenamefont {Zhang}, \citenamefont {Li}, \citenamefont {Shen},
  \citenamefont {Qi}, \citenamefont {Yang}, \citenamefont {Jena} \emph
  {et~al.}}]{shi2021charge}%
  \BibitemOpen
  \bibfield  {author} {\bibinfo {author} {\bibfnamefont {Wujun}\ \bibnamefont
  {Shi}}, \bibinfo {author} {\bibfnamefont {Benjamin~J}\ \bibnamefont
  {Wieder}}, \bibinfo {author} {\bibfnamefont {Holger~L}\ \bibnamefont
  {Meyerheim}}, \bibinfo {author} {\bibfnamefont {Yan}\ \bibnamefont {Sun}},
  \bibinfo {author} {\bibfnamefont {Yang}\ \bibnamefont {Zhang}}, \bibinfo
  {author} {\bibfnamefont {Yiwei}\ \bibnamefont {Li}}, \bibinfo {author}
  {\bibfnamefont {Lei}\ \bibnamefont {Shen}}, \bibinfo {author} {\bibfnamefont
  {Yanpeng}\ \bibnamefont {Qi}}, \bibinfo {author} {\bibfnamefont {Lexian}\
  \bibnamefont {Yang}}, \bibinfo {author} {\bibfnamefont {Jagannath}\
  \bibnamefont {Jena}},  et~al.,\ }\bibfield  {title} {\enquote
  {\bibinfo {title} {A charge-density-wave topological semimetal},}\ }\href
  {\doibase 10.1038/s41567-020-01104-z} {\bibfield  {journal} {\bibinfo
  {journal} {Nature Physics}\ }\textbf {\bibinfo {volume} {17}},\ \bibinfo
  {pages} {381--387} (\bibinfo {year} {2021})}\BibitemShut {NoStop}%
\bibitem [{\citenamefont {Zhang}\ et~al. (2019)\citenamefont {Zhang},
  \citenamefont {Jiang}, \citenamefont {Song}, \citenamefont {Huang},
  \citenamefont {He}, \citenamefont {Fang}, \citenamefont {Weng},\ and\
  \citenamefont {Fang}}]{zhang2019catalogue}%
  \BibitemOpen
  \bibfield  {author} {\bibinfo {author} {\bibfnamefont {Tiantian}\
  \bibnamefont {Zhang}}, \bibinfo {author} {\bibfnamefont {Yi}~\bibnamefont
  {Jiang}}, \bibinfo {author} {\bibfnamefont {Zhida}\ \bibnamefont {Song}},
  \bibinfo {author} {\bibfnamefont {He}~\bibnamefont {Huang}}, \bibinfo
  {author} {\bibfnamefont {Yuqing}\ \bibnamefont {He}}, \bibinfo {author}
  {\bibfnamefont {Zhong}\ \bibnamefont {Fang}}, \bibinfo {author}
  {\bibfnamefont {Hongming}\ \bibnamefont {Weng}}, \ and\ \bibinfo {author}
  {\bibfnamefont {Chen}\ \bibnamefont {Fang}},\ }\bibfield  {title} {\enquote
  {\bibinfo {title} {Catalogue of topological electronic materials},}\ }\href
  {\doibase 10.1038/s41586-019-0944-6} {\bibfield  {journal} {\bibinfo
  {journal} {Nature}\ }\textbf {\bibinfo {volume} {566}},\ \bibinfo {pages}
  {475--479} (\bibinfo {year} {2019})}\BibitemShut {NoStop}%
\bibitem [{\citenamefont {Tasker}(1979)}]{tasker1979stability}%
  \BibitemOpen
  \bibfield  {author} {\bibinfo {author} {\bibfnamefont {P~W}\ \bibnamefont
  {Tasker}},\ }\bibfield  {title} {\enquote {\bibinfo {title} {The stability of
  ionic crystal surfaces},}\ }\href {\doibase 10.1088/0022-3719/12/22/036}
  {\bibfield  {journal} {\bibinfo  {journal} {Journal of Physics C: Solid State
  Physics}\ }\textbf {\bibinfo {volume} {12}},\ \bibinfo {pages} {4977--4984}
  (\bibinfo {year} {1979})}\BibitemShut {NoStop}%
\bibitem [{\citenamefont {Otani}\ and\ \citenamefont
  {Sugino}(2006)}]{PhysRevB.73.115407}%
  \BibitemOpen
  \bibfield  {author} {\bibinfo {author} {\bibfnamefont {M.}~\bibnamefont
  {Otani}}\ and\ \bibinfo {author} {\bibfnamefont {O.}~\bibnamefont {Sugino}},\
  }\bibfield  {title} {\enquote {\bibinfo {title} {First-principles
  calculations of charged surfaces and interfaces: A plane-wave nonrepeated
  slab approach},}\ }\href {\doibase 10.1103/PhysRevB.73.115407} {\bibfield
  {journal} {\bibinfo  {journal} {Phys. Rev. B}\ }\textbf {\bibinfo {volume}
  {73}},\ \bibinfo {pages} {115407} (\bibinfo {year} {2006})}\BibitemShut
  {NoStop}%
\end{thebibliography}
\onecolumngrid
\section*{List of references}

\end{bibunit}
\end{document}